\documentclass{aa}

\usepackage{graphicx}
\usepackage{txfonts}
\usepackage{csquotes}
\usepackage{hyperref}
\usepackage{threeparttable}

\begin{document}

\title{XUE: The CO$_2$-rich terrestrial planet-forming region\\ of an externally irradiated Herbig disk}

\author{Jenny Frediani
          \inst{1} 
        \and
        Arjan Bik
        \inst{1} 
        \and
        María Claudia Ramírez-Tannus
        \inst{2}
        \and
        Rens Waters\inst{3}
        \and
        Konstantin V. Getman\inst{4}
        \and
        Eric D. Feigelson\inst{4,5}
        \and
        Bayron Portilla-Revelo\inst{4,5}
        \and
        Benoît Tabone\inst{6}
        \and
        Thomas J. Haworth\inst{7}
        \and
        Andrew Winter\inst{2}
        \and
        Thomas Henning\inst{2}
        \and
        Giulia Perotti\inst{2}
        \and
        Alexis Brandeker\inst{1}
        \and
        Germán Chaparro\inst{8}
        \and
        Pablo Cuartas-Restrepo\inst{8}
        \and
        Sebastián Hernández\inst{8}
        \and
        Michael A. Kuhn\inst{9}
        \and
        Thomas Preibisch\inst{10}
        \and
        Veronica Roccatagliata\inst{11,12}
        \and
        Sierk E. van Terwisga\inst{13}
        \and
        Peter Zeidler\inst{14}
}  
   \institute{Department of Astronomy, Stockholm University, AlbaNova University Centre, 106 91 Stockholm, Sweden
   \and
   Max-Planck-Institut für Astronomie,Königstuhl 17, D-69117 Heidelberg, Germany
   \and
   Department of Astrophysics/IMAPP, Radboud University, PO Box 9010, 6500 GL Nijmegen, The Netherlands
   \and
   Department of Astronomy \& Astrophysics, Pennsylvania State University, 525 Davey Laboratory, University Park, PA 16802, USA
   \and
   Center for Exoplanets and Habitable Worlds, Pennsylvania State University, 525 Davey Laboratory, University Park, PA 16802, USA
   \and
   Institut d’Astrophysique Spatiale, Université Paris-Saclay, CNRS, Bâtiment 121, F-91405 Orsay Cedex, France
   \and
   Astronomy Unit, School of Physics and Astronomy, Queen Mary University of London, London E1 4NS, UK
   \and
   FACom, Instituto de Física—FCEN, Universidad de Antioquia, Calle 70 No. 52-21, Medellín 050010, Colombia
   \and
   Centre for Astrophysics Research, Department of Physics, Astronomy, and Mathematics, University of Hertfordshire, Hatfield, AL10 9UW, UK
   \and
   Universit\"ats-Sternwarte M\"unchen, 
Ludwig-Maximilians-Universit\"at, Scheinerstr.~1, 81679 M\"unchen, Germany
\and
Alma Mater Studiorum, Università di Bologna, Dipartimento di Fisica e Astronomia (DIFA), Via Gobetti 93/2, 40129 Bologna, Italy
\and
 INAF-Osservatorio Astrofisico di Arcetri, Largo E. Fermi 5, 50125 Firenze, Italy
 \and
 Space Research Institute, Austrian Academy of Sciences, Schmiedlstr. 6, 8042 Graz, Austria
 \and
 AURA for the European Space Agency, Space Telescope Science Institute, Baltimore, MD, USA
}

\titlerunning{XUE 10: The CO$_2$-rich terrestrial planet-forming region of an externally irradiated Herbig disk}
\authorrunning{Frediani, Bik et al.}

\abstract 
{}
{We investigate the \textit{James Webb Space Telescope} (JWST) MIRI MRS gas molecular content of an externally irradiated Herbig disk, the F-type XUE 10 source, in the context of the eXtreme UV Environments (XUE) program. XUE 10 belongs to the massive star cluster NGC 6357 (1.69 kpc), where it is exposed to an external far-ultraviolet (FUV) radiation $\approx$ 10$^3$ times stronger than in the solar neighborhood.}
{We modeled the molecular features in the mid-infrared spectrum
with local thermodynamic equilibrium (LTE) 0D slab models. We derived basic parameters of the stellar host from a VLT FORS2 optical spectrum using PHOENIX stellar templates.}
{We detected bright CO$_2$ gas with the first simultaneous detection ($>$ 5$\sigma$) of four isotopologues ($^{12}$CO$_2$, $^{13}$CO$_2$, $^{16}$O$^{12}$C$^{18}$O, $^{16}$O$^{12}$C$^{17}$O) in a protoplanetary disk. We also detected faint CO emission (2$\sigma$) and the HI Pf$\alpha$ line (8$\sigma$). We placed strict upper limits on the water content, finding a total column density of $\lesssim$ 10$^{18}$ cm$^{-2}$.  
The CO$_2$ species trace low gas temperatures (300--370 K) with a range of column densities of 7.4 $\times$ 10$^{17}$ cm$^{-2}$ ($^{16}$O$^{12}$C$^{17}$O)--1.3 $\times$ 10$^{20}$ cm$^{-2}$ ($^{12}$CO$_2$) in an equivalent emitting radius of 1.15 au.
The emission of $^{13}$CO$_2$ is likely affected by line optical depth effects.
The $^{16}$O$^{12}$C$^{18}$O and $^{16}$O$^{12}$C$^{17}$O abundances may be isotopically anomalous compared to the $^{16}$O/$^{18}$O and $^{16}$O/$^{17}$O ratios measured in the interstellar medium and the Solar System.}
{We propose that the mid-infrared spectrum of XUE 10 is explained by H$_2$O removal either via advection or strong photo-dissociation by stellar UV irradiation and enhanced local CO$_2$ gas phase production. Outer disk truncation supports the observed CO$_2$--H$_2$O dichotomy. A CO$_2$ vapor enrichment in $^{18}$O and $^{17}$O can be explained by means of external UV irradiation and early (10$^{4-5}$ yr) delivery of isotopically anomalous water ice to the inner disk.}


\maketitle

\section{Introduction} \label{sec:introduction}
\begin{figure*}[htp!]
    \centering
    \includegraphics[width=\linewidth]{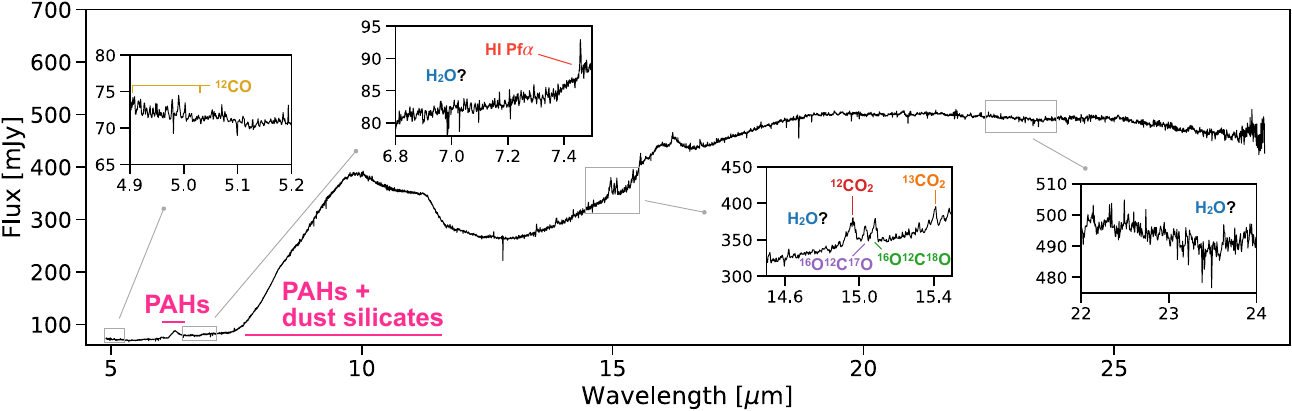}
    \caption{Full JWST/MIRI MRS extinction-corrected spectrum of XUE 10. The identified dust and gas phase polycyclic aromatic hydrocarbon (PAH) features are labeled and highlighted with a solid line. The insets zoom into various wavelength regions with the detection of carbon monoxide ($^{12}$CO), four carbon dioxide isotopic species ($^{12}$CO$_{2}$, $^{13}$CO$_{2}$, $^{16}$O$^{12}$C$^{18}$O, $^{16}$O$^{12}$C$^{17}$O), atomic hydrogen (HI Pf$\alpha$ line), and the tentative detection of water (H$_{2}$O). }
    \label{fig:MIRI_spec}
\end{figure*}
The discovery of more than 5000 exo-planets in our Galaxy, a number that steadily increases,\footnote{\url{1 https://exoplanetarchive.ipac.caltech.edu/}} and the growing evidence for several classes of them (e.g., super-Earths, hot Jupiters) differing from the Solar System architecture demand a deep understanding of the properties of protoplanetary disks and their connection to the observed exoplanet demography \citep{winn_2015, Mordasini_2012, dawson_2018}. Our knowledge on the formation of planetary systems is mostly based on surveys of protoplanetary disks located in nearby ($D \lesssim$ 400 pc), relatively isolated star-forming regions \citep[e.g., Taurus;][]{Manara_2023}, where it is possible to achieve the high angular and spectral resolution coupled with high sensitivity needed to study the gas and the dust components of disks in detail as well as their physical structure. 

Nevertheless, observations \citep[e.g.,][]{Stolte_2010, Guarcello_2016} and theory \citep{Fatuzzo_2008, Winter_2020} have come to agree that the majority of stars and planetary systems actually form in massive clusters \citep[e.g.,][]{Lada_2003, krumholz_2019}. In these star-forming regions, the protoplanetary disks are exposed to the external far-ultraviolet (FUV) radiation of massive young stars of spectral type O and B present in the surroundings \citep{winter_2022, Haworth_2025}. The typical FUV flux (measured between 912 $\AA$ and 2400 $\AA$) of irradiated protoplanetary disks is in the range of $\approx 10^{3-5} \, G_0$ \citep{winter_2022}, where $1\, G_0$ $\approx$ 1.6 $\times$ 10$^{-3}$ erg s$^{-1}$ cm$^{-2}$ is the average value measured in the solar neighborhood \citep{Habing_1968}. The external photoevaporation induced by FUV photons is thought to be the dominant dispersal mechanism of protoplanetary disks in massive star-forming regions \citep{Scally_2001, Guarcello_2016, Winter_2018}. This process drives significant disk mass-loss rates \citep{Adams_2004, Facchini_2016, 2019MNRAS.485.3895H, Haworth_2018,  2023MNRAS.526.4315H}, up to about 10$^{-6}$ $M$$_{\odot}$ yr$^{-1}$ in the Orion Nebula Cluster \citep[ONC;][]{Johnstone_1998, Henney_1999, Ballering_2023, Aru_2024}, reducing the gas mass, shortening its dissipation timescale, and resulting in a rapid outer disk dispersal \citep{Adams_2004, Concha_2019, Winter_2020, Gavin_2022}. Depending on the disk structure, consequent steep surface density gradients in the outer disc may result in rapid inward drift of large dust grains, while small dust grains may be depleted by entrainment in the photoevaporative wind \citep{Facchini_2016, Sellek_2020, Qiao_2023, Garate_2024, Paine_2025}. Gas and dust depletion may alter the assembly of both rocky and gaseous types of planets, although it might also instigate planet formation by streaming instability, enhancing the local dust-to-gas ratio in the outer disk \citep{Throop_2005}.

There is compelling evidence for outer disk depletion by external photoevaporation in Orion \citep[e.g.,][]{Mann_2014, Ansdell_2017, vanTerwisga_2019, vanTerwisga_2020} and in other close by massive clusters \citep{Stolte_2010, Roccatagliata_2011, Guarcello_2016, Richert_2018}. Theoretical works have illustrated how chemistry is expected to be affected in the outer and inner disk regions \citep{Walsh_2013, Ndugu_2024, Keyte_2025, Gross_2025}.
We identify the inner disk as the region extending out to 10 au from the central pre-main-sequence star (PMS). Given the typical gas ($\sim$ 100--6000 K) and dust temperatures ($\sim$ 200--1500 K) of the inner disk \citep{Kamp_2004}, near- to mid-infrared (IR) observations are needed to probe this region.
Ground-based near-IR, the \textit{Infrared Space Observatory}, the \textit{Spitzer Space Telescope}, and now JWST spectroscopic surveys of nearby T Tauri disks \citep[see, e.g.,][]{Salyk_2008, Pontoppidan_2010, Pascucci_2013, Dishoeck_2023, Henning_2024, Arulanantham_2025} have revealed a plethora of ro-vibrational and pure rotational emission lines of several molecular species, including water, carbon monoxide (CO), carbon dioxide (CO$_2$), hydroxide (OH), and molecular hydrogen, which are all connected to the physical structure of a protoplanetary disk, its evolution, and the mass budget for forming planets.  
Thanks to its superior sensitivity and spatial resolution compared to \textit{Spitzer}, JWST makes it possible to extend this kind of spectroscopic study to disk populations located at distances $D$ $\gtrsim$ 1.5 kpc, therefore including a large sample of typical massive star-forming regions. In the ONC, JWST observations toward the low-mass d203-506 proplyd, i.e., a cometary-shaped photo-evaporating disk, reported the first ever detection of the methyl cation molecule \citep[CH$_{3}$$^{+}$;][]{Berne_2023, Zannese_2025}, which was later discovered in the non-irradiated T Tauri TW Hya disk \citep{Henning_2024}. In the same proplyd, evidence was also found for the reprocessing of water into OH by means of strong UV photo-dissociation at the disk surface \citep{Zannese_2024}. 
However, it is still unclear how the external photoevaporation influences the inner disk from an evolutionary and observational perspective. 

The eXtreme UV Environments (XUE) program was proposed in order to characterize the physical and chemical properties of a sample of 12 externally irradiated protoplanetary disks in the NGC 6357 massive cluster, taking advantage of pivotal mid-IR JWST Cycle 1 observations obtained with the Medium Resolution Spectrometer (MRS) of the Mid-InfraRed Instrument \citep[MIRI;][]{Ramirez-Tannus_2021}. 
NGC 6357 represents one of the youngest (0.8--1.4 Myr) and closest (1.69 kpc) massive clusters now accessible with JWST \citep{Fang_2012, Getman_2014, Ramirez_2020, Ramirez-Tannus_2023}, hosting more than 20 O stars and one of the most massive stars in our Galaxy: Pis24 1 \citep[$M$$_{\star}$ $\approx$ 74 $M$$_{\odot}$, O4III(f+)+O3.5If*;][]{Massey_2001, Walborn_2002}. NGC 6357 comprises three coeval subclusters (Pismis 24, G353.1+0.6, and G353.2+0.7) located at the same distance, but it is possible to probe different ranges of FUV field strengths, from approximately 10$^3$ $G$$_0$ (in G353.2+0.7) to  10$^6$ $G$$_0$ (in Pismis 24). \citet{Ramirez-Tannus_2023} presented the first results on the XUE 1 solar analog, which retains similar inner disk properties to nearby non-irradiated T Tauri. \citet{Bayron_2025} recently validated these findings with a thermo-chemical disk model and found evidence for a truncated disk with outer gas depletion. \citet{Ramirez-Tannus_2025} recently reviewed the general properties of the XUE sources with respect to other samples, while a forthcoming work will present radiation thermo-chemical simulations tailored to the UV field in NGC 6357 (Hernández et al., in prep.). 

Except for XUE 1, the XUE sample is constituted by disks surrounding intermediate mass T Tauri (IMTTs) stars. These are characterized by higher stellar masses than T~Tauri sources ($M$$_{\star}$ $\sim$ 1.5--10 $M$$_{\odot}$), with a late spectral type F and G, and they will eventually become Herbig A stars \citep{Valegard_2021, Brittain_2023}. Herbig stars have an earlier spectral type, A and B, and they represent the prototypical formation sites of giant planets \citep{Johnson_2010, Reffert_2015}. This work is dedicated to the spectral analysis of the XUE 10 disk around a early F-type IMTT star with an effective temperature of about 7000 K and a stellar mass of $\approx$ 3 $M_{\odot}$. This makes XUE 10 a borderline case between an IMTT and an Herbig star, as shown by \citet{Valegard_2021}. We classify XUE 10 as a disk surrounding an Herbig star. The source has coordinates R.A. (J2000) = 17$^{\rm h}$25$^{\rm m}$55$^{\rm s}$.274 and Dec. (J2000) = --34$^{\circ}$15'47''.76 and is located in the G353.2+0.7 subcluster of NGC 6357 with an estimated FUV exposure of 5.2 $\times$ 10$^{3}$ G$_0$ \citep{Ramirez-Tannus_2025}. 

The remainder of the manuscript is organized as follows. The data used for the analysis in this work and the adopted data reduction are presented in Section \ref{sec:observations}. In Section \ref{sec:results} we present the results on the gas molecular chemistry and the central star.  
In Section \ref{subsec:gas} we determine the main physical parameters of gas phase molecules by means of 0D local thermodynamic equilibrium (LTE) slab models. In Section \ref{sec:discussion} we place our findings in context with the available disk samples with similar infrared characterization, and we discuss different scenarios to explain the observed CO$_2$-rich mid-IR spectrum of XUE 10. The conclusions are summarized in Section \ref{sec:conclusions}.
\begin{figure*}[htp!]
    \centering
    \includegraphics[width=0.8\linewidth]{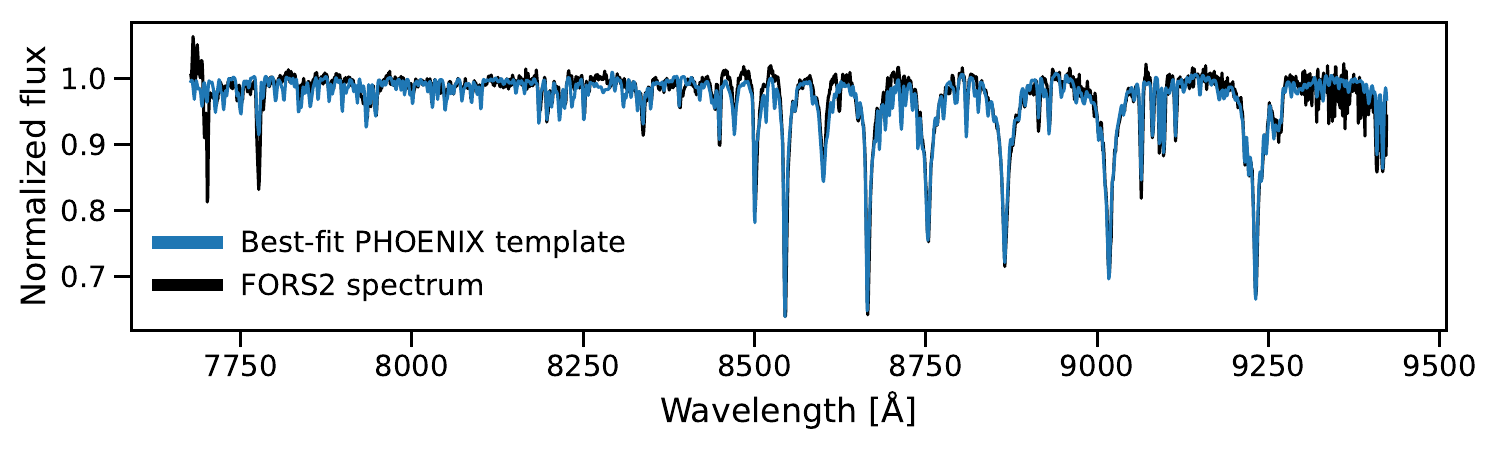}
    \caption{Chi-square best-fit PHOENIX template (blue) overlaid on the telluric-corrected observed optical spectrum (black) after normalization.}
    \label{fig:FORS_spec}
\end{figure*}
\section{Observations and data reduction}\label{sec:observations}
\subsection{JWST/MIRI MRS IFU spectroscopy}\label{subsec:JWSTobs}
XUE 10 was observed with the MIRI MRS instrument \citep{Rieke_2015, Wells_2015, Wright_2015} on the 13 September 2023 as part of the XUE project in Cycle 1 \citep[GO-1759;][]{Ramirez-Tannus_2021}. 
All three MIRI wavelength settings (SHORT, MEDIUM, and LONG) were used for the observations. A four-point dither optimized for a point source was performed in the negative direction. The observations were obtained in the FASTR1 readout mode with a total exposure time of 2700 seconds. No dedicated background observations were taken, as the background nebula in the observations is bright and highly spatially variable. 

We processed the XUE 10 data using the JWST Science Calibration Pipeline (v. 1.14.0), with context 1225 of the Calibration Reference Data System \citep[CRDS; ][]{Bushouse_2023}. We run the \texttt{Detector1} step without major modifications. For \texttt{Spec2} we used the background subtraction procedure described in \citet{Ramirez-Tannus_2023}. Apart from that, we run it with residual fringe correction and bad pixel removal. We run \texttt{Spec3} with the 1D residual fringe correction option, and combined the cubes using the \texttt{drizzle} algorithm. We extracted the source with a modified aperture linearly increasing from 0$\farcs$3 at 5 $\mu$m to 1.28\arcsec\ at 22 $\mu$m, and applied aperture corrections corresponding to these extraction apertures. The final reduced spectrum between 4.9--28 $\mu$m is shown in Figure~\ref{fig:MIRI_spec}. We measured the noise ($\sigma_{\rm RMS}$) in line free wavelength ranges across the spectrum (see Figure~\ref{fig:continuum_normalization}). The 1$\sigma$ noise level of the spectral ranges from 0.7 mJy at the short wavelength part of the spectrum, to 3.3 mJy at the reddest wavelengths.
\subsection{VLT/FORS2 slit spectroscopy}\label{subsec:FORS2obs}
The FORS2 instrument at the VLT, working in the 3300--11000 $\AA$ wavelength range, was used in Long-Slit (LSS) spectroscopic mode to observe XUE 10 on the 20 April 2024 as part of program 113.26CR (P.I.: A.~Bik). The observations were obtained with an airmass of 1.06 using a 0$\farcs$7 wide slit and the Grism 1028z (GRIS\_1028z+29), which observes at 7730--9480 $\AA$ with a resolving power $R$ $\sim$ 2560.

The FORS2 spectrum of XUE 10 was reduced using \texttt{PipeIt} \citep[v. 1.15.0;][]{pypeit:joss_arXiv}, a Python package for semi-automated reduction of astronomical spectroscopic data. The reduction consists of a basic calibration including a bias correction, flat fielding and a wavelength calibration using He\,I, Ar\,I+Ar\,II, and Ne\,I+Ne\,II arc lamps. The observations were taken in nodding mode, with 4 separate integrations of 450 seconds each, where the star was each time in a different position on the slit.
We reduced the data in staring mode, extracted the spectrum of XUE 10 in each position separately, and combined the extracted spectra to improve the signal-to-noise ratio. We corrected the reduced 1D spectrum for atmospheric telluric absorption using the ESO standard LTT~7987, with a final signal-to-noise ratio of $\sim$ 190 (see Figure~\ref{fig:FORS_spec}).
\section{Spectral inventory}\label{sec:results}
In this section we present the general dust and gas properties of the JWST mid-IR spectrum. After that, we derive the physical properties for the central star of XUE 10 from the FORS2 optical spectrum. 
\subsection{MIRI MRS spectrum} 
At the mid-infrared wavelengths covered by the MIRI MRS instrument, the emission shown in Figure~\ref{fig:MIRI_spec} comes from the inner disk regions of XUE 10.
The spectrum shows a rising dust continuum, indicative of the re-emission at infrared wavelengths of heated dust grains in the upper layers of the disk. We detect emission from gas phase polycyclic aromatic hydrocarbon (PAH) particles at 6.2 $\mu$m and 11.3 $\mu$m. At the latter wavelength, the PAH feature overlaps with an emission band of forsterite \citep{Boekel_2005}. There is also evidence for the PAH complex at 7.7--8.6 $\mu$m, but this coincides with the blue shoulder of the silicate band and therefore does not clearly stand out.
We also detect a broad emission feature between 15.5 and 17 $\mu$m, whose origin remains unclear. We discuss this feature further in Appendix \ref{subsec:dust}.

Regarding the molecular gas phase component, we detect unusually bright CO$_2$ emission. We detect 
for the first time in a protoplanetary disk spectrum four isotopic species, i.e., isotopologues of CO$_2$, between 12.9 and 17.6 $\mu$m at high significance ($>$ 5$\sigma$) above the continuum level: $^{12}$CO$_2$, $^{13}$CO$_2$, $^{16}$O$^{12}$C$^{18}$O, and $^{16}$O$^{12}$C$^{17}$O. We do not only detect the fundamental $\mathit{Q}$-branch of these isotopes (corresponding to the 000~$\rightarrow$~010 transition of the $\upsilon_2$ bending mode), but also hot $\mathit{Q}$-branch bandheads (e.g., 011~$\rightarrow$~111) and blended $\mathit{P}$- and $\mathit{R}$-branch lines (see Figure \ref{fig:13-18mu_slabmodels}). CO$_2$ is a commonly detected species in protoplanetary disks, and is one of the crucial diagnostics for their chemo-physical structure and the processes shaping them \citep[e.g.,][]{Oberg_2011, Pontoppidan_2014}. The isotopes are naturally less abundant, and therefore fainter and harder to detect. 
Besides XUE 10, the $^{13}$CO$_2$ isotopologue has been observed by now in several nearby T~Tauri disks \citep{Grant_2023, Gasman_2025}. \citet{Vasblom_2024} were instead the first to report a tentative detection of $^{16}$O$^{12}$C$^{18}$O in CX Tau (T Tauri), yet finding no evidence of the $^{16}$O$^{12}$C$^{17}$O isotopologue. Most recently, \citet{Salyk_2025} claimed the first firm detection of $^{16}$O$^{12}$C$^{18}$O, and also a tentative detection of $^{16}$O$^{12}$C$^{17}$O in MY Lup, another T Tauri disk. At 4.9--6 $\mu$m, the MIRI instrument covers the upper side of the ro-vibrational $\mathit{P}$-branch ladder ($\mathit{J}$ $\geq$ 25) of CO, the primary gas molecular tracer in protoplanetary disks. In this wavelength range we detect weak CO emission with a 2$\sigma$ significance on the brightest lines (see Figure \ref{fig:46mu_slabmodels}). We do not find instead robust evidence of water emission in XUE 10.
The mid-IR ro-vibrational (5--8 $\mu$m) and rotational transitions ($>$ 10 $\mu$m) \citep{Meijerink_2009} of water are thought to probe progressively further regions in the inner disk, with hot and lukewarm components ($>$ 400 K) at the shorter wavelengths, and a cold component ($\sim$ 200--400 K) arising at longer wavelengths \citep{Banzatti_2020, Banzatti_2023}. 
We only place an upper limit on water lines at 6--8 $\mu$m, $\sim$ 17.2 $\mu$m and $\sim$ 23.9 $\mu$m (see Figure \ref{fig:1825mu_slabmodels}). In the 6--8~$\mu$m window we serendipitously detect (8$\sigma$) the Pfund $\alpha$ line of atomic hydrogen. No other hydrogen lines or other species (e.g., OH, C$_2$H$_2$, HCN) are identified in the mid-IR spectrum of XUE 10.
\subsection{Stellar host properties}\label{subsec:results_star}
The normalized optical spectrum of XUE 10 is presented in Figure~\ref{fig:FORS_spec}. The spectrum shows calcium triplet (CaT) as well as hydrogen absorption lines. Fainter metal absorption lines from e.g., OI are detected as well, but no emission lines.

We derived the stellar parameters of XUE 10 by comparing the observed FORS2 spectrum to a grid of PHOENIX stellar atmosphere models \citep{Phoenix},\footnote{\url{http://phoenix.astro.physik.uni-goettingen.de}} as described in Appendix~\ref{app:spectral_classification}. We used a minimized chi-square test to fit $T_{\rm eff}$, log$_{10}$(g), and [Fe/H], and benchmarked the statistical errors on the fit quantities with a Markov chain Monte Carlo (MCMC) simulation (see the corner plot in Figure~\ref{fig:cornerplots}). The derived stellar parameters are summarized in Table~\ref{tab:stellar_host}. In particular, the [Fe/H] solutions could be degenerate, having the highest peak centered at 0, and another peak at [Fe/H] $\geq$ 0.5 (see top panel in Figure~\ref{fig:cornerplots}). The near-solar metallicity value of XUE 10 is in line with near-infrared observations of protostars in other star-forming regions, e.g., Orion and Lupus \citep[][]{Biazzo_2011, Biazzo_2017}. 
Given the derived effective temperature, XUE 10 corresponds to a PMS with spectral type F0--F1 V \citep{Gray_2009}. 
We constrained the visual extinction to an $A$$_{\rm V}$ $\sim$ 4.8 using the Virtual Observatory SED Analyzer (VOSA) tool \citep{Vosa}\footnote{\url{http://svo2.cab.inta-csic.es/theory/vosa/}} and the archival optical and infrared photometric data on the source (see also Appendix~\ref{app:spectral_classification}).
From the measurement of the VVV (Variables in the Vía Láctea) VISTA (Visible and Infrared Survey Telescope for Astronomy) apparent J-band magnitude $m$$_{\rm J}$ = 12.004 $\pm$ 0.001, we derived a bolometric luminosity of $L$$_{\rm bol}$ = 69 $\pm$ 5 $L$$_{\odot}$. For this purpose, we adopted the bolometric correction in J-band for a F1 PMS from \citet{Pecaut_2013}, the constrained $A$$_{\rm V}$ (with an uncertainty of $\pm$ 0.2 from \textit{Gaia}), and a total-to-selective extinction factor $R$$_{\rm V}$ = 3.3, as measured in NGC 6357 \citep{Massi_2015, Russeil_2017, Fouesneau_2022}.
The derived luminosity places XUE 10 above the main sequence, consistent with a PMS nature, as expected for the young age of NGC 6357. 
The measured \textit{Gaia} DR3 parallax also places XUE 10 in the cluster \citep{Mike_2019}, suggesting that it has the same age as the other members. 
Assuming a PARSEC \citep{Parsec} isochrone of 1 Myr, which takes into account extinction, XUE 10 is consistent with a PMS of stellar mass $M$$_{\star}$ = 2.5--3.0 $M$$_{\sun}$. 
\begin{table}[htp!]
    \caption{Summary of the stellar properties of XUE 10.}\label{tab:stellar_host}
    \centering
    \resizebox{\columnwidth}{!}{
    {
    \begin{tabular}{ccccccc}
\hline
\hline
\noalign{\vskip 1mm}
SpT & $T$$_{\rm eff}$ & $M$$_{\star}$ & $L$$_{\rm bol}$ & $A$$_{\rm V}$ & log(g) & [Fe/H]\\
& [K] & [$M$$_{\odot}$] & [$L$$_{\odot}$] & & & \\
\noalign{\vskip 1mm}
\hline
\noalign{\vskip 0.5mm}
F0--F1 & 7000$^{+400}_{-300}$ & 2.5--3.0 & 60--70 & 4.6--5.0 & 4.0$^{+0.5}_{-0.5}$ & +0.5$^{+0.4}_{-0.5}$\\
\noalign{\vskip 1mm}
    \hline
    \end{tabular}
    }}
    \begin{tablenotes}
    \footnotesize
    \item \textbf{Notes.} The uncertainty evaluation is detailed in Appendix~\ref{app:spectral_classification}.
    \end{tablenotes}
    \end{table}
\section{0D LTE gas modeling}\label{subsec:gas}
In this section we characterize the chemical composition and the main molecular physical parameters, i.e., total column density ($N$$_{\rm tot}$), gas 
temperature ($T$$_{\rm gas}$), and equivalent emitting radius ($R$$_{\rm em}$) of the detected molecules.
Prior to modeling molecular features in the MIRI spectrum, we subtracted the local dust continuum 
as described in Appendix \ref{app:normalization}. In the continuum-subtracted spectrum, we employed 0D gas slab radiative transfer models computed assuming LTE gas \citep{Tabone_2023}.
\begin{table}[htp!]
    \caption{Best-fit $\chi^2_{\rm red}$ parameters from 0D gas slab modeling of the detected emission of CO (4.9--5.2 $\mu$m) and CO$_2$ (12.9--17.6 $\mu$m).} \label{tab1}
    \centering
    {%
    \begin{tabular}{lcccc}
    \hline
    \hline
    \noalign{\vskip 1mm}
    Species & log$_{10}$(\textit{N}$_{\rm tot}$) & \textit{T}$_{\rm gas}$ & \textit{R}$_{\rm em}$ & $\chi^2_{\rm red, ~min}$\\
    & [cm$^{-2}$] & [K] & [au] & \\
    \noalign{\vskip 1mm}
    \hline
    \noalign{\vskip 0.5mm}
    $^{12}$CO & 14.0$^{+1.4}_{-0}$ & 1850$_{-1150}^{+750}$ & 10$^{+0}_{-5}$ & 1.1\\
    \noalign{\vskip 0.5mm}
    \hline
    \noalign{\vskip 0.5mm}
    \multicolumn{5}{c}{Fixed emitting radius}\\
    \noalign{\vskip 0.5mm}
    \hline
    \noalign{\vskip 0.5mm}
    $^{12}$CO$_2$ & 20.11$_{-1.69}^{+2.89}$ & 365$_{-186}^{+438}$ & 1.15$_{-0.14}^{+1.89}$ & 1.6\\
    \noalign{\vskip 0.8mm}
    $^{13}$CO$_2$ & 19.9$_{-2.9}^{+2.7}$ & 304$_{-54}^{+896}$ & 1.15 & 1.8\\
    \noalign{\vskip 0.8mm}
    $^{16}$O$^{12}$C$^{18}$O & 18.3$_{-1.3}^{+0.8}$ & 365$_{-41}^{+362}$ & 1.15 & 1.6\\
    \noalign{\vskip 0.8mm}
    $^{16}$O$^{12}$C$^{17}$O & 17.87$_{-1.29}^{+1.03}$ & 324$_{-51}^{+570}$ & 1.15 & 1.2\\
    \noalign{\vskip 0.5mm}
    \hline
    \noalign{\vskip 0.5mm}
    \multicolumn{5}{c}{Free emitting radius}\\
    \noalign{\vskip 0.5mm}
    \hline
    \noalign{\vskip 0.5mm}
    $^{13}$CO$_2$ & 18.7$^{+4.3}_{-3.7}$ & 609$^{+591}_{-475}$ & 0.49$^{+0.01}_{-0.4}$ & 1.6\\
    \noalign{\vskip 0.8mm}
    $^{16}$O$^{12}$C$^{18}$O & 18.2$_{-3.2}^{+2.4}$ & 344$_{-206}^{+540}$ & 1.2$_{-0.7}^{+8.8}$ & 1.6\\
    \noalign{\vskip 0.8mm}
    $^{16}$O$^{12}$C$^{17}$O & 17.67$_{-2.70}^{+2.04}$ & 467$_{-333}^{+378}$ & 0.8$_{-0.3}^{+9.2}$ & 1.04\\
    \noalign{\vskip 1mm}
    \hline
    \end{tabular}
    }
    \begin{tablenotes}
    \footnotesize
    \item \textbf{Notes.} Column (5) reports the minimum reduced chi-square found for each fit. The $\pm$ error bars, where indicated, refer to the 1$\sigma$ chi-square contours.
    \end{tablenotes}
\end{table}
We took into account the presence of various molecular species and modeled their emission in several wavelength windows, exploring the parameter spaces as listed in Table~\ref{tab:fitting_ranges}. In particular, we considered the effect of using different grids or values of equivalent emitting radius on the column density and the temperature measured for the CO$_2$ isotopologues and CO (see Table \ref{tab1}).
The molecular data (e.g., transition energies, Einstein coefficients) are retrieved from the HITRAN database \citep{Gordon_2022}. We assumed a Gaussian line profile with an intrinsic broadening of $\Delta$V = 4.7 km s$^{-1}$, whose exact value has little effect on the resulting synthetic spectra \citep{Salyk_2008}. 

A reduced chi-square ($\chi^2_{\rm red}$) fit was adopted to find the best-fit parametric values for each identified molecular species. Further details can be found in Appendix~\ref{app:fitting_tests}. Following the procedure described in \citet{Grant_2023}, we start by identifying the molecular species exhibiting the strongest emission. This is typically H$_2$O; however, in XUE 10 spectrum, CO$_2$ dominates. After subtracting the best-fit model of this primary species, we proceed to fit the next most prominent molecular emission.
We included the treatment for mutual shielding from adjacent lines for all species, as discussed in \citet{Tabone_2023}. The best-fitting models correspond to the minimum $\chi^{2}_{\rm red}$ found, and their parameters are summarized in Table~\ref{tab1} with $\pm$1$\sigma$ error bars, where present, taken from the reduced chi-square maps (see Appendix~\ref{app:chi-square}). The 1$\sigma$, 2$\sigma$, and 3$\sigma$ contours in the maps are calculated as ($\chi^2-\chi^2_{\rm min})^{1/2}$, adding respectively $\chi^2_{\rm min}$+ 2.3, $\chi^2_{\rm min}$+ 6.2, and $\chi^2_{\rm min}$+ 11.8. 
As it is evident in the shape of the contour areas in the chi-square maps (see Figure~\ref{fig:chisquaremaps}), the errors on the measured quantities can be correlated due to the parameter degeneracies intrinsic of the 0D slab modeling. In other words, there is a range of combinations of $N$$_{\rm tot}$--$T$$_{\rm gas}$--$R$$_{\rm em}$ that fit equally well the spectrum for a given molecular species, for example with a high/low column density and a small/large emitting radius, as these have a direct effect onto the line fluxes and the shape of the observed molecular features. These degeneracies can be alleviated by fitting a mix of optically thin and optically thick features \citep{Grant_2023}, as also done in this work, or with the aid of more refined (e.g., 1D) thermo-chemical models.
\subsection{CO$_2$}\label{subsub:CO2}
\begin{figure*}[htp!]
    \centering
    \includegraphics[width=0.8\linewidth]{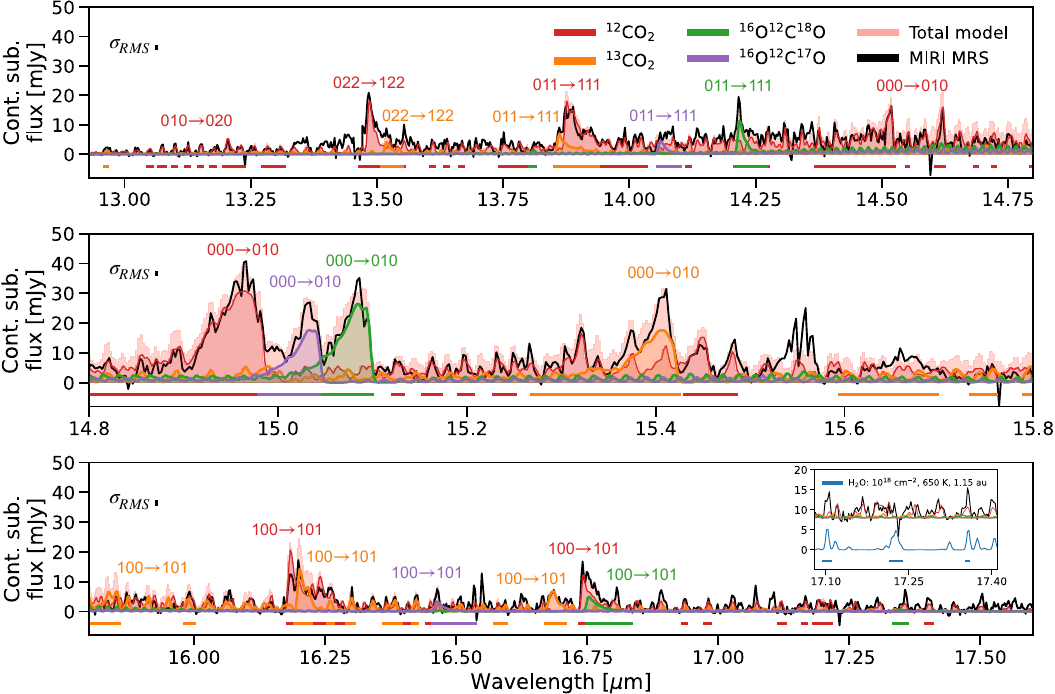}
    \caption{Continuum-subtracted MIRI spectrum of XUE 10 (black) with overlaid best-fit slab models of the identified carbon dioxide isotopologues, $^{12}$CO$_2$ (red), $^{13}$CO$_2$ (orange), $^{16}$O$^{12}$C$^{18}$O (green), and $^{16}$O$^{12}$C$^{17}$O (purple) between 12.93 $\mu$m and 17.6 $\mu$m (from top to bottom). The vibrational quantum numbers $(\upsilon_1 \upsilon_2 \upsilon_3)$ corresponding to the fundamental $\upsilon_2$ $\mathit{Q}$-branch, and its associated hot bands are labeled for each species. The inset in the bottom panel shows the fiducial LTE slab model of H$_2$O that matches the observed line luminosity between 17.08--17.4 $\mu$m. The colored horizontal bars indicate the fit wavelength ranges for CO$_2$ listed in Table~\ref{tab:fitting_ranges}. The spectral uncertainty is indicated in each panel.}
    \label{fig:13-18mu_slabmodels}
\end{figure*}
We computed the slab models for $^{12}$CO$_2$, $^{13}$CO$_2$, $^{16}$O$^{12}$C$^{18}$O, and $^{16}$O$^{12}$C$^{17}$O using all identified emission lines (blended) and bands per species listed in Table~\ref{tab:fitting_ranges}. We built the model grid varying $T$$_{\rm gas}$ from 100 to 1200 K, log$_{10}$($N$$_{\rm tot}$) from 15 to 22, and $R$$_{\rm em}$ from 0.01 au to 10 au (in log$_{10}$-space).
We first fit the $^{12}$CO$_2$ species, selecting the wavelength ranges where only $^{12}$CO$_2$ is predominant, and derived the best fitting values for $N$$_{\rm tot}$, $T$$_{\rm gas}$ and $R$$_{\rm em}$. 
Then we subtracted its best fitting model, and proceeded with the fit of $^{13}$CO$_2$, then $^{16}$O$^{12}$C$^{18}$O, and lastly $^{16}$O$^{12}$C$^{17}$O in the same fashion. 
We assumed that they all emit from the same equivalent emitting area, by fixing the emitting radius to that derived from the $^{12}$CO$_2$ fit (R$_{\rm em}$ = 1.15 au; \enquote{Fixed emitting radius} in Table \ref{tab1}). 
The isotopologues of $^{12}$CO$_2$ are expected to be less abundant and more optically thin than the main species, as evident for CO \citep[][]{Dartois_2003, van_Zadelhoff}.

Figure \ref{fig:13-18mu_slabmodels} shows the resulting best-fit models in the continuum-subtracted spectrum, and the respective $\chi^2_{\rm red}$ maps are reported in Figure~\ref{fig:chisquaremaps}. The inferred column densities for $^{12}$CO$_2$ (1.3 $\times$ 10$^{20}$ cm$^{-2}$) and $^{13}$CO$_2$ (8.3 $\times$ 10$^{19}$ cm$^{-2}$) are the highest ever found in the terrestrial planet-forming region of a protoplanetary disk, with a factor 10 to 100 higher than in previous works \citep{Grant_2023, Gasman_2023, Vasblom_2024, Salyk_2025}. 
In comparison, the probed gas temperatures ($\sim$ 300--370 K) are similar within uncertainties.
A major difference is observed in the derived emitting radius (0.05--0.6 au in the literature), which may be own to the higher luminosity of the XUE 10 star, inducing a larger warm disk surface where CO$_2$ is emitting.  
For the $^{16}$O$^{12}$C$^{18}$O and $^{16}$O$^{12}$C$^{17}$O isotopologues, the best-fit models yield relatively high column densities of 1.9 $\times$ 10$^{18}$ cm$^{-2}$ ($^{16}$O$^{12}$C$^{18}$O), and 7.4 $\times$ 10$^{17}$ cm$^{-2}$ ($^{16}$O$^{12}$C$^{17}$O). These species seem to trace a similar gas temperature ($\sim$ 320--370 K) to $^{12}$CO$_2$ and $^{13}$CO$_2$. This could indicate, assuming that the LTE assumption holds, that the three optically thinner isotopologues are tracing the same emitting region as $^{12}$CO$_2$.
\begin{figure*}[htp!]
    \centering
    \includegraphics[width=0.8\linewidth]{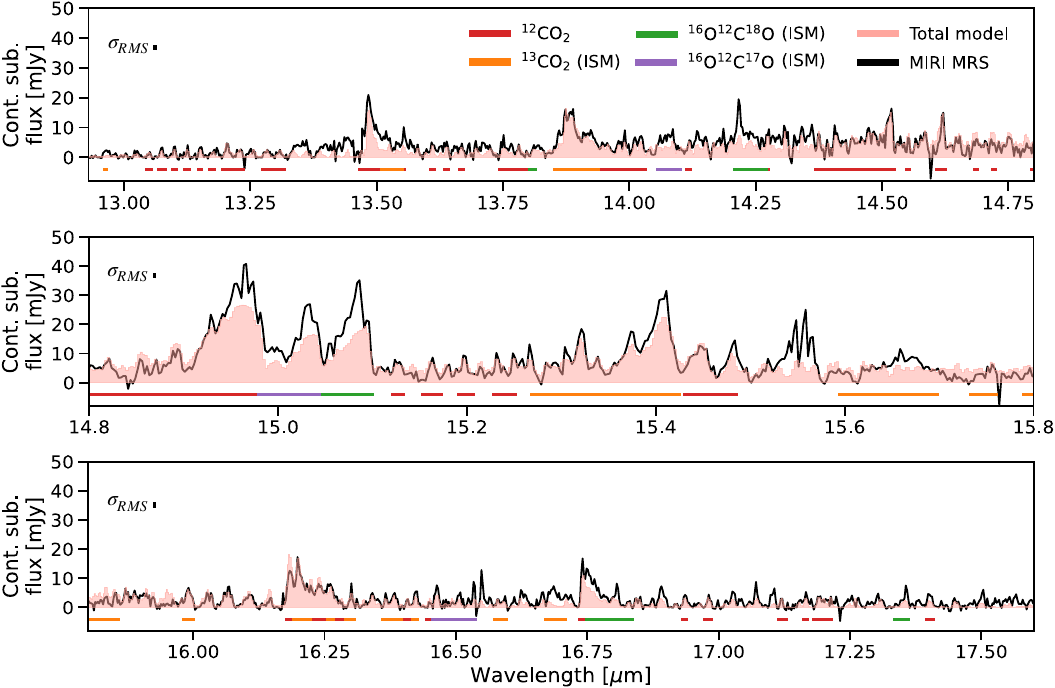}
    \caption{Continuum-subtracted MIRI spectrum of XUE 10 (black) overlaid with the total slab model of $^{12}$CO$_2$, $^{13}$CO$_2$, $^{16}$O$^{12}$C$^{18}$O, and $^{16}$O$^{12}$C$^{17}$O between 12.93 $\mu$m and 17.6 $\mu$m (from top to bottom panel) assuming ISM isotopic ratios in the column density. The colored horizontal bars indicate the fit wavelength ranges per species listed in Table~\ref{tab:fitting_ranges}. The spectral uncertainty is indicated in each panel.}
    \label{fig:1318mu_ISMmodels}
\end{figure*}
\begin{figure*}[htp!]
    \centering
    \includegraphics[width=0.8\linewidth]{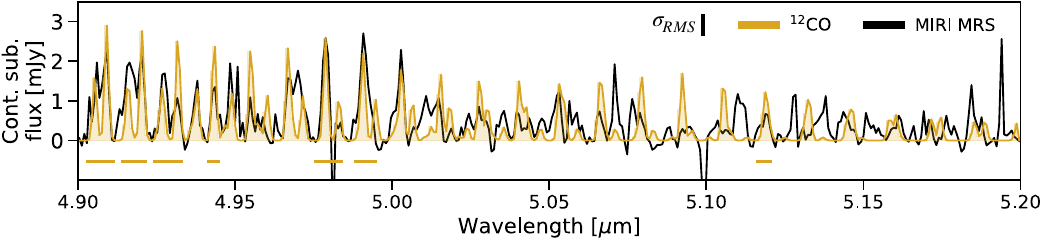}
    \caption{Continuum-subtracted MIRI spectrum of XUE 10 (black) overlaid with the best-fit slab models of carbon monoxide ($^{12}$CO; gold). The colored horizontal bars indicate the fit wavelength ranges listed in Table~\ref{tab:fitting_ranges}. The spectral uncertainty is labeled in figure.}
    \label{fig:46mu_slabmodels}
\end{figure*}
\begin{figure*}[htp!]
    \centering
    \includegraphics[width=0.8\linewidth]{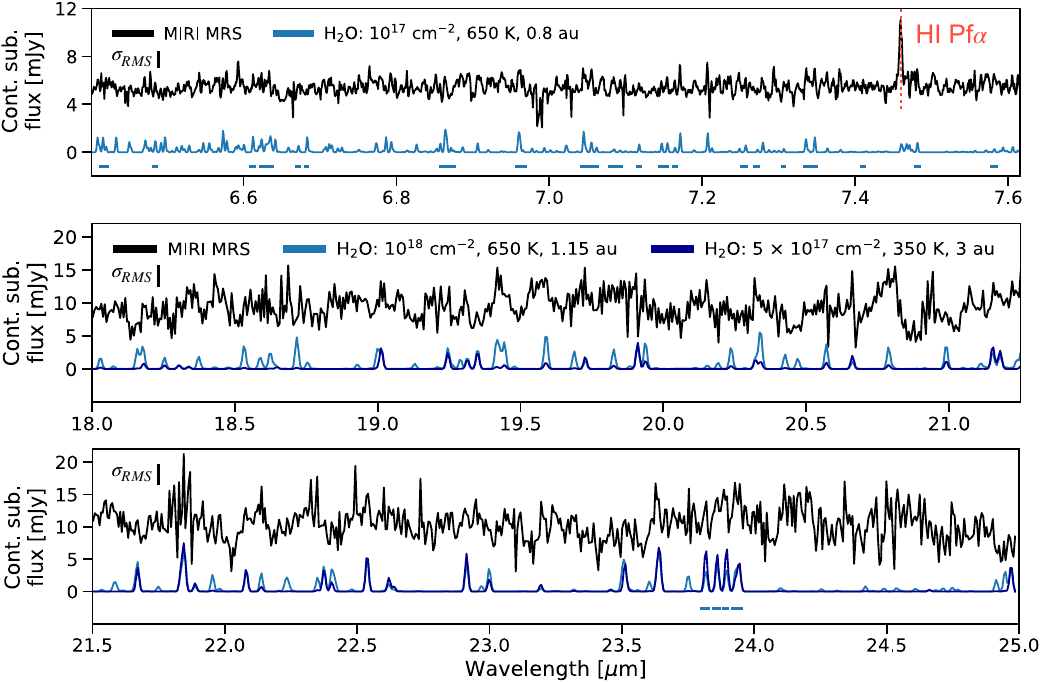}
    \caption{Continuum-subtracted MIRI spectrum of XUE 10 (black) and fiducial LTE slab models of H$_2$O between 6.4 $\mu$m and 7.6 $\mu$m (top panel) and between 18 $\mu$m and 25 $\mu$m (middle and bottom panels). The observed spectrum is vertically offset for visual clarity. The vertical dotted line marks the detected H$\alpha$ line of the Pfund series. The colored horizontal bars indicate where H$_2$O lines are tentatively identified. The spectral uncertainty is indicated in each panel.}
    \label{fig:1825mu_slabmodels}
\end{figure*}
\subsection{Unusual C- and O- isotopic ratios in CO$_2$}\label{subsec:isotopic_ratios}
When looking at the column density ratios between the different detected CO$_2$ isotopologues, one can see that they are not consistent with the $^{12}$C/$^{13}$C, $^{16}$O/$^{18}$O, and $^{16}$O/$^{17}$O isotopic ratios observed in the interstellar medium (ISM) if they are interpreted as abundance ratios in the disk (see Table~\ref{tab:isotopic_ratios}). In particular, the column densities of $^{16}$O$^{12}$C$^{18}$O and $^{16}$O$^{12}$C$^{17}$O seem to be enhanced by a factor of four to six and that of $^{13}$CO$_2$ is enhanced by a factor of more than ten compared to the ordinary values. 

This may be due to line optical depth ($\tau_{line}$) effects, where the emission of the isotopologues is optically thinner than $^{12}$CO$_2$, hence tracing deeper disk layers, corresponding to an apparently larger column of gas along the line of sight. The isotopologues may also trace disk layers at larger radii.

The observed spectrum of each isotopologue is actually a mix of optically thin and optically thick emission lines.
In fact, the measured optical depth at the line center (hereafter line opacity) in the slab models peaks in an optically thick regime ($\tau_{line}$ $\gg$ 1) around the fundamental $\upsilon_2$ $\mathit{Q}$-branch of each isotopologue, with maximum averaged line optical depth of 90 ($^{12}$CO$_2$), 55 ($^{13}$CO$_2$), 4 ($^{16}$O$^{12}$C$^{18}$O), and 3 ($^{16}$O$^{12}$C$^{17}$O), corresponding to the strongest $\mathit{Q}$-branch transitions. The line opacity then gradually decreases into an optically thin regime ($\tau_{line}$ $<$ 1), corresponding to the $\mathit{P}$- and $\mathit{R}$-branch lines and hot $\mathit{Q}$-branch bandheads at wavelengths shorter than 14 $\mu$m and longer than 16 $\mu$m. We can therefore hypothesize that the bulk of each isotopologue emission is primarily governed either by the characteristics of the optically thick $\mathit{Q}$-branches, or by those of the optically thinner features.
If the optically thick emission dominates, the CO$_2$ isotopologues must trace roughly the same gas reservoir, and the difference in $\mathit{Q}$-branch line optical depth can explain the enhanced column densities reported in the top part of Table \ref{tab1}. 
To test this hypothesis, we simultaneously fit $^{12}$CO$_2$, $^{13}$CO$_2$, $^{16}$O$^{12}$C$^{18}$O, and $^{16}$O$^{12}$C$^{17}$O at fixed $^{12}$CO$_2$ temperature (365 K) and emitting radius (1.15 au). We varied the column density of $^{12}$CO$_2$ in log$_{10}$--space between 17 and 22, scaling the column density of the isotopes by their ISM ratios (see Table~\ref{tab:isotopic_ratios}). In this case, a single slab model takes into account the contribution of all four species.
The resulting fit is shown in Figure~\ref{fig:1318mu_ISMmodels}. The best-fit ($\chi^2_{red}$ = 2.2) column densities are 2.2 $\times$ 10$^{18}$ cm$^{-2}$ ($^{13}$CO$_2$), 5.3 $\times$ 10$^{17}$ cm$^{-2}$ ($^{16}$O$^{12}$C$^{18}$O) and 1.5 $\times$ 10$^{17}$ cm$^{-2}$ ($^{16}$O$^{12}$C$^{17}$O). We note that these values are within the derived 1$\sigma$ error bars in Table~\ref{tab1}. One can see in Figure \ref{fig:1318mu_ISMmodels}, however, that while the spectrum of $^{13}$CO$_2$ is well reproduced in this case, the observed features of the other species are underestimated, including the optically thinner hot bands. The shape of their $\mathit{Q}$-branches is also not well reproduced. This finding suggests that $^{13}$CO$_2$ emission is well characterized by its optically thick features, and that it comes from the same gas reservoir as $^{12}$CO$_2$. This way, a column density of $\sim$ 10$^{19}$ cm$^{-2}$ can be explained by the smaller line optical depth of its $\mathit{Q}$-branch in comparison to $^{12}$CO$_2$. Instead, the emission of $^{16}$O$^{12}$C$^{18}$O and $^{16}$O$^{12}$C$^{17}$O must be dominated by their optically thin features, therefore tracing disk layers of different gas temperature and/or equivalent radius than $^{12}$CO$_2$. 

In order to evaluate the dependence of $N$$_{\rm tot}$ on the equivalent emitting radius, we independently modeled $^{13}$CO$_2$, $^{16}$O$^{12}$C$^{18}$O, and $^{16}$O$^{12}$C$^{17}$O leaving $R$$_{\rm em}$ as a free parameter (\enquote{Free emitting radius} in Table \ref{tab1} and Figure~\ref{fig:chisquaremaps}). As expected from the intrinsic parameter degeneracies of the slab modeling approach, in this case a larger number of models of different parameter combinations can reproduce equally well the spectrum. $^{13}$CO$_2$ is consistent with tracing a closer (0.49 au) and hotter (609 K) gas reservoir than $^{12}$CO$_2$, but at ISM-like column density (5 $\times$ 10$^{18}$ cm$^{-2}$). Therefore, it is not possible to determine whether $^{12}$CO$_2$ and $^{13}$CO$_2$ are co-spatial, which hinders our ability to draw strong conclusions when comparing the $^{12}$C/$^{13}$C abundance ratio to the ISM value (see Table~\ref{tab:isotopic_ratios}). 
Emission from $^{16}$O$^{12}$C$^{17}$O is instead consistent with originating at an equivalent radius of 0.8 au and a gas temperature of 467 K, while $^{16}$O$^{12}$C$^{18}$O seems to emit from a larger radius of 1.2 au, further out than $^{12}$CO$_2$, at a lower temperature of 344 K. These gas temperatures are consistent with those inferred from the slab models with fixed emitting radius, while the column densities remain aligned with non-ISM oxygen isotopic ratios, i.e., 1.6 $\times$ 10$^{18}$ cm$^{-2}$ for $^{16}$O$^{12}$C$^{18}$O, and 0.5 $\times$ 10$^{18}$ cm$^{-2}$ for $^{16}$O$^{12}$C$^{17}$O respectively. Given the overall optically thin emission of these isotopologues, though, the derived column densities (and emitting radii) may represent lower limits. We conclude that $^{16}$O$^{12}$C$^{18}$O and $^{16}$O$^{12}$C$^{17}$O trace optically thin layers in the disk not co-spatial with $^{12}$CO$_2$, and where the oxygen isotopic composition may be anomalous. However, 0D gas slab models do not allow one to discriminate between line optical depth effects and abundance patterns.
The most plausible conclusion is that we are affected by line optical depth effects also toward $^{16}$O$^{12}$C$^{18}$O and $^{16}$O$^{12}$C$^{17}$O. A 2D disk thermochemical analysis, e.g., using ProDiMo \citep{Kamp2010, Woitke_2016, Rab2018, Woitke2024} or DALI \citep[e.g.,][]{Bruderer_2012, Bruderer_2013} models may be employed to derive the emitting regions in the disk of the different isotopologues and study their abundance profile in the radial and vertical disk direction. To do so, however, the chemical networks of $^{16}$O$^{12}$C$^{18}$O and $^{16}$O$^{12}$C$^{17}$O are needed to be implemented in these frameworks.
\begin{table}[htp!]
    \caption{Isotopic ratios of C and O measured in the ISM (from CO) and in this work (from CO$_2$).}\label{tab:isotopic_ratios}
    \centering
    {
    \begin{tabular}{llll}
\hline
\hline
\noalign{\vskip 1mm}
Ratio & ISM & Observed & Reference\\
\noalign{\vskip 1mm}
\hline
\noalign{\vskip 1mm}
$^{12}$C/$^{13}$C & 67 $\pm$ 15 & 4$^{a}$ $\pm$ 4 & \citet{Milam_2005}\\
$^{16}$O/$^{18}$O & 557$^{b}$ $\pm$ 15 & 65 $\pm$ 92 & \citet{Wilson_1994}\\
$^{16}$O/$^{17}$O & 2005$^{b}$ & 174 $\pm$ 247 & \citet{Wilson_1994}\\
$^{18}$O/$^{17}$O & 3.2 $\pm$ 0.2 & 3 $\pm$ 4 & \citet{Penzias_1981}\\
\noalign{\vskip 1mm}
    \hline
    \end{tabular}
    }
    \begin{tablenotes}
    \footnotesize
    \item \textbf{Notes.} Column (3) reports the column density ratio from Table~\ref{tab1}. Column (4) report the references for the measured ratios in the ISM. $^{a}$ This ratio is likely due to line optical depth effects and not a true abundance anomaly (see text). $^{b}$We consider half of these ratios, counting one $^{18}$O/$^{17}$O atom in CO$_2$.
    \end{tablenotes}
    \end{table}
\subsection{CO and H$_2$O}\label{subsub:CO}
We observe faint CO emission at 4.9--5.2 $\mu$m, with a detection at 2$\sigma$ significance above noise level on the brightest lines (Figure~\ref{fig:46mu_slabmodels}). We characterized the emission calculating the slab models varying $T$$_{\rm gas}$ from 100 to 3400 K, and log$_{10}$($N$$_{\rm tot}$) from 14 to 20. The $\chi^{2}_{red}$ map in Figure~\ref{fig:chisquaremaps} points toward a column density $\lesssim$ 2.5 $\times$ 10$^{15}$ cm$^{-2}$ and a gas temperature of about 1850 K within an equivalent emitting radius of 5--10 au (see Table~\ref{tab1}). As the CO emission is optically thin, we only constrain a lower limit on the column density and on the emitting radius. 
The fit gas temperature is within the range typically observed in T Tauri disks \citep[e.g.,][]{Ramirez-Tannus_2023, Schwarz_2024, Temmink_CO}. However, it shall be noted that JWST/MIRI observations only cover the high-$\mathit{J}$ transitions of the CO $\mathit{P}$-branch, which tend to trace hotter and less dense gas than the low-$\mathit{J}$ transitions (i.e., visible in the near-IR range). As a consequence, fitting the CO emission using only the high-$\mathit{J}$ transitions yields a higher excitation temperature than the actual gas kinetic temperature \citep{Temmink_CO}. 

From Figure \ref{fig:46mu_slabmodels} one can notice that the best-fit models predict bright CO lines between $\sim$ 5.05 and 5.15 $\mu$m, which in the data are undetected or $\lesssim$ 1$\sigma$ above the noise level. We find no evidence of contamination from stellar spectral features at the wavelengths of the CO lines, with the exception of the transition at 5.1332 $\mu$m, which may be hampered by the stellar spectrum. Therefore, at the resolution of MIRI MRS we may be observing a blend of CO lines in the $\upsilon$ = 1--0, $\upsilon$ = 2--1 and $\upsilon$ = 3--2 vibrational states, where the 2--1 and 3--2 vibrational transitions are (partially) excited in non-LTE conditions. As a result, our LTE slab model overpredicts the line flux of these transitions, which must not emit at the same temperature of the 1--0 transitions \citep{Ramirez-Tannus_2023, Grant_2023, Temmink_CO}. Follow-up observations at higher sensitivity, including the near-IR coverage of the lower hand of the $\mathit{P}$-branch ladder, coupled with non-LTE modeling are needed to increase the signal-to-noise ratio of the individual CO lines, and deliver unbiased physical parameters of its emission in XUE 10.

Apart from the detected CO$_2$ and CO emission, we only tentatively detected H$_2$O. In particular, the brightest water lines expected in the $\sim$ 6--8 $\mu$m and 18--25 $\mu$m regions lie at $<$ 2$\sigma$ above the noise level (see Figure \ref{fig:1825mu_slabmodels}), whereas the 13--18 $\mu$m wavelength range is dominated by carbon dioxide (see Figure \ref{fig:13-18mu_slabmodels}). As a result, the brightest water lines near 17 $\mu$m can only be identified at 1$\sigma$ in the residual spectrum of CO$_2$. We therefore could not perform a reliable modeling of this species in the same fashion as for the others. Instead, we computed fiducial slab models of water (plotted in Figure \ref{fig:13-18mu_slabmodels} and Figure \ref{fig:1825mu_slabmodels}) with the aim of reproducing the lines tentatively identified in each wavelength range. The fiducial model for the 13--18 $\mu$m region was calibrated to match the H$_2$O line luminosity measured in the residual spectrum of CO$_2$ between 17.04 and 17.39 $\mu$m (see inset in Figure \ref{fig:13-18mu_slabmodels}), assuming the same equivalent emitting radius of CO$_2$ (1.15 au). In this way, we were able to estimate an upper limit on the total column density of water of $N$$_{\rm tot}$ $\approx$ 10$^{18}$ cm$^{-2}$.

This model is consistent with the warm (400--600 K) component of water typically detected around the 17 $\mu$m wavelength region in non-irradiated T Tauri disks \citep{Banzatti_2025}. The 6--8 $\mu$m water lines, however, are better reproduced by a model, at equal temperature (650 K), with lower column density (10$^{17}$ cm$^{-2}$) within a smaller radius (0.8 au). At longer wavelengths, between 18--25 $\mu$m, the water spectrum may be explained by the combination of a cold (350 K) and a warm (650 K) component, consistent with mounting evidence that multiple physical components are required to reproduce the mid-infrared water spectrum \citep{Temmink_water}. Considering the 2$\sigma$ H$_2$O lines in the 7 $\mu$m region, the upper limit on the total column density would be even lower (N$_{\rm tot}$ $\approx$ 10$^{17}$ cm$^{-2}$). 
We note that the quadruplet of water lines expected at 23.80--23.96 $\mu$m (see bottom panel in Figure~\ref{fig:1825mu_slabmodels}) does not exhibit the line flux ratio characteristic of a strong cool water component, recently used as a proxy of ongoing ice sublimation near the water snowline \citep{Banzatti_2023, Vasblom_2024}. By comparing the detection of the CO$_2$ $\mathit{Q}$-branch at 15 $\mu$m with the upper limit on H$_2$O from our fiducial model computed at 17 $\mu$m, we find a lower limit for the CO$_2$/H$_2$O column density ratio of $N$$_{\rm tot}$ (CO$_2$/H$_2$O) $\approx$ 130. However, the lack of robust water line detections can only be partially explained by overlapping CO$_2$ emission. The dust continuum is not a viable alternative explanation, as optically thick dust would hide both water and CO$_2$. A seemingly low water abundance has previously been reported in other Herbig disks \citep[e.g.,][]{Fedele_2011}. These observations are in contrast to theoretical expectations \citep{Walsh_2015, Antonellini_2016, Marcelino_2018}, which predict a water vapor column density of the order of 10$^{19-20}$ cm$^{-2}$ in Herbig disks, comparable to the majority of known T Tauri sources. We assess the origin of a water-poor infrared spectrum in disks around intermediate-mass stars in the following discussion.
\section{Discussion}\label{sec:discussion}
\subsection{The peculiar spectrum of XUE 10}\label{discussion:spectrum}
\begin{figure*}[htp!]
\centering
\includegraphics[width=0.7\linewidth]{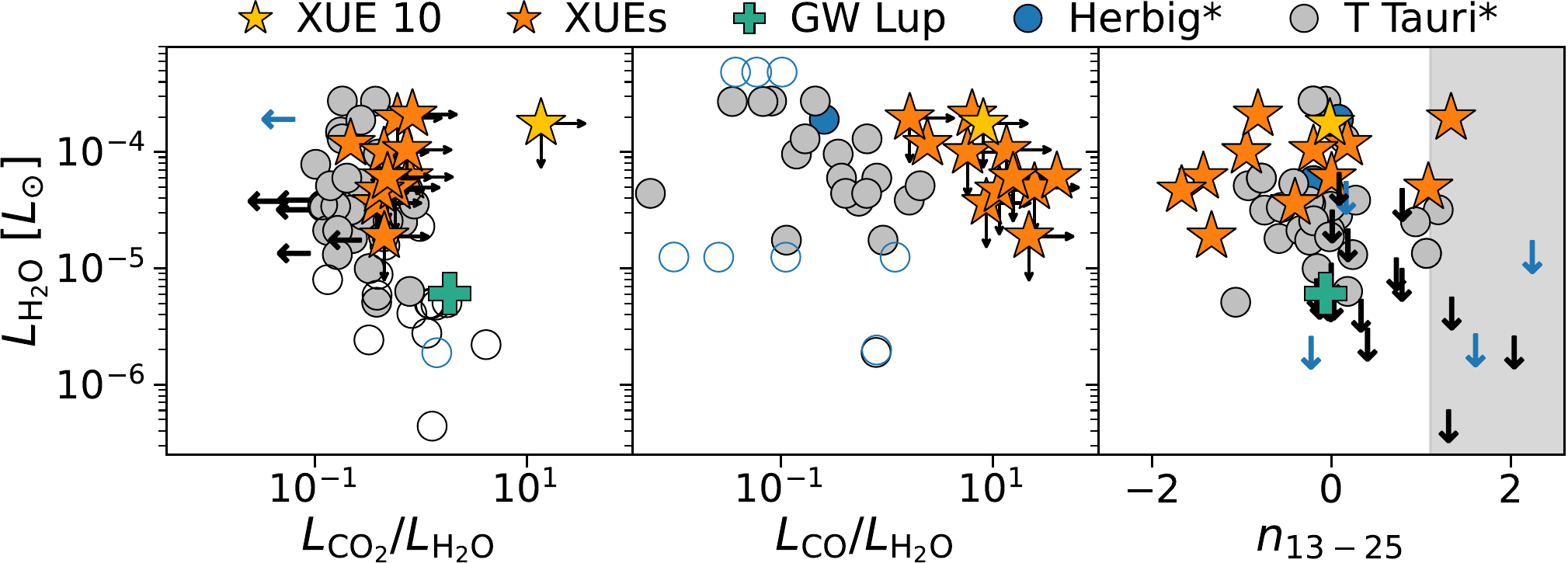}
\caption{Comparison between the gas phase molecular line luminosities and 13--25 $\mu$m spectral indices measured in the XUE 10 (yellow star) and the rest of the XUE sample \citep[orange stars;][]{Ramirez-Tannus_2025}, GW Lup \citep[][]{Grant_2023}, and the \textit{Spitzer} sample analyzed in \citet{Banzatti_2020, Banzatti_2022} (blue and gray markers), labeled with an asterisk. The y-axis displays the H$_2$O luminosity; from left to right, the x-axis displays the $^{12}$CO$_2$ and CO luminosities, plus the estimated 13--25 $\mu$m spectral index. The upper limits are indicated with arrows, and the open circles show the sources detected with upper limits in both the x- and y-axes. The shaded region in the right-most panel flags where \textit{n}$_{13-25}$$>$ 1.1, indicative of a dust-depleted inner disk.
}\label{fig:line_luminosities}
\end{figure*}
Among all the MIRI MRS spectra of protoplanetary disks published so far \citep[see, e.g., the MINDS and JDISC programs:][]{Henning_2024, Arulanantham_2025}, XUE 10 is the first Herbig disk located in a massive star-forming region to be analyzed.
The comparison with MINDS and JDISC published sources, including the ones discussed in Sec. \ref{sec:results} (i.e., GW Lup, CX Tau, Sz 98, MY Lup) must be taken with caution, since they all are low-mass to solar-mass T~Tauri disks located in low-mass isolated star-forming regions. Besides the striking simultaneous detection of four CO$_2$ isotopic species in emission, the mid-IR spectrum of XUE 10 appears to be \enquote{line-poor}, i.e, with non-detections of H$_2$O, OH, C$_2$H$_2$, HCN etc., in contrast to the typical \enquote{line-rich} mid-IR spectrum of T Tauri sources. The \enquote{line-poor} aspect of XUE 10 is in common with other known IMTT and Herbig disk sources. The origin of this discrepancy between T Tauri and disks surrounding intermediate-mass stars remains unknown. However, we do have samples of Herbig disks in isolation and in massive star-forming regions \citep[e.g.,][]{Stapper_2022, Stapper_2025}, but we lack the JWST/MIRI spectra to do a proper comparison, and establish if there are systematic differences in the gas content of these two groups of disks. Moreover, the brightness of IMTT or Herbig young stars, which tend to saturate in nearby regions, and the consequent higher disk infrared continuum, have often limited the observational analysis of the inner disk regions to the discussion of dust features rather than of the gas phase molecular chemistry. 

The main references for the latter are studies of nearby ($<$ 700 pc) Herbig disks observed at 1--5 $\mu$m with VLT/CRIRES \citep{Fedele_2011}, 4.5--5.2 $\mu$m with the iSHELL instrument at the Infrared Telescope Facility \citep{Banzatti_2022}, and with \textit{Spitzer} at 10--36 $\mu$m \citep{Banzatti_2020}. 
The general finding in these samples is the lack of the chemical richness that is seen in the atmosphere of T~Tauri disks, with mostly upper limits or detections with line fluxes $<$ 10$^{-14}$ erg cm$^{-2}$ s$^{-1}$ of H$_2$O, OH, $^{12}$CO, $^{13}$CO, CO$_2$, HCN and C$_2$H$_2$. Interestingly, at the longer wavelengths probed by \textit{Herschel} (50--200 $\mu$m), H$_2$O and OH are detected with similarly low column densities of 10$^{14}$--10$^{15}$ cm$^{-2}$ at large emitting radii ($>$ 10 au), pointing at an outer--inner depletion of water molecules in Herbig disks by means of the UV irradiation from the central massive protostars \citep{Fedele_2011, Heays_2017}.
XUE 10 therefore looks similar to other IMTT or Herbig disks in the sense of a bright dust continuum with PAH and silicate dust emission and because of the lack of a \enquote{variety} of bright gas phase molecular species. 

Nonetheless, XUE 10 remains unique in its CO$_2$ chemistry. Archival \textit{Spitzer} observations of disks around intermediate-mass T Tauri and Herbig stars show no comparable CO$_2$ emission features, and this absence cannot be fully attributed to sensitivity limits. In fact, the CO$_2$ isotopologue $\mathit{Q}$-branches seen in XUE 10’s mid-infrared spectrum would be strong enough to be detected with \textit{Spitzer}. A similar infrared CO$_2$ spectrum has only been observed in two post-AGB (asymptotic giant branch) stars, HR 4049 \citep{Cami_2001}, and HD 52961 \citep{Gielen_2009}. Some post-AGB stars, in fact, can be characterized by the presence of a circumstellar envelope of material originated by the mass loss of the evolved star \citep[e.g.,][]
{Van_Winckel_2003}. We firmly exclude the hypothesis of a post-AGB nature of XUE 10 based on its spectral classification, age constraint, parallax, and also bolometric luminosity, which would be significantly higher for a post-AGB star placed at the same distance of the NGC 6357 cluster. 

In Figure~\ref{fig:line_luminosities} we compare the line luminosities of H$_2$O, CO$_2$, CO and the 13--25 $\mu$m spectral index observed in XUE 10 with the rest of the XUE sample \citep{Ramirez-Tannus_2025}, the JWST observations of the CO$_2$-rich GW Lup T~Tauri disk in the nearby Lupus I cloud \citep[$\sim$ 155 pc;][]{Grant_2023}, and with \textit{Spitzer} measurements of other T~Tauri and Herbig protoplanetary disks located in nearby ($<$ 200 pc), low-mass star-forming regions ($<$ 3 Myr), adopting the samples from \citet{Banzatti_2020, Banzatti_2022}. The line luminosities are estimated from the observed spectra at 17.08--17.4 $\mu$m (H$_2$O), where for XUE 10 we derive an upper limit in the CO$_2$ residual spectrum, 14--15 $\mu$m ($^{12}$CO$_2$), and 4.9--5.2 $\mu$m (CO). We find that XUE 10 stands out for the highest $^{12}$CO$_2$/H$_2$O line luminosity ratio, which must be interpreted as a lower limit. For CO, XUE 10 is at the high end of the distribution, but not such a strong outlier, especially in comparison to the other XUE sources. Following the approach of \citet{Ramirez-Tannus_2023}, which was originally presented in \citet{Perotti_2023}, we also computed the continuum spectral index between 13 and 25 $\mu$m, as a proxy of the level of dust depletion of the inner disk. The rightmost panel of Figure~\ref{fig:line_luminosities} suggests that the inner disk of XUE 10 is not depleted in $\mu$m-size dust grains, similarly to the bulk of surveyed T~Tauri and Herbig disks, albeit a large spread is present. Dust depletion of pebble-size grains is instead common among disks observed around Herbig stars \citep[e.g.,][]{Banzatti_2018}, but with the present data this cannot be assessed for the XUE sample.

Given the relatively strong external radiation field incident on XUE 10, it is also worth comparing our results with the d203-506 proplyd in the ONC. Notably, we do not detect the ro-vibrational band of CH$_{3}$$^{+}$ at 7.15 $\mu$m, which has first been detected in the photo-evaporating wind of d203-506 \citep{Berne_2023}, and later in the non-irradiated TW Hya disk \citep{Henning_2024}. This non-detection may indicate that XUE 10 lacks a photo-dissociation region (PDR), which has been proposed as the origin of CH$_{3}$$^{+}$ emission observed toward d203-506, i.e., in the extended PDR surrounding the disk, as well as in the irradiated edge of the Orion Bar \citep{Zannese_2025}. This may be explained by the difference in FUV exposure between XUE 10 ($\approx$ 10$^3$ G$_0$) and d203-506 ($\approx$ 10$^4$ G$_0$), or shielding of XUE 10 from external irradiation, or the possibility that external photoevaporation has already removed most of its outer disk \citep{Ramirez-Tannus_2025}. In the non-irradiated case of TW Hya, instead, the photo-dissociation region producing CH$_{3}$$^{+}$ is interpreted as a gas depleted cavity, creating the PDR-like emission region. Additionally, OH emission, which has been proposed to signpost the destruction of water by means of FUV radiation in d203-506 \citep{Zannese_2024}, is also absent in the spectrum of XUE 10. A more detailed discussion of the OH non-detection is provided in the following sections.
\subsection{Inner disk chemistry scenarios}\label{discussion:scenarios}
Here we consider possible scenarios which could give rise to the peculiar mid-IR spectrum of XUE 10. CO$_2$ and H$_2$O represent two of the main oxygen carriers in protoplanetary disks. The relative abundance of these species is therefore highly informative.

We can summarize the presence and replenishment of CO$_2$ and H$_2$O in the inner disk of protoplanetary disks via two mechanisms: ice sublimation of the mantle from radially drifting dust grain at the H$_2$O and CO$_2$ ice lines and local gas phase formation through hydroxide (OH). In particular, interstellar and cometary ices are known to be rich in CO$_2$, with measured abundances of 15--25\% compared to H$_2$O ice, or $\sim$ 10$^{-5}$ with respect to the total gas density \citep[e.g.,][]{deGraauw_1996, Bergin_2005, Boogert_2015, LeRoy_2015}. We neglect grain-surface formation of CO$_2$ in the inner disk, which is negligible compared to gas phase production \citep{Bosman_2017}. Considering the latter, OH can react with CO to form CO$_2$ via: CO + OH $\rightarrow$ CO$_2$ + H \citep{Smith_2004}, which has a negligible activation energy and is most effective at gas temperatures up to $\sim$ 400 K, after which destruction channels for OH and CO$_2$ start to become important. In fact, OH can also react with molecular hydrogen to form water: OH + H$_2$  $\rightarrow$ H$_2$O + H \citep{Glassgold_2009}. This reaction becomes effective at $\gtrsim$ 200--300 K, due to a high activation barrier of nearly 1700 K, necessary to form OH first \citep{McElroy_2013}. Given the exothermic (spontaneous) nature of the CO$_2$ and H$_2$O formation channels, and the availability of gas phase H$_2$ and OH reactants in the warm disk atmosphere (200--1000 K), H$_2$O should be more abundant than CO$_2$ in the inner disk but on condition of neglecting photochemical effects and vertical or turbulent mixing. This was first evident from \textit{Spitzer} observations, where a low CO$_2$ abundance, 10$^{-4}$ with respect to H$_2$O ($\sim$ 10$^{-8}$ with respect to the total gas density) was found \citep{Salyk_2008, Pontoppidan_2014}, and it has been repeatedly confirmed by JWST observations in nearby regions \citep[e.g.,][]{Banzatti_2025}. 

There is, however, mounting evidence of CO$_2$-rich inner disk spectra, at least among T~Tauri sources
\citep{Gasman_2023, Grant_2023, Vasblom_2024, Salyk_2025}, alongside modeling efforts \citep{Bosman_2022, Sellek_2024} investigating the origin of variations in the spectra of CO$_2$ in contrast to water. XUE 10 is a special case, exhibiting the highest contrast between CO$_2$ and H$_2$O mid-infrared emission to date. It is currently also the only Herbig disk with a CO$_2$-rich mid-infrared spectrum. We therefore need to explain why XUE 10 lacks bright H$_2$O vapor emission, and how CO$_2$ vapor can become so abundant and dominate over H$_2$O at the disk gas temperatures and locations probed at mid-infrared wavelengths. Additionally, we aim to investigate the origin of the potential detection of oxygen isotopic anomalies in CO$_2$ vapor in the inner disk, and contextualize our findings with the evolution of disks around young, intermediate-mass stars ($M$$_{\star}$ $>$ 1.5 $M$$_{\odot}$), and the external environment. We propose the following two explanatory scenarios and discuss their application to the case of XUE 10: 
\begin{itemize}
    \item The molecular emission observable in the inner disk is determined by in situ processes, i.e., stellar UV irradiation and local gas phase chemistry.
    \item The observable inner disk emission strongly depends on radial transport processes of gas and dust from the outer disk, whose outcome is not reset by in situ processes. These include radial drift of ice-rich dust grains and gas advection, but also processes that can impede their radial transport, specifically outer disk truncation (e.g., via external UV photoevaporation), and dust traps/gas cavities.
\end{itemize}
\subsubsection{\textit{In situ chemistry}}\label{discussion:scenario1}
The harsher UV irradiation coming from a central Herbig star, compared to a cooler T Tauri star, can drive substantial photo-dissociation of water according to the reaction: H$_2$O + $h\nu$ $\rightarrow$ OH + H \citep{Heays_2017}. If this reaction proceeds on relatively short timescales compared to vertical or turbulent mixing in the disk and compared to the local gas phase H$_2$O production, water is unable to self-shield and its abundance lowers down, while the freed OH radicals can be sequestrated by CO to form CO$_2$. CO$_2$ itself can undergo photo-dissociation, releasing CO and O via: CO$_2$ + $h\nu$ $\rightarrow$ CO + O \citep{Heays_2017}, and O$_2$ secondarily \citep{Lu_2014}. If the UV photo-dissociation rate of H$_2$O and CO$_2$ molecules were of the same order of magnitude, both species would efficiently self-shield. In this sense, the CO$_2$--H$_2$O dichotomy could be explained by the disk thermal structure. In other words, H$_2$O and CO$_2$ (and OH) would get efficiently destroyed in the upper warm disk layers by stellar irradiation, while in the deeper and cooler layers atomic O and OH molecules would rather form CO$_2$ instead of water \citep{Bosman_2018}. However, both H$_2$O and OH (with much fewer detections of CO$_2$) have been observed toward several disks around older A- and B-type Herbig stars in nearby regions \citep{Fedele_2011, Banzatti_2020}, where the central stars have comparable luminosities and likely emit stronger UV radiation than the F-type star in XUE 10. We therefore speculate that CO$_2$ photo-dissociation is not as significant as H$_2$O photo-dissociation in XUE 10, and that CO$_2$ but not water may self-shield in its inner regions. This implies that alternatively to thermal effects, a high column density of CO$_2$ can be achieved in the inner disk through gas chemistry by actively removing water. 
Interestingly, this possibility agrees with a thermochemical DALI synthetic spectrum where water self-shielding is not included \citep{Bosman_2022}, pointing toward a deep atmospheric layer of CO$_2$. The gas temperature derived in this work for $^{12}$CO$_2$ ($\approx$ 370 K; see Table~\ref{tab1}) together with the observation of its hot $\upsilon_2$ $\mathit{Q}$-branch bandheads at 13.9 and 16.2 $\mu$m (excited only in very dense gas), further support the hypothesis of a deep CO$_2$ layer in the disk atmosphere of XUE 10. We note that $^{12}$CO$_2$ may be partially non-LTE excited, particularly in the $\mathit{Q}$-branch. Non-LTE excitation conditions may arise due to infrared pumping fields, which are known to affect other species such as CO and hydrogen cyanide (HCN) at infrared wavelengths \citep[e.g.,][]{Thi2011, Bruderer_2015}. This would imply a lower gas temperature and a larger emitting area than estimated under LTE assumptions. A colder gas reservoir ($T$$_{\rm gas}$ $<$ 300 K) would in turn strengthen the hypothesis of an efficient CO$_2$ gas phase production in the inner disk at the expense of water \citep{Bosman_2017, Salyk_2025}.

It is important to underline that the production of CO$_2$ vapor through its gas phase formation channel crucially depends on the availability of OH molecules. This radical primarily forms through the reaction O + H$_2$ $\rightarrow$ OH + H, with an activation barrier of $\sim$ 1700 K \citep{McElroy_2013}, through water photo-dissociation, and in a minor fraction via H + O$_2$ $\rightarrow$ OH + H, which is characterized by a much higher activation energy of $\sim$ 8000 K \citep{Baulch_1992}. 
The non-detection of OH in XUE 10 further suggests a low water column density in its inner disk, which may limit the photo-dissociation-driven replenishment of this radical \citep{Zannese_2024}. The rotationally excited emission lines of OH may also remain undetectable with JWST due to limited sensitivity above the dust continuum.

This in situ scenario can explain a low H$_2$O column density, the non-detection of OH, and the observation of a high column density of CO$_2$, and therefore the detection of bright CO$_2$ isotopologues in emission. However, as water must be removed to allow for efficient CO$_2$ production, this scenario critically depends on the relative timescales of H$_2$O photo-dissociation and of vertical mixing in the inner disk of XUE 10. Concerning the latter, observational evidence suggests weak turbulence in the well-studied Herbig disk HD 163296 \citep{Flaherty_2015}, with no significant differences in turbulent mixing strength reported between Herbig and T~Tauri disks \citep{Mulders_2012}. The oddity might therefore regard the timescale over which H$_2$O photo-dissociates, which must be unusually short in XUE 10 in comparison to Herbig disks of earlier spectral type (A and B), and more evolved ($\gtrsim$ 3 Myr).
Moreover, the presence of potential oxygen isotopic anomalies in CO$_2$ vapor in the inner disk cannot be explained solely by local gas phase channels, unless a substantial dust shielding is present in the inner disk, preventing the UV photo-dissociation of the optically thinner CO$_2$ isotopologues forming in the gas phase. We propose that a more realistic scenario should include the impact of radial transport processes of gas and ice-rich dust grains from the outer to the inner disk, which we consider in the following section.
\subsubsection{\textit{Transport-dependent chemistry}}\label{discussion:scenario2}
We start by considering the process of gas advection, connected to the dust drift, which leads to viscous gas accretion onto the star. While the $\mu$m-size dust grains are well coupled to the gas and follow its motion, the bigger grains, so-called pebbles, will decouple from the gas and radially drift inward, settling toward the disk midplane. Due to the gradient in temperature across the disk, and accounting for some vertical mixing, these pebbles will release part of their icy mantle in the gas phase via sublimation when crossing the ice line of their mantle components
\citep{Ciesla_2006, Lambrechts_2012, Banzatti_2020, Banzatti_2023}. The main ice lines in protoplanetary disks correspond, in order of increasing radial distance from the star, to H$_2$O, CO$_2$, CO, and methane (CH$_4$), but their exact radial and vertical positions are determined by the detailed thermal structure of the disk \citep{Oberg_2011, Piso_2015}.

We may therefore be observing the CO$_2$-rich gas phase of XUE 10 that follows the advection of the bulk of water vapor and its viscous accretion onto the star, i.e., while the CO$_2$ gas is still drifting inward. This may be reasonable, considering that the advection timescale is expected to be shorter in IMTT or Herbig disks compared to lower-mass counterparts, as the accretion rates are generally higher, and the gas temperatures may enhance the disk viscosity and therefore the inward advection \citep[e.g.,][]{Dullemond_2004, Calvet_2004, Lopez_2006}. However, the VLT/KMOS infrared spectrum of XUE 10 \citep{Ramirez_2020} does not show evidence of the Br$\gamma$ line, one of the main proxies for stellar accretion, suggesting that XUE 10 has a low accretion luminosity, and therefore gas advection may not be very efficient. There is also evidence that midplane cosmic-ray UV fields can boost OH and favor CO$_2$-rich ice formation in the inner disk. As these pebbles drift inward and their volatiles desorb, the gas can absorb~30\% of the cosmic-ray UV flux, throttling new H$_2$O photo-dissociation and locking in the CO$_2$ excess \citep{Chaparro_1, Chaparro_2}.

Even if the water vapor was being effectively advected past the snow line, we might be insensitive to an enhancement in column density of gas phase H$_2$O in the upper disk layers with infrared observations. In fact, the influx of dust that accompanies the inward drift of water-ice may obscure the enhanced emission of water vapor near the snow line \citep{Sellek_2024, Houge_2025}. This is because the dry dust grains inside the snow line are thought to fragment more easily; the resulting smaller grains will couple to the gas, producing a \enquote{traffic jam} of dust \citep{Pinilla_2016}, and increase the dust opacity ($\tau_{\rm dust}$) in the inner disk. Outside the H$_2$O snow line, instead, the icy dust grains are predicted to have a higher fragmentation velocity, i.e., they fragment less easily \citep{Blum_2008, Gundlach_2015}. Therefore, the CO$_2$ emission coming from the ice line further out in the disk will be less affected by this increase of dust opacity. 

\citet{Sellek_2024} proposed that the observed CO$_2$ column density, and particularly the CO$_2$/H$_2$O ratio, can be used to trace the release of volatiles from icy grains drifting inward in protoplanetary disks. The lower limit for the CO$_2$/H$_2$O column density ratio found in XUE 10 ($N$$_{\rm tot}$ (CO$_2$/H$_2$O) $\approx$ 130) is three orders of magnitude higher than that observed in other disks, such as the CO$_2$-rich T Tauri GW Lup disk \citep[0.7;][]{Grant_2023}. Such a high ratio may indicate efficient delivery of both CO$_2$ and H$_2$O ices past their respective sublimation fronts. The apparent absence of water vapor in the disk region observable with JWST could therefore be due to the higher mid-IR optical depth ($\tau_{\rm dust}$ = 1 mid-IR surface) inside the water snowline, obscuring the residual H$_2$O vapor located closer-in the CO$_2$-emitting region.
\begin{figure}[htp!]
\centering
\includegraphics[width=0.9\columnwidth]{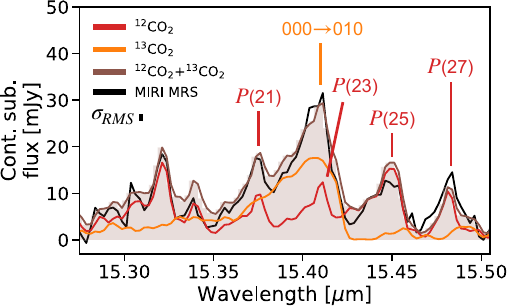}
\caption{Comparison of the $^{13}$CO$_2$ fundamental $\upsilon_2$ $\mathit{Q}$-branch (orange) with $^{12}$CO$_2$ $\mathit{P}$-branch lines (red) in the continuum-subtracted spectrum of XUE 10 (black). The orange and red curves refer respectively to the $^{13}$CO$_2$ and $^{12}$CO$_2$ best-fit slab models. In brown with shaded filling, we indicate the sum of the two models. The spectral uncertainty is labeled in the figure.
}\label{fig:CO2iceline}
\end{figure}
Following the discussion in \citet{Bosman_2017}, we highlight that the line ratio between the $^{12}$CO$_2$ and $^{13}$CO$_2$ isotopic species is informative of the ice sublimation at the CO$_2$ ice line. 
As the $\mathit{Q}$-branch of $^{13}$CO$_2$ visible at 15.42 $\mu$m is optically thinner than $^{12}$CO$_2$, it may trace deeper in the disk and therefore be more sensitive to the abundance structure of CO$_2$. We note that a similar argument applies to other optically thinner isotopic species, i.e., $^{16}$O$^{12}$C$^{18}$O and $^{16}$O$^{12}$C$^{17}$O in this context. At present, however, these isotopic species are more challenging to compare using line ratios, given their unique simultaneous detections in XUE 10, and their expected much lower abundance and band strength. Moreover, it is unclear whether these isotopologues are tracing deeper disk layers than $^{12}$CO$_2$ (see Section \ref{subsec:isotopic_ratios}).

Focusing on $^{13}$CO$_2$, its $\mathit{Q}$-branch largely dominates over the nearby contributions from the CO$_2$ $\mathit{P}$-branch in the spectral region of XUE 10 of Figure~\ref{fig:CO2iceline}, with a peak line flux of 18 mJy. Similarly to \citet{Grant_2023}, in XUE 10 the CO$_2$ $\mathit{P}$(25) $\mathit{P}$-branch line (15 mJy) at 15.45 $\mu$m is stronger than the $\mathit{P}$(23) line (12 mJy) at 15.41 $\mu$m, due to the contribution of a hot band of the $\upsilon_2$ $\mathit{Q}$-branch that arises only at high temperatures or high column densities. We therefore considered the $\mathit{P}$(27) line (7 mJy) at 15.48 $\mu$m as representative of individual $\mathit{P}$-branch lines with similar $\mathit{J}$ levels. From our best-fit models, we estimate a peak line ratio between the $^{13}$CO$_2$ $\mathit{Q}$-branch and the $\mathit{P}$(27) line of $^{12}$CO$_2$ to be 2.37 $\pm$ 1.05, which is comparable within error bar to the ratio found in GW Lup \citep[1.4;][]{Grant_2023}, and MY Lup \citep[2.24 $\pm$ 0.24,][]{Salyk_2025}. 
The non-LTE CO$_2$ excitation model computed with DALI by \citet{Bosman_2017} predicts that a different initial inner/outer abundance ($x_{\rm in}$/$x_{\rm out}$) of CO$_2$ (with respect to the total gas density), or a different
gas-to-dust ratio ($g/d$) can impact the observed line ratios between $^{12}$CO$_2$ $\mathit{P}$- and $\mathit{R}$- lines, and between the $^{13}$CO$_2$ $\mathit{Q}$-branch and a representative $^{12}$CO$_2$ $\mathit{P}$- line. The $\mathit{Q}_{\rm ^{13}CO_2}$-branch/$\mathit{P}_{{\rm ^{12}CO_2}}$(27) line ratio we measure toward XUE 10 is consistent within uncertainties with a wide range of models \citep[see Figure 9 of][]{Bosman_2017}. This includes the case of GW Lup, which points at a low outer ($x_{\rm out}$ = 10$^{-8}$) and high inner ($x_{\rm in}$ = 10$^{-6}$) disk CO$_2$ abundance at enhanced gas-to-dust ratio ($g/d$ = 10\,000). This finding suggests that the difference in optical depth between $^{12}$CO$_2$ and $^{13}$CO$_2$ measured toward XUE 10 is not enough to break the degeneracy between different DALI models. A non-LTE treatment of these molecular species in the mid-IR spectrum of XUE 10 may also be required.

In this second scenario, we assumed so far a smooth transport of gas and dust through the inner disk, where the H$_2$O vapor is removed by advection and by stellar UV photo-dissociation, while CO$_2$ vapor remains abundant in the gas phase and can be strongly enhanced by local gas phase production from OH and CO in cold gas. Nevertheless, there is a number of processes that can slow down or impede this inward transport of gas and dust, and they may play a relevant role in this scenario. We focus on the presence of a dust trap or a gas cavity in the inner or outer disk, and outer disk truncation by photoevaporation.

The presence of a inner dust trap or a gas cavity between the H$_2$O and the CO$_2$ ice lines has already been proposed to explain a gas phase H$_2$O--CO$_2$ dichotomy in the inner disk \citep{Gasman_2023, Grant_2023, Vasblom_2024}. While the H$_2$O vapor will be suppressed across the gas cavity, or while the water ice-rich grains will be trapped beyond the snow line in the dust trap, CO$_2$ will still be abundant in the lukewarm gas phase, and its line flux will be less affected, because the contributions to its emission come from larger disk radii \citep{Walsh_2015, Anderson_2021, Sellek_2024}. Dust traps and gas cavities may also be present in the outer disk (if not truncated), and similarly act against the delivery of gas and water ice-rich dust grains to the inner disk. Other gap-opening mechanisms may be the internal X-ray photoevaporation driven by the central star, and forming planets in the disk \citep{Lienert_2024}. While we cannot draw any conclusion regarding the latter, internal X-ray photoevaporation is unlikely to play a role here, given that XUE 10 is undetected in X-ray with \textit{Chandra} ($L$$_{X\rm }$ $<$ 10$^{29}$ erg s$^{-1}$). 

The spectral energy distribution of XUE 10 is consistent with Group I Herbig \enquote{flaring} disks \citep{Ramirez-Tannus_2025}, which are often characterized by large cavities \citep[e.g.,][]{Banzatti_2018}, favoring the hypothesis of the presence of substructures in this disk. However, the location of traps/cavities must be compared to the location of the ice lines. Due to the higher stellar luminosity of IMTT and Herbig stars compared to T~Tauris, the ice lines of H$_2$O and CO$_2$ will be pushed significantly outward in a disk similar to XUE 10. In a flaring disk passively irradiated by a central star with the luminosity of XUE 10 ($\approx$ 69 $L$$_{\odot}$), we estimate that the H$_2$O and CO$_2$ ice lines should be roughly located at 21 au and 85 au in radius respectively \citep{Chiang_1997}. For comparison, \citet{Grant_2023} estimate in GW Lup ice line locations at $\sim$ 0.4 au (H$_2$O) and at a slightly larger radius than 1 au (CO$_2$). Therefore, the trap/cavity hypothesis may explain the CO$_2$--rich spectrum of XUE 10, but requiring that the outer disk is not truncated below $\sim$ 100 au, which may actually be the case for the XUE sample. High angular resolution observations are needed to spatially resolve the inner and outer disk and look for substructures, which are common among nearby Herbig disks \citep[e.g.,][]{Andrews_2018}. This type of observation is almost unfeasible toward XUE 10 with the current generation of telescopes, but it will become possible in the near future with the Extremely Large Telescope (ELT),\footnote{\url{https://elt.eso.org}} and with the ALMA Wideband Sensitivity Upgrade (WSU).\footnote{\url{https://www.almaobservatory.org/en/scientists/alma-2030-wsu/wsu-program/}} 

If the outer disk was to be truncated by the external FUV irradiation, similar to what is observed in disks in the ONC \citep[e.g.,][]{Boyden_2020, Boyden_2023}, this could result in gas depletion \citep{Keyte_2025}, and also in a reduced influx of dust into the inner disk, cutting down the reservoir of drifting ice-rich dust grains \citep{Facchini_2016, Qiao_2023, Garate_2024, Paine_2025}. Although the influx of CO$_2$ ice (and gas) would be reduced in a similar way to water, the combination of an outer drainage of gas and dust with the in situ chemistry discussed in Section \ref{discussion:scenario1} may reproduce the observed CO$_2$-rich spectrum of XUE 10. The recently developed model by \citet{Sellek_2024}, which considers the 1D inner disk chemical evolution resulting from gas viscous evolution and dust radial drift, also in presence of traps/cavities, represents an optimal way to further benchmark the transport of H$_2$O and CO$_2$ in IMTT or Herbig disks similar to XUE 10. Hereby, the difference in advection timescale, dust opacity, and depletion mechanism of H$_2$O compared to T~Tauri sources may crucially impact on the H$_2$O and CO$_2$ observables in the inner regions of IMTT and Herbig disks, and validate one of our proposed explanations. 

We must stress that with the current data collected on the XUE sample, we only have indirect evidence of disk truncation by photoevaporation, while truncation by other means, e.g., stellar encounters, seems unlikely \citep{Ramirez-Tannus_2025}. Follow-up sub-millimeter ALMA observations are needed to make a detection of the outer disk, and to place constrains on the disk masses of gas and dust. As recently shown by \citet{Bayron_2025}, 2D thermochemical disk models can now provide indirect inference of gas depletion and outer disk truncation from the observed JWST/MIRI spectra.

We tested whether the observed inner disk chemistry could be due to a high disk inclination, as it has been proposed for the CO$_2$-rich T Tauri disk MY Lup, which is known to have a high inclination of $\sim$ 73$^{\circ}$ \citep{Salyk_2025}. We compared ProDiMo models of an oxygen-rich (C/O = 0.01) disk with an inner radius of 0.6 au and a cavity extending between 1--2.5 au for different inclination values ($i$ = 45$^{\circ}$, 60$^{\circ}$, 65$^{\circ}$, 68$^{\circ}$, 70$^{\circ}$). In order to remain consistent with the derived stellar photometry, the disk inclination is constrained to be $<$ 68$^{\circ}$. Inclination values between 45$^{\circ}$--65$^{\circ}$ imply a reduction of the line strength up to 11\% for CO$_2$ (at 15 $\mu$m), and up to 57\% for H$_2$O (at 17 $\mu$m). Therefore, even though a more favorable disk geometry would perhaps allow us to obtain a higher signal-to-noise ratio on water lines, we would still need to explain the observed high line fluxes of CO$_2$. However, higher inclination values could still be feasible with a different disk setup, e.g., a more compact disk, or a disk with stronger dust settling. More robust conclusions on this topic would require a dedicated thermo-chemical disk model.
\begin{figure}[t]
    \centering
    \includegraphics[width=0.45\textwidth]{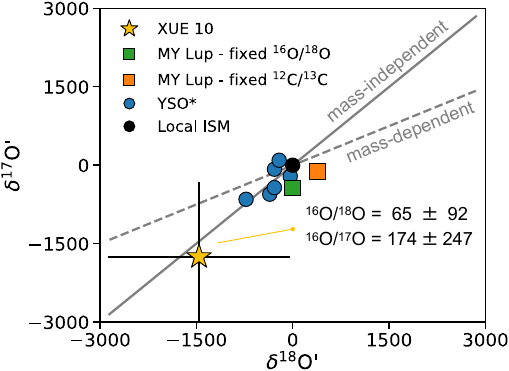}
    \caption{Oxygen isotope fractionation of protostellar systems relative to the ISM (black circle). Solid and dashed lines have a slope of 0.5 and 1, as expected respectively for mass-independent and mass-dependent isotopic fractionation. XUE 10 is shown with a star marker and error bars, and the derived ratios are also labeled with uncertainties. The green and orange square markers refer to the values derived from CO$_2$ emission toward MY Lup with different assumptions \citep{Salyk_2025}. The blue circles indicate the isotopic ratios derived from CO emission toward the sample of nearby young solar analogs (YSOs) from \citet{Smith_2015}.}
    \label{fig:isotopic_ratios}
\end{figure}
\subsection{Oxygen isotopic ratios in relation with Solar System meteorites and the ISM}\label{discussion:isotopic_ratios}
In  Section \ref{subsec:isotopic_ratios} we concluded that $^{13}$CO$_2$ is most likely emitting at $^{12}$C/$^{13}$C ISM ratio. Hence, in the following we focus on the interpretation of potential oxygen isotopic anomalies detected in $^{16}$O$^{12}$C$^{18}$O and $^{16}$O$^{12}$C$^{17}$O in a protoplanetary disk. 

The extensive literature of over five decades discusses oxygen isotopic anomalies in meteorites, i.e., solids that condensed in the Sun’s protoplanetary disk about 4.53 Gyr ago \citep{Clayton93}. Meteoritic solids are thought to have interacted with gaseous reservoirs in the solar nebula, when the protosolar disk was still present. The isotopic anomalies found in meteorites are often too large to be explained by atomic mass-dependent fractionation, i.e., according to the mass of the isotope, or aqueous processing in the disk. A mass-independent isotopic fractionation in the gas reservoir is required. 
In particular, the reservoir of oxygen isotopes in the solar nebula appears to have experienced complex spatial and temporal variations, as observed in various classes of carbonaceous chondrites, i.e., chondrule-poor stony meteorites \citep{Taylor65, Clayton73, Clayton99}, and early and late crystallizing minerals \citep{Clayton93}. The mobility of meteoritic components within the disk may have also played a role in the solar nebula; for example, meteoritic components have been ejected from inner to outer regions in an X-ray wind, as discussed by \citet{Shu00}. 

Various processes have been proposed to coherently explain the meteoritic findings, e.g., isotopic inheritance from different parts of the ISM, supernova ejecta, spallation by solar flare protons in the disk, and disequilibrium chemistry. \citet{Thiemans83} were the first to make a connection between a source of UV radiation, particularly the young Sun, with the mass-independent fractionation of $^{18}$O/$^{17}$O oxygen atoms in the O$_2$ molecule.
Later on, \citet{Lyons05} proposed the selective photo-dissociation of CO, rather than O$_2$, caused by the UV irradiation of a young O or B star external to the solar nebula to explain the mass-independent fractionation of oxygen atoms in meteorites. The scenario of CO photo-dissociation due to the UV irradiation of the central star has been validated for protoplanetary disks by \citet{Visser_2009} and \citet{Miotello_2014}.

Specifically, in the outer regions of gas-rich disks, at $N_{\rm tot}$ $>$ 10$^{15}$ cm$^{-2}$ CO can self-shield against UV irradiation, leading to a depletion of $^{12}$C$^{18}$O and $^{12}$C$^{17}$O relative to $^{12}$C$^{16}$O in the gas phase. This is because the UV radiation penetrates deeper in the optically thinner absorbing lines of the less abundant isotopologues. The relatively high $^{12}$C$^{16}$O/$^{12}$C$^{18}$O ratio observed in CO toward a small sample of nearby T~Tauri disks \citep{Brittain_2005, Smith_2009} may be explained by this isotopic-selective CO self-shielding. In XUE 10, though, we find enhanced chemical abundances of the $^{16}$O$^{12}$C$^{18}$O and $^{16}$O$^{12}$C$^{17}$O isotopologues of CO$_2$ instead of CO. A transport-dependent chemical scenario can account for the observed oxygen isotopic anomaly in CO$_2$ in the inner disk of a protoplanetary disk.
Assuming that CO$_2$ ices have a standard (ISM) isotopic composition in oxygen, the isotopic anomaly can be transferred to CO$_2$ from CO gas and H$_2$O ices via photochemistry. In particular, isotope-selective UV photo-dissociation of CO in the outer disk releases $^{18}$O and $^{17}$O atoms that can be vertically mixed downward in the disk through the \enquote{cold finger} effect. These oxygen atoms can then participate in surface reactions on dust grains in the midplane, forming H$_2^{18}$O- and H$_2^{17}$O-enriched water ices \citep{Stevenson_1988, Meijerink_2009}. These isotopically enriched water ices then need to be transported inward, possibly via radial drift of dust grains. Near the snow line, accounting for vertical mixing, the ices sublimate and water photo-dissociation can release $^{18}$OH and $^{17}$OH atoms into the gas phase, allowing in the end for the formation of large amounts of $^{16}$O$^{12}$C$^{18}$O and $^{16}$O$^{12}$C$^{17}$O vapor.

For this whole process to work, the mixing and inward drift of isotopically anomalous H$_2$O ices must occur on timescales shorter than the timescale of photoevaporation, or in general dissipation, of the outer disk. This condition is consistent with photoevaporating disk models based on the FRIED grid \citep{Sellek_2020}, which predict a disk photoevaporation timescale $>$ 1 Myr. In contrast, the peak flux of inwardly drifting dust is predicted to occur within 10$^{5}$ yr for an impinging irradiation of 1000 G$_0$. Additionally, \citet{Lyons05} show that H$_2$$^{18}$O and H$_2$$^{17}$O ice reservoirs can quickly form in the disk midplane within 10$^{4-5}$ yr, although a vigorous vertical mixing ($\alpha$ $\approx$ 10$^{-2}$) is required. 

Alternatively, CO$_2$ ices may be intrinsically anomalous in their oxygen composition, and thus may not reflect an anomaly in H$_2$O ice. Accounting for inward dust drift, these CO$_2$ ices will eventually cross the CO$_2$ ice line, releasing $^{16}$O$^{12}$C$^{18}$O and $^{16}$O$^{12}$C$^{17}$O into the gas phase and enriching the inner disk reservoir of O isotopes locked in CO$_2$. Yet, the abundance of icy CO$_2$ transported inward is expected to be significantly lower than that of H$_2$O ice. Consequently, the enrichment of the inner disk gas reservoir following ice line sublimation is likely dominated by water, unless it has already been accreted onto the central star \citep{Sellek_2024}. Nevertheless, an anomalous strength of infrared molecular bands may not reflect anomalies in the abundance structure of a given molecular species in the disk, as different spectral features may emit from different disk layers. A meaningful translation between CO$_2$ infrared band strengths and abundance pattern requires a detailed isotope-sensitive thermochemical-structural model for this source. 

Given all these caveats, we can place XUE 10 in an oxygen-isotope fractionation plot, as shown in Figure~\ref{fig:isotopic_ratios}, in relation to other protoplanetary disks and to the ISM \citep{Penzias_1981, Wilson_1994}, which resembles the Solar System \citep{Ireland_2020}. Here, $\delta$$^{17}$O$'$ and $\delta$$^{18}$O$'$ are defined as: 10$^{3}$ ln($R$$_{\rm obs}$/$R$$_{\rm ISM}$), where $R$$_{\rm obs}$ is the observed $^{16}$O/$^{17}$O and $^{16}$O/$^{18}$O isotopic (abundance) ratio, while $R$$_{\rm ISM}$ is the corresponding observed isotopic (abundance) ratio in the ISM \citep{Smith_2009}. The derived oxygen isotopic ratios for XUE 10 labeled in figure correspond to the column density ratios of $^{12}$CO$_2$/$^{16}$O$^{12}$C$^{18}$O and $^{12}$CO$_2$/$^{16}$O$^{12}$C$^{17}$O (in Table~\ref{tab1}) with the errors propagated from the $\chi^2_{\rm red}$ 1$\sigma$ ranges per species. The error bar on the $\delta$$^{17}$O$'$/$\delta$$^{18}$O$'$ ratio is computed by propagating the uncertainties on the quantities in the definition of $\delta$$^{17/18}$O$'$. The uncertainty on the measured $^{16}$O/$^{18}$O and $^{16}$O/$^{17}$O ratios in XUE 10 is clearly too large to distinguish between different fractionation scenarios, and the interpretation of the isotopic ratio remains difficult.
As pointed out by \citet{Salyk_2025}, though, the detection of multiple CO$_2$ isotopologues, made now possible with JWST, is promising for future avenue aimed to trace the evolution of molecular chemistry through isotopy at different planet formation stages in analogy to the Solar System.
\section{Conclusions} 
\label{sec:conclusions}
We have presented JWST/MIRI MRS observations of the inner part of the Herbig disk XUE 10 ($M$$_{\star}$ $\approx$ 3 $M$$_{\odot}$) located in the massive NGC~6357 cluster (1.69 kpc) and which has an estimated external FUV irradiation exposure of about 5 $\times$ 10$^3$ $G_{\rm 0}$ stronger than the solar neighborhood and nearby star-forming regions. We summarize our conclusions as follows:
\begin{enumerate}
    \item We report the first simultaneous detection ($>$ 5$\sigma$) of four isotopic species of CO$_2$ ($^{12}$CO$_{2}$, $^{13}$CO$_{2}$, $^{16}$O$^{12}$C$^{18}$O, $^{16}$O$^{12}$C$^{17}$O) in a protoplanetary disk, a faint CO emission (2$\sigma$), atomic hydrogen HI Pf$\alpha$ (8$\sigma$), and an upper limit on the H$_2$O content ($N$$_{\rm tot}$ $\lesssim$ 10$^{18}$ cm$^{-2}$). 
    \item The species $^{16}$O$^{12}$C$^{18}$O and $^{16}$O$^{12}$C$^{17}$O may be isotopically enhanced compared to the oxygen isotopic ratios ($^{16}$O/$^{18}$O and $^{16}$O/$^{17}$O) found in the ISM and in the Solar System, though line optical depth effects cannot be ruled out.
    \item We detected PAH emission at 6.2, 7.7--8.6, and 11 $\mu$m and of silicate aggregates at 10--11 $\mu$m, which is consistent with other observations of IMTT and Herbig disks, except for a 15.5--17 $\mu$m feature whose origin remains unclear. 
    \item We propose that the observed chemistry in XUE 10 depends on transport processes. Water is removed either via advection or strong photo-dissociation by stellar UV irradiation. CO$_2$ is locally enhanced in the gas phase via the CO + OH $\rightarrow$ CO$_2$ + H reaction, facilitated by water removal in relatively cold gas ($T$$_{\rm gas}$ $<$ 300 K). External FUV irradiation supports the observed CO$_2$--H$_2$O dichotomy and the potential detection of CO$_2$ oxygen isotopic anomalies in the inner disk. 
\end{enumerate}

A controlled infrared sample of IMTT and Herbig disks both in isolated and irradiated environments is needed at JWST resolution to perform a statistical comparison of the disk properties around young intermediate-mass stars and to corroborate trends as a function of the stellar mass. Major modeling efforts are also needed to understand whether the \enquote{line-poor} evidence in disks around intermediate-mass stars is merely an observational bias due to a limitation in sensitivity against the bright dust continuum or whether it is symptomatic of underlying crucial physical differences between IMTT or Herbig disks and the lower-mass counterparts. The detection of several CO$_2$ isotopologues in XUE 10 highlights the possibility of investigating the isotopic composition of the molecular carriers of the CHNOPS life elements at the protoplanetary disk stage. 
\begin{acknowledgements}
The authors thank the anonymous referee for their insightful comments and suggestions that improved the quality of this manuscript. The authors also thank Marc Chaussidon for useful discussions on isotopic anomalies in planet formation.\\ 

This work is based on observations made with the NASA/ESA/CSA James Webb Space Telescope (JWST), and on observations collected at the European Southern Observatory under ESO programme 113.26CR. The JWST data were obtained from the Mikulski Archive for Space Telescopes (MAST) at the Space Telescope Science Institute (STScI), which is operated by the Association of Universities for Research in Astronomy, Inc., under NASA contract NAS 5-03127 for JWST. These observations are associated with program \#1759.\\

JF and AB acknowledge support from the Swedish National Space Agency (2022-00154).

MCRT acknowledges support by the German Aerospace Center (DLR) and the Federal Ministry for Economic Affairs and Energy (BMWi) through program 50OR2314 ‘Physics and Chemistry of Planet-forming disks in extreme environments’.

BT acknowledges the support by the Programme National PCMI of CNRS/
INSU with INC/INP cofunded by CEA and CNES.

TJH acknowledges UKRI guaranteed funding for a Horizon Europe ERC consolidator grant (EP/Y024710/1) and a Royal Society Dorothy Hodgkin Fellowship.

BPR's research was supported by NASA STScI grant JWST-GO-01759.002-A.

The work of TP was partly supported by the  Excellence Cluster 
ORIGINS which is funded by the Deutsche Forschungsgemeinschaft (DFG, 
German Research Foundation) under Germany's Excellence Strategy - 
EXC-2094 - 390783311.

VR acknowledges the support of the European Union’s Horizon 2020 research and
innovation program and the European Research Council via the ERC Synergy
Grant “ECOGAL” (project ID 855130).
\end{acknowledgements}
\bibliographystyle{aa}
\bibliography{aa55718-25}

\begin{thebibliography}{183}
\expandafter\ifx\csname natexlab\endcsname\relax\def\natexlab#1{#1}\fi

\bibitem[{{Adams} {et~al.}(2004){Adams}, {Hollenbach}, {Laughlin}, \& {Gorti}}]{Adams_2004}
{Adams}, F.~C., {Hollenbach}, D., {Laughlin}, G., \& {Gorti}, U. 2004, \apj, 611, 360

\bibitem[{{Agúndez} {et~al.}(2018){Agúndez}, {Roueff, Evelyne}, {Le Petit, Franck}, \& {Le Bourlot, Jacques}}]{Marcelino_2018}
{Agúndez}, {Roueff, Evelyne}, {Le Petit, Franck}, \& {Le Bourlot, Jacques}. 2018, A\&A, 616, A19

\bibitem[{{Allen} {et~al.}(2025){Allen}, {Anania}, {Andersen}, {Aru}, {Ballabio}, {Ballering}, {Beccari}, {Bern{\'e}}, {Bik}, {Boyden}, {Coleman}, {D{\'\i}az-Berrios}, {Eatson}, {Frediani}, {Forbrich}, {Gkimisi}, {Goicoechea}, {Gupta}, {Guarcello}, {Haworth}, {Henney}, {Isella}, {Itrich}, {Keyte}, {Kim}, {Kuhn}, {Le Petit}, {Luo}, {Manara}, {Mauco}, {Meshaka}, {Millstone}, {Owen}, {Paine}, {Parker}, {Peake}, {Peatt}, {Pinilla}, {Qiao}, {Ram{\'\i}rez-Tannus}, {Ramsay}, {Reiter}, {Rogers}, {Rosotti}, {Schroetter}, {Sellek}, {Testi}, {van Terwisga}, {Vicente}, {Walsh}, {Winter}, {Wright}, \& {Zeidler}}]{Haworth_2025}
{Allen}, M., {Anania}, R., {Andersen}, M., {et~al.} 2025, OJAp, 8, 54

\bibitem[{Anderson {et~al.}(2017)Anderson, Bergin, Blake, Ciesla, Visser, \& Lee}]{Anderson_2017}
Anderson, D.~E., Bergin, E.~A., Blake, G.~A., {et~al.} 2017, \apj, 845, 13

\bibitem[{Anderson {et~al.}(2021)Anderson, Blake, Cleeves, Bergin, Zhang, Schwarz, Salyk, \& Bosman}]{Anderson_2021}
Anderson, D.~E., Blake, G.~A., Cleeves, L.~I., {et~al.} 2021, \apj, 909, 55

\bibitem[{Andrews {et~al.}(2018)Andrews, Huang, Pérez, Isella, Dullemond, Kurtovic, Guzmán, Carpenter, Wilner, Zhang, Zhu, Birnstiel, Bai, Benisty, Hughes, Öberg, \& Ricci}]{Andrews_2018}
Andrews, S.~M., Huang, J., Pérez, L.~M., {et~al.} 2018, \apjl, 869, L41

\bibitem[{{Ansdell} {et~al.}(2017){Ansdell}, {Williams}, {Manara}, {Miotello}, {Facchini}, {van der Marel}, {Testi}, \& {van Dishoeck}}]{Ansdell_2017}
{Ansdell}, M., {Williams}, J.~P., {Manara}, C.~F., {et~al.} 2017, \apj, 153, 240

\bibitem[{{Antonellini} {et~al.}(2016){Antonellini}, {Kamp, I.}, {Lahuis, F.}, {Woitke, P.}, {Thi, W.-F.}, {Meijerink, R.}, {Aresu, G.}, {Spaans, M.}, {Güdel, M.}, \& {Liebhart, A.}}]{Antonellini_2016}
{Antonellini}, {Kamp, I.}, {Lahuis, F.}, {et~al.} 2016, A\&A, 585, A61

\bibitem[{{Arabhavi} {et~al.}(2024){Arabhavi}, {Kamp}, {Henning}, {van Dishoeck}, {Christiaens}, {Gasman}, {Perrin}, {G{\"u}del}, {Tabone}, {Kanwar}, {Waters}, {Pascucci}, {Samland}, {Perotti}, {Bettoni}, {Grant}, {Lagage}, {Ray}, {Vandenbussche}, {Absil}, {Argyriou}, {Barrado}, {Boccaletti}, {Bouwman}, {Caratti o Garatti}, {Glauser}, {Lahuis}, {Mueller}, {Olofsson}, {Pantin}, {Scheithauer}, {Morales-Calder{\'o}n}, {Franceschi}, {Jang}, {Pawellek}, {Rodgers-Lee}, {Schreiber}, {Schwarz}, {Temmink}, {Vlasblom}, {Wright}, {Colina}, \& {{\"O}stlin}}]{Arabhavi_2024}
{Arabhavi}, A.~M., {Kamp}, I., {Henning}, T., {et~al.} 2024, Science, 384, 1086

\bibitem[{{Aru} {et~al.}(2024){Aru}, {Mauc{\'o}}, {Manara}, {Haworth}, {Facchini}, {McLeod}, {Miotello}, {Petr-Gotzens}, {Robberto}, {Rosotti}, {Vicente}, {Winter}, \& {Ansdell}}]{Aru_2024}
{Aru}, M.~L., {Mauc{\'o}}, K., {Manara}, C.~F., {et~al.} 2024, \aap, 687, A93

\bibitem[{Arulanantham {et~al.}(2025)Arulanantham, Salyk, Pontoppidan, Banzatti, Zhang, Öberg, Long, Carr, Najita, Pascucci, Colmenares, Xie, Huang, Green, Andrews, Blake, Bergin, Pinilla, Vioque, Dahl, Raul, Krijt, \& Collaboration}]{Arulanantham_2025}
Arulanantham, N., Salyk, C., Pontoppidan, K., {et~al.} 2025, \apj, 170, 67

\bibitem[{Ballering {et~al.}(2023)Ballering, Cleeves, Haworth, Bally, Eisner, Ginsburg, Boyden, Fang, \& Kim}]{Ballering_2023}
Ballering, N.~P., Cleeves, L.~I., Haworth, T.~J., {et~al.} 2023, \apj, 954, 127

\bibitem[{{Banzatti} {et~al.}(2018){Banzatti}, {Garufi, A.}, {Kama, M.}, {Benisty, M.}, {Brittain, S.}, {Pontoppidan, K. M.}, \& {Rayner, J.}}]{Banzatti_2018}
{Banzatti}, {Garufi, A.}, {Kama, M.}, {et~al.} 2018, A\&A, 609, L2

\bibitem[{Banzatti {et~al.}(2022)Banzatti, Abernathy, Brittain, Bosman, Pontoppidan, Boogert, Jensen, Carr, Najita, Grant, Sigler, Sanchez, Kern, \& Rayner}]{Banzatti_2022}
Banzatti, A., Abernathy, K.~M., Brittain, S., {et~al.} 2022, \apj, 163, 174

\bibitem[{{Banzatti} {et~al.}(2020){Banzatti}, {Pascucci}, {Bosman}, {Pinilla}, {Salyk}, {Herczeg}, {Pontoppidan}, {Vazquez}, {Watkins}, {Krijt}, {Hendler}, \& {Long}}]{Banzatti_2020}
{Banzatti}, A., {Pascucci}, I., {Bosman}, A.~D., {et~al.} 2020, \apj, 903, 124

\bibitem[{Banzatti {et~al.}(2023)Banzatti, Pontoppidan, Chávez, Salyk, Diehl, Bruderer, Herczeg, Carmona, Pascucci, Brittain, Jensen, Grant, van Dishoeck, Kamp, Bosman, Öberg, Blake, Meyer, Gaidos, Boogert, Rayner, \& Wheeler}]{Banzatti_2023}
Banzatti, A., Pontoppidan, K.~M., Chávez, J.~P., {et~al.} 2023, \apj, 165, 72

\bibitem[{Banzatti {et~al.}(2025)Banzatti, Salyk, Pontoppidan, Carr, Zhang, Arulanantham, Krijt, Öberg, Cleeves, Najita, Pascucci, Blake, Romero-Mirza, Bergin, Cieza, Pinilla, Long, Mallaney, Xie, Waggoner, Kaeufer, \& the JDISCS~collaboration}]{Banzatti_2025}
Banzatti, A., Salyk, C., Pontoppidan, K.~M., {et~al.} 2025, \apj, 169, 165

\bibitem[{{Baulch} {et~al.}(1992){Baulch}, {Cobos}, {Cox}, {Esser}, {Frank}, {Just}, {Kerr}, {Pilling}, {Troe}, {Walker}, \& {Warnatz}}]{Baulch_1992}
{Baulch}, D.~L., {Cobos}, C.~J., {Cox}, R.~A., {et~al.} 1992, J. Phys. Chem. Ref. Data, 21, 411

\bibitem[{{Bayo} {et~al.}(2008){Bayo}, {Rodrigo}, {Barrado Y Navascu{\'e}s}, {Solano}, {Guti{\'e}rrez}, {Morales-Calder{\'o}n}, \& {Allard}}]{Vosa}
{Bayo}, A., {Rodrigo}, C., {Barrado Y Navascu{\'e}s}, D., {et~al.} 2008, \aap, 492, 277

\bibitem[{Bergin {et~al.}(2005)Bergin, Melnick, Gerakines, Neufeld, \& Whittet}]{Bergin_2005}
Bergin, E.~A., Melnick, G.~J., Gerakines, P.~A., Neufeld, D.~A., \& Whittet, D. C.~B. 2005, \apj, 627, L33

\bibitem[{{Bern{\'e}} {et~al.}(2023){Bern{\'e}}, {Martin-Drumel}, {Schroetter}, {Goicoechea}, {Jacovella}, {Gans}, {Dartois}, {Coudert}, {Bergin}, {Alarcon}, {Cami}, {Roueff}, {Black}, {Asvany}, {Habart}, {Peeters}, {Canin}, {Trahin}, {Joblin}, {Schlemmer}, {Thorwirth}, {Cernicharo}, {Gerin}, {Tielens}, {Zannese}, {Abergel}, {Bernard-Salas}, {Boersma}, {Bron}, {Chown}, {Cuadrado}, {Dicken}, {Elyajouri}, {Fuente}, {Gordon}, {Issa}, {Kannavou}, {Khan}, {Lacinbala}, {Languignon}, {Le Gal}, {Maragkoudakis}, {Meshaka}, {Okada}, {Onaka}, {Pasquini}, {Pound}, {Robberto}, {R{\"o}llig}, {Schefter}, {Schirmer}, {Sidhu}, {Tabone}, {Van De Putte}, {Vicente}, \& {Wolfire}}]{Berne_2023}
{Bern{\'e}}, O., {Martin-Drumel}, M.-A., {Schroetter}, I., {et~al.} 2023, \nat, 621, 56

\bibitem[{{Biazzo} {et~al.}(2017){Biazzo}, {Frasca, A.}, {Alcalá, J. M.}, {Zusi, M.}, {Covino, E.}, {Randich, S.}, {Esposito, M.}, {Manara, C. F.}, {Antoniucci, S.}, {Nisini, B.}, {Rigliaco, E.}, \& {Getman, F.}}]{Biazzo_2017}
{Biazzo}, {Frasca, A.}, {Alcalá, J. M.}, {et~al.} 2017, A\&A, 605, A66

\bibitem[{{Biazzo} {et~al.}(2011){Biazzo}, {Randich, S.}, {Palla, F.}, \& {Briceño, C.}}]{Biazzo_2011}
{Biazzo}, {Randich, S.}, {Palla, F.}, \& {Briceño, C.} 2011, A\&A, 530, A19

\bibitem[{Blum \& Wurm(2008)}]{Blum_2008}
Blum, J. \& Wurm, G. 2008, ARA\&A, 46, 21

\bibitem[{{Boersma} {et~al.}(2010){Boersma}, {Bauschlicher}, {Allamandola}, {Ricca}, {Peeters}, \& {Tielens}}]{Boersma_2010}
{Boersma}, C., {Bauschlicher}, C.~W., {Allamandola}, L.~J., {et~al.} 2010, \aap, 511, A32

\bibitem[{Boogert {et~al.}(2015)Boogert, Gerakines, \& Whittet}]{Boogert_2015}
Boogert, A.~A., Gerakines, P.~A., \& Whittet, D.~C. 2015, Annu. Rev. Astron., 53, 541–581

\bibitem[{{Bosman} {et~al.}(2017){Bosman}, {Bruderer, Simon}, \& {van Dishoeck, Ewine F.}}]{Bosman_2017}
{Bosman}, {Bruderer, Simon}, \& {van Dishoeck, Ewine F.} 2017, A\&A, 601, A36

\bibitem[{Bosman {et~al.}(2022)Bosman, Bergin, Calahan, \& Duval}]{Bosman_2022}
Bosman, A.~D., Bergin, E.~A., Calahan, J.~K., \& Duval, S.~E. 2022, \apjl, 933, L40

\bibitem[{{Bosman} {et~al.}(2018){Bosman}, {Walsh}, \& {van Dishoeck}}]{Bosman_2018}
{Bosman}, A.~D., {Walsh}, C., \& {van Dishoeck}, E.~F. 2018, \aap, 618, A182

\bibitem[{Boyden \& Eisner(2020)}]{Boyden_2020}
Boyden, R.~D. \& Eisner, J.~A. 2020, \apj, 894, 74

\bibitem[{{Boyden} \& {Eisner}(2023)}]{Boyden_2023}
{Boyden}, R.~D. \& {Eisner}, J.~A. 2023, \apj, 947, 7

\bibitem[{{Bressan} {et~al.}(2012){Bressan}, {Marigo}, {Girardi}, {Salasnich}, {Dal Cero}, {Rubele}, \& {Nanni}}]{Parsec}
{Bressan}, A., {Marigo}, P., {Girardi}, L., {et~al.} 2012, \mnras, 427, 127

\bibitem[{Brittain {et~al.}(2023)Brittain, Kamp, Meeus, Oudmaijer, \& Waters}]{Brittain_2023}
Brittain, S.~D., Kamp, I., Meeus, G., Oudmaijer, R.~D., \& Waters, L. B. F.~M. 2023, Space Sci. Rev., 219

\bibitem[{Brittain {et~al.}(2005)Brittain, Rettig, Simon, \& Kulesa}]{Brittain_2005}
Brittain, S.~D., Rettig, T.~W., Simon, T., \& Kulesa, C. 2005, \apj, 626, 283

\bibitem[{{Bruderer}(2013)}]{Bruderer_2013}
{Bruderer}. 2013, A\&A, 559, A46

\bibitem[{{Bruderer} {et~al.}(2015){Bruderer}, {Harsono}, \& {van Dishoeck}}]{Bruderer_2015}
{Bruderer}, S., {Harsono}, D., \& {van Dishoeck}, E.~F. 2015, \aap, 575, A94

\bibitem[{{Bruderer} {et~al.}(2012){Bruderer}, {van Dishoeck}, {Doty}, \& {Herczeg}}]{Bruderer_2012}
{Bruderer}, S., {van Dishoeck}, E.~F., {Doty}, S.~D., \& {Herczeg}, G.~J. 2012, \aap, 541, A91

\bibitem[{{Bushouse} {et~al.}(2023){Bushouse}, {Eisenhamer}, {Dencheva}, {Davies}, {Greenfield}, {Morrison}, {Hodge}, {Simon}, {Grumm}, {Droettboom}, {Slavich}, {Sosey}, {Pauly}, {Miller}, {Jedrzejewski}, {Hack}, {Davis}, {Crawford}, {Law}, {Gordon}, {Regan}, {Cara}, {MacDonald}, {Bradley}, {Shanahan}, {Jamieson}, {Teodoro}, \& {Williams}}]{Bushouse_2023}
{Bushouse}, H., {Eisenhamer}, J., {Dencheva}, N., {et~al.} 2023, {JWST Calibration Pipeline}

\bibitem[{Calvet {et~al.}(2004)Calvet, Muzerolle, Briceño, Hernández, Hartmann, Saucedo, \& Gordon}]{Calvet_2004}
Calvet, N., Muzerolle, J., Briceño, C., {et~al.} 2004, \apj, 128, 1294

\bibitem[{{Cami} \& {Yamamura}(2001)}]{Cami_2001}
{Cami} \& {Yamamura}. 2001, A\&A, 367, L1

\bibitem[{{Chiang} \& {Goldreich}(1997)}]{Chiang_1997}
{Chiang}, E.~I. \& {Goldreich}, P. 1997, \apj, 490, 368

\bibitem[{{Ciesla} \& {Cuzzi}(2006)}]{Ciesla_2006}
{Ciesla}, F.~J. \& {Cuzzi}, J.~N. 2006, \icarus, 181, 178

\bibitem[{{Clayton}(1993)}]{Clayton93}
{Clayton}, R.~N. 1993, Annu. Rev. Earth Planet. Sci., 21, 115

\bibitem[{{Clayton} {et~al.}(1973){Clayton}, {Grossman}, \& {Mayeda}}]{Clayton73}
{Clayton}, R.~N., {Grossman}, L., \& {Mayeda}, T.~K. 1973, Science, 182, 485

\bibitem[{{Clayton} \& {Mayeda}(1999)}]{Clayton99}
{Clayton}, R.~N. \& {Mayeda}, T.~K. 1999, \gca, 63, 2089

\bibitem[{{Coleman} \& {Haworth}(2022)}]{Gavin_2022}
{Coleman}, G. A.~L. \& {Haworth}, T.~J. 2022, \mnras, 514, 2315

\bibitem[{{Concha-Ram{\'\i}rez} {et~al.}(2019){Concha-Ram{\'\i}rez}, {Wilhelm}, {Portegies Zwart}, \& {Haworth}}]{Concha_2019}
{Concha-Ram{\'\i}rez}, F., {Wilhelm}, M. J.~C., {Portegies Zwart}, S., \& {Haworth}, T.~J. 2019, \mnras, 490, 5678

\bibitem[{{Dartois} {et~al.}(2003){Dartois}, {Dutrey}, \& {Guilloteau}}]{Dartois_2003}
{Dartois}, E., {Dutrey}, A., \& {Guilloteau}, S. 2003, \aap, 399, 773

\bibitem[{{Dawson} \& {Johnson}(2018)}]{dawson_2018}
{Dawson}, R.~I. \& {Johnson}, J.~A. 2018, \araa, 56, 175

\bibitem[{{de Graauw} {et~al.}(1996){de Graauw}, {Whittet}, {Gerakines}, {Bauer}, {Beintema}, {Boogert}, {Boxhoorn}, {Chiar}, {Ehrenfreund}, {Feuchtgruber}, {Helmich}, {Heras}, {Huygen}, {Kester}, {Kunze}, {Lahuis}, {Leech}, {Lutz}, {Morris}, {Prusti}, {Roelfsema}, {Salama}, {Schaeidt}, {Schutte}, {Spoon}, {Tielens}, {Valentijn}, {Vandenbusshe}, {van Dishoeck}, {Wesselius}, {Wieprecht}, \& {Wright}}]{deGraauw_1996}
{de Graauw}, T., {Whittet}, D.~C.~B., {Gerakines}, P.~A., {et~al.} 1996, \aap, 315, L345

\bibitem[{Dishoeck {et~al.}(2023)Dishoeck, Grant, Tabone, Gelder, Francis, Tychoniec, Bettoni, Arabhavi, Gasman, Kavanagh, Nazari, Vlasblom, Christiaens, Klaassen, Beuther, Henning, \& Kamp}]{Dishoeck_2023}
Dishoeck, E., Grant, S., Tabone, B., {et~al.} 2023, Faraday Discussions, 245

\bibitem[{{Dullemond} \& {Dominik}(2004)}]{Dullemond_2004}
{Dullemond}, C.~P. \& {Dominik}, C. 2004, \aap, 421, 1075

\bibitem[{{Facchini} {et~al.}(2016){Facchini}, {Clarke}, \& {Bisbas}}]{Facchini_2016}
{Facchini}, S., {Clarke}, C.~J., \& {Bisbas}, T.~G. 2016, \mnras, 457, 3593

\bibitem[{{Fang} {et~al.}(2012){Fang}, {van Boekel}, {King}, {Henning}, {Bouwman}, {Doi}, {Okamoto}, {Roccatagliata}, \& {Sicilia-Aguilar}}]{Fang_2012}
{Fang}, M., {van Boekel}, R., {King}, R.~R., {et~al.} 2012, \aap, 539, A119

\bibitem[{Fatuzzo \& Adams(2008)}]{Fatuzzo_2008}
Fatuzzo, M. \& Adams, F.~C. 2008, \apj, 675, 1361

\bibitem[{Fedele {et~al.}(2011)Fedele, Pascucci, Brittain, Kamp, Woitke, Williams, Dent, \& Thi}]{Fedele_2011}
Fedele, D., Pascucci, I., Brittain, S., {et~al.} 2011, \apj, 732, 106

\bibitem[{Flaherty {et~al.}(2015)Flaherty, Hughes, Rosenfeld, Andrews, Chiang, Simon, Kerzner, \& Wilner}]{Flaherty_2015}
Flaherty, K.~M., Hughes, A.~M., Rosenfeld, K.~A., {et~al.} 2015, \apj, 813, 99

\bibitem[{Foreman-Mackey {et~al.}(2013)Foreman-Mackey, Hogg, Lang, \& Goodman}]{emcee}
Foreman-Mackey, D., Hogg, D.~W., Lang, D., \& Goodman, J. 2013, PASP, 125, 306–312

\bibitem[{{Fouesneau} {et~al.}(2022){Fouesneau}, {Andrae, R.}, {Dharmawardena, T.}, {Rybizki, J.}, {Bailer-Jones, C. A. L.}, \& {Demleitner, M.}}]{Fouesneau_2022}
{Fouesneau}, {Andrae, R.}, {Dharmawardena, T.}, {et~al.} 2022, A\&A, 662, A125

\bibitem[{{Garcia Lopez} {et~al.}(2006){Garcia Lopez}, {Natta}, {Testi}, \& {Habart}}]{Lopez_2006}
{Garcia Lopez}, R., {Natta}, A., {Testi}, L., \& {Habart}, E. 2006, \aap, 459, 837

\bibitem[{{Gasman} {et~al.}(2025){Gasman}, {Temmink, Milou}, {van Dishoeck, Ewine F.}, {Kurtovic, Nicolas T.}, {Grant, Sierra L.}, {Sellek, Andrew}, {Tabone, Benoît}, {Henning, Thomas}, {Kamp, Inga}, {Güdel, Manuel}, {Barrado, David}, {Caratti o Garatti, Alessio}, {Glauser, Adrian M.}, {Waters, Laurens B. F. M.}, {Arabhavi, Aditya M.}, {Jang, Hyerin}, {Kanwar, Jayatee}, {Lienert, Julia L.}, {Perotti, Giulia}, {Schwarz, Kamber}, \& {Vlasblom, Marissa}}]{Gasman_2025}
{Gasman}, {Temmink, Milou}, {van Dishoeck, Ewine F.}, {et~al.} 2025, A\&A, 694, A147

\bibitem[{{Gasman} {et~al.}(2023){Gasman}, {van Dishoeck, Ewine F.}, {Grant, Sierra L.}, {Temmink, Milou}, {Tabone, Benoît}, {Henning, Thomas}, {Kamp, Inga}, {Güdel, Manuel}, {Lagage, Pierre-Olivier}, {Perotti, Giulia}, {Christiaens, Valentin}, {Samland, Matthias}, {Arabhavi, Aditya M.}, {Argyriou, Ioannis}, {Abergel, Alain}, {Absil, Olivier}, {Barrado, David}, {Boccaletti, Anthony}, {Bouwman, Jeroen}, {Caratti o Garatti, Alessio}, {Geers, Vincent}, {Glauser, Adrian M.}, {Guadarrama, Rodrigo}, {Jang, Hyerin}, {Kanwar, Jayatee}, {Lahuis, Fred}, {Morales-Calderón, Maria}, {Mueller, Michael}, {Nehmé, Cyrine}, {Olofsson, Göran}, {Pantin, Éric}, {Pawellek, Nicole}, {Ray, Tom P.}, {Rodgers-Lee, Donna}, {Scheithauer, Silvia}, {Schreiber, Jürgen}, {Schwarz, Kamber}, {Vandenbussche, Bart}, {Vlasblom, Marissa}, {Waters, Rens L. B. F. M.}, {Wright, Gillian}, {Colina, Luis}, {Greve, Thomas R.}, \& {Östlin, Göran}}]{Gasman_2023}
{Gasman}, {van Dishoeck, Ewine F.}, {Grant, Sierra L.}, {et~al.} 2023, A\&A, 679, A117

\bibitem[{Getman {et~al.}(2014)Getman, Feigelson, Kuhn, Broos, Townsley, Naylor, Povich, Luhman, \& Garmire}]{Getman_2014}
Getman, K.~V., Feigelson, E.~D., Kuhn, M.~A., {et~al.} 2014, \apj, 787, 108

\bibitem[{{Gielen} {et~al.}(2009){Gielen}, {Van Winckel, H.}, {Matsuura, M.}, {Min, M.}, {Deroo, P.}, {Waters, L. B. F. M.}, \& {Dominik, C.}}]{Gielen_2009}
{Gielen}, {Van Winckel, H.}, {Matsuura, M.}, {et~al.} 2009, A\&A, 503, 843

\bibitem[{Glassgold {et~al.}(2009)Glassgold, Meijerink, \& Najita}]{Glassgold_2009}
Glassgold, A.~E., Meijerink, R., \& Najita, J.~R. 2009, \apj, 701, 142

\bibitem[{{Gordon} {et~al.}(2022){Gordon}, {Rothman}, {Hargreaves}, {Hashemi}, {Karlovets}, {Skinner}, {Conway}, {Hill}, {Kochanov}, {Tan}, {Wcis{\l}o}, {Finenko}, {Nelson}, {Bernath}, {Birk}, {Boudon}, {Campargue}, {Chance}, {Coustenis}, {Drouin}, {Flaud}, {Gamache}, {Hodges}, {Jacquemart}, {Mlawer}, {Nikitin}, {Perevalov}, {Rotger}, {Tennyson}, {Toon}, {Tran}, {Tyuterev}, {Adkins}, {Baker}, {Barbe}, {Can{\`e}}, {Cs{\'a}sz{\'a}r}, {Dudaryonok}, {Egorov}, {Fleisher}, {Fleurbaey}, {Foltynowicz}, {Furtenbacher}, {Harrison}, {Hartmann}, {Horneman}, {Huang}, {Karman}, {Karns}, {Kassi}, {Kleiner}, {Kofman}, {Kwabia-Tchana}, {Lavrentieva}, {Lee}, {Long}, {Lukashevskaya}, {Lyulin}, {Makhnev}, {Matt}, {Massie}, {Melosso}, {Mikhailenko}, {Mondelain}, {M{\"u}ller}, {Naumenko}, {Perrin}, {Polyansky}, {Raddaoui}, {Raston}, {Reed}, {Rey}, {Richard}, {T{\'o}bi{\'a}s}, {Sadiek}, {Schwenke}, {Starikova}, {Sung}, {Tamassia}, {Tashkun}, {Vander Auwera}, {Vasilenko}, {Vigasin}, {Villanueva}, {Vispoel}, {Wagner}, {Yachmenev}, \&
  {Yurchenko}}]{Gordon_2022}
{Gordon}, I.~E., {Rothman}, L.~S., {Hargreaves}, R.~J., {et~al.} 2022, \jqsrt, 277, 107949

\bibitem[{Grant {et~al.}(2023)Grant, van Dishoeck, Tabone, Gasman, Henning, Kamp, Güdel, Lagage, Bettoni, Perotti, Christiaens, Samland, Arabhavi, Argyriou, Abergel, Absil, Barrado, Boccaletti, Bouwman, o~Garatti, Geers, Glauser, Guadarrama, Jang, Kanwar, Lahuis, Morales-Calderón, Mueller, Nehmé, Olofsson, Pantin, Pawellek, Ray, Rodgers-Lee, Scheithauer, Schreiber, Schwarz, Temmink, Vandenbussche, Vlasblom, Waters, Wright, Colina, Greve, Justannont, \& Östlin}]{Grant_2023}
Grant, S.~L., van Dishoeck, E.~F., Tabone, B., {et~al.} 2023, \apjl, 947, L6

\bibitem[{{Gray} \& {Corbally}(2009)}]{Gray_2009}
{Gray}, R.~O. \& {Corbally}, J., C. 2009, {Stellar Spectral Classification} (Princeton University Press)

\bibitem[{{Gross} \& {Cleeves}(2025)}]{Gross_2025}
{Gross}, R.~E. \& {Cleeves}, L.~I. 2025, \apj, 980, 189

\bibitem[{Guarcello {et~al.}(2023)Guarcello, Drake, Wright, Albacete-Colombo, Clarke, Ercolano, Flaccomio, Kashyap, Micela, Naylor, Schneider, Sciortino, \& Vink}]{Guarcello_2016}
Guarcello, M.~G., Drake, J.~J., Wright, N.~J., {et~al.} 2023, ApJS, 269, 13

\bibitem[{{Gundlach} \& {Blum}(2015)}]{Gundlach_2015}
{Gundlach}, B. \& {Blum}, J. 2015, \apj, 798, 34

\bibitem[{{Gárate} {et~al.}(2024){Gárate}, {Pinilla, Paola}, {Haworth, Thomas J.}, \& {Facchini, S.}}]{Garate_2024}
{Gárate}, {Pinilla, Paola}, {Haworth, Thomas J.}, \& {Facchini, S.} 2024, A\&A, 681, A84

\bibitem[{{Habing}(1968)}]{Habing_1968}
{Habing}, H.~J. 1968, \bain, 19, 421

\bibitem[{{Haworth} \& {Clarke}(2019)}]{2019MNRAS.485.3895H}
{Haworth}, T.~J. \& {Clarke}, C.~J. 2019, \mnras, 485, 3895

\bibitem[{Haworth {et~al.}(2018)Haworth, Clarke, Rahman, Winter, \& Facchini}]{Haworth_2018}
Haworth, T.~J., Clarke, C.~J., Rahman, W., Winter, A.~J., \& Facchini, S. 2018, \mnras, 481, 452

\bibitem[{{Haworth} {et~al.}(2023){Haworth}, {Coleman}, {Qiao}, {Sellek}, \& {Askari}}]{2023MNRAS.526.4315H}
{Haworth}, T.~J., {Coleman}, G. A.~L., {Qiao}, L., {Sellek}, A.~D., \& {Askari}, K. 2023, \mnras, 526, 4315

\bibitem[{{Heays} {et~al.}(2017){Heays}, {Bosman}, \& {van Dishoeck}}]{Heays_2017}
{Heays}, A.~N., {Bosman}, A.~D., \& {van Dishoeck}, E.~F. 2017, \aap, 602, A105

\bibitem[{{Henney} \& {O'Dell}(1999)}]{Henney_1999}
{Henney}, W.~J. \& {O'Dell}, C.~R. 1999, \apj, 118, 2350

\bibitem[{{Henning} {et~al.}(2024){Henning}, {Kamp}, {Samland}, {Arabhavi}, {Kanwar}, {van Dishoeck}, {G{\"u}del}, {Lagage}, {Waelkens}, {Abergel}, {Absil}, {Barrado}, {Boccaletti}, {Bouwman}, {Caratti o Garatti}, {Geers}, {Glauser}, {Lahuis}, {Mueller}, {Nehm{\'e}}, {Olofsson}, {Pantin}, {Ray}, {Scheithauer}, {Vandenbussche}, {Waters}, {Wright}, {Argyriou}, {Christiaens}, {Franceschi}, {Gasman}, {Grant}, {Guadarrama}, {Jang}, {Morales-Calder{\'o}n}, {Pawellek}, {Perotti}, {Rodgers-Lee}, {Schreiber}, {Schwarz}, {Tabone}, {Temmink}, {Vlasblom}, {Colina}, {Greve}, \& {{\"O}stlin}}]{Henning_2024}
{Henning}, T., {Kamp}, I., {Samland}, M., {et~al.} 2024, \pasp, 136, 054302

\bibitem[{Houge {et~al.}(2025)Houge, Krijt, Banzatti, Blake, Pinilla, Pontoppidan, Trapman, Williams, \& Zhang}]{Houge_2025}
Houge, A., Krijt, S., Banzatti, A., {et~al.} 2025, \mnras, 537

\bibitem[{{Husser} {et~al.}(2013){Husser}, {Wende-von Berg}, {Dreizler}, {Homeier}, {Reiners}, {Barman}, \& {Hauschildt}}]{Phoenix}
{Husser}, T.~O., {Wende-von Berg}, S., {Dreizler}, S., {et~al.} 2013, \aap, 553, A6

\bibitem[{{Ireland} {et~al.}(2020){Ireland}, {Avila}, {Greenwood}, {Hicks}, \& {Bridges}}]{Ireland_2020}
{Ireland}, T.~R., {Avila}, J., {Greenwood}, R.~C., {Hicks}, L.~J., \& {Bridges}, J.~C. 2020, \ssr, 216, 25

\bibitem[{Johnson {et~al.}(2010)Johnson, Aller, Howard, \& Crepp}]{Johnson_2010}
Johnson, J.~A., Aller, K.~M., Howard, A.~W., \& Crepp, J.~R. 2010, \pasp, 122, 905

\bibitem[{{Johnstone} {et~al.}(1998){Johnstone}, {Hollenbach}, \& {Bally}}]{Johnstone_1998}
{Johnstone}, D., {Hollenbach}, D., \& {Bally}, J. 1998, \apj, 499, 758

\bibitem[{{Jones} {et~al.}(2023){Jones}, {{\'A}lvarez-M{\'a}rquez}, {Sloan}, {Kavanagh}, {Argyriou}, {Law}, {Labiano}, {Patapis}, {Mueller}, {Larson}, {Bright}, {Klaassen}, {Fox}, {Gasman}, {Geers}, {Glauser}, {Guillard}, {Nayak}, {Noriega-Crespo}, {Ressler}, {Sargent}, {Temim}, {Vandenbussche}, \& {Garc{\'\i}a Mar{\'\i}n}}]{Jones_2023}
{Jones}, O.~C., {{\'A}lvarez-M{\'a}rquez}, J., {Sloan}, G.~C., {et~al.} 2023, \mnras, 523, 2519

\bibitem[{{Kaeufer} {et~al.}(2024){Kaeufer}, {Min, M.}, {Woitke, P.}, {Kamp, I.}, \& {Arabhavi, A. M.}}]{Kaeufer_2024}
{Kaeufer}, {Min, M.}, {Woitke, P.}, {Kamp, I.}, \& {Arabhavi, A. M.} 2024, A\&A, 687, A209

\bibitem[{{Kamp} \& {Dullemond}(2004)}]{Kamp_2004}
{Kamp}, I. \& {Dullemond}, C.~P. 2004, \apj, 615, 991

\bibitem[{{Kamp} {et~al.}(2010){Kamp}, {Tilling}, {Woitke}, {Thi}, \& {Hogerheijde}}]{Kamp2010}
{Kamp}, I., {Tilling}, I., {Woitke}, P., {Thi}, W.~F., \& {Hogerheijde}, M. 2010, \aap, 510, A18

\bibitem[{{Kanwar} {et~al.}(2024){Kanwar}, {Kamp}, {Jang}, {Waters}, {van Dishoeck}, {Christiaens}, {Arabhavi}, {Henning}, {G{\"u}del}, {Woitke}, {Absil}, {Barrado}, {Caratti o Garatti}, {Glauser}, {Lahuis}, {Scheithauer}, {Vandenbussche}, {Gasman}, {Grant}, {Kurtovic}, {Perotti}, {Tabone}, \& {Temmink}}]{Kanwar_2024}
{Kanwar}, J., {Kamp}, I., {Jang}, H., {et~al.} 2024, \aap, 689, A231

\bibitem[{{Keller} {et~al.}(2008){Keller}, {Sloan}, {Forrest}, {Ayala}, {D'Alessio}, {Shah}, {Calvet}, {Najita}, {Li}, {Hartmann}, {Sargent}, {Watson}, \& {Chen}}]{Keller_2008}
{Keller}, L.~D., {Sloan}, G.~C., {Forrest}, W.~J., {et~al.} 2008, \apj, 684, 411

\bibitem[{{Keyte} \& {Haworth}(2025)}]{Keyte_2025}
{Keyte}, L. \& {Haworth}, T.~J. 2025, \mnras, 537, 598

\bibitem[{{Kress} {et~al.}(2010){Kress}, {Tielens}, \& {Frenklach}}]{Kress_2010}
{Kress}, M.~E., {Tielens}, A. G.~G.~M., \& {Frenklach}, M. 2010, ASR, 46, 44

\bibitem[{{Krumholz} {et~al.}(2019){Krumholz}, {McKee}, \& {Bland-Hawthorn}}]{krumholz_2019}
{Krumholz}, M.~R., {McKee}, C.~F., \& {Bland-Hawthorn}, J. 2019, \araa, 57, 227

\bibitem[{{Kuhn} {et~al.}(2019){Kuhn}, {Hillenbrand}, {Sills}, {Feigelson}, \& {Getman}}]{Mike_2019}
{Kuhn}, M.~A., {Hillenbrand}, L.~A., {Sills}, A., {Feigelson}, E.~D., \& {Getman}, K.~V. 2019, \apj, 870, 32

\bibitem[{Lada \& Lada(2003)}]{Lada_2003}
Lada, C.~J. \& Lada, E.~A. 2003, ARA\&A, 41, 57

\bibitem[{{Lambrechts} \& {Johansen}(2012)}]{Lambrechts_2012}
{Lambrechts}, M. \& {Johansen}, A. 2012, \aap, 544, A32

\bibitem[{{Le Roy} {et~al.}(2015){Le Roy}, {Altwegg, Kathrin}, {Balsiger, Hans}, {Berthelier, Jean-Jacques}, {Bieler, Andre}, {Briois, Christelle}, {Calmonte, Ursina}, {Combi, Michael R.}, {De Keyser, Johan}, {Dhooghe, Frederik}, {Fiethe, Björn}, {Fuselier, Stephen A.}, {Gasc, Sébastien}, {Gombosi, Tamas I.}, {Hässig, Myrtha}, {Jäckel, Annette}, {Rubin, Martin}, \& {Tzou, Chia-Yu}}]{LeRoy_2015}
{Le Roy}, {Altwegg, Kathrin}, {Balsiger, Hans}, {et~al.} 2015, A\&A, 583, A1

\bibitem[{{Lienert} {et~al.}(2024){Lienert}, {Bitsch}, \& {Henning}}]{Lienert_2024}
{Lienert}, J.~L., {Bitsch}, B., \& {Henning}, T. 2024, \aap, 691, A72

\bibitem[{Long {et~al.}(2025)Long, Pascucci, Houge, Banzatti, Pontoppidan, Najita, Krijt, Xie, Williams, Herczeg, Andrews, Bergin, Blake, Colmenares, Harsono, Romero-Mirza, Li, Lu, Pinilla, Wilner, Vioque, Zhang, \& the JDISCS~collaboration}]{long2024first}
Long, F., Pascucci, I., Houge, A., {et~al.} 2025, \apjl, 978, L30

\bibitem[{Lu {et~al.}(2014)Lu, Chang, Yin, Ng, \& Jackson}]{Lu_2014}
Lu, Z., Chang, Y.~C., Yin, Q.-Z., Ng, C.~Y., \& Jackson, W.~M. 2014, Science, 346, 61

\bibitem[{{Lyons} \& {Young}(2005)}]{Lyons05}
{Lyons}, J.~R. \& {Young}, E.~D. 2005, \nat, 435, 317

\bibitem[{{Manara} {et~al.}(2023){Manara}, {Ansdell}, {Rosotti}, {Hughes}, {Armitage}, {Lodato}, \& {Williams}}]{Manara_2023}
{Manara}, C.~F., {Ansdell}, M., {Rosotti}, G.~P., {et~al.} 2023, in Astronomical Society of the Pacific Conference Series, Vol. 534, Protostars and Planets VII, ed. S.~{Inutsuka}, Y.~{Aikawa}, T.~{Muto}, K.~{Tomida}, \& M.~{Tamura}, 539

\bibitem[{Mann {et~al.}(2014)Mann, Di~Francesco, Johnstone, Andrews, Williams, Bally, Ricci, Hughes, \& Matthews}]{Mann_2014}
Mann, R.~K., Di~Francesco, J., Johnstone, D., {et~al.} 2014, \apj, 784, 82

\bibitem[{{Massey} {et~al.}(2001){Massey}, {DeGioia-Eastwood}, \& {Waterhouse}}]{Massey_2001}
{Massey}, P., {DeGioia-Eastwood}, K., \& {Waterhouse}, E. 2001, \apj, 121, 1050

\bibitem[{{Massi} {et~al.}(2015){Massi}, {Giannetti, A.}, {Di Carlo, E.}, {Brand, J.}, {Beltrán, M. T.}, \& {Marconi, G.}}]{Massi_2015}
{Massi}, {Giannetti, A.}, {Di Carlo, E.}, {et~al.} 2015, A\&A, 573, A95

\bibitem[{{McElroy} {et~al.}(2013){McElroy}, {Walsh}, {Markwick}, {Cordiner}, {Smith}, \& {Millar}}]{McElroy_2013}
{McElroy}, D., {Walsh}, C., {Markwick}, A.~J., {et~al.} 2013, \aap, 550, A36

\bibitem[{{Meeus} {et~al.}(2001){Meeus}, {Waters, L. B. F. M.}, {Bouwman, J.}, {van den Ancker, M. E.}, {Waelkens, C.}, \& {Malfait, K.}}]{Meeus_2001}
{Meeus}, {Waters, L. B. F. M.}, {Bouwman, J.}, {et~al.} 2001, A\&A, 365, 476

\bibitem[{Meijerink {et~al.}(2009)Meijerink, Pontoppidan, Blake, Poelman, \& Dullemond}]{Meijerink_2009}
Meijerink, R., Pontoppidan, K.~M., Blake, G.~A., Poelman, D.~R., \& Dullemond, C.~P. 2009, \apj, 704, 1471

\bibitem[{Milam {et~al.}(2005)Milam, Savage, Brewster, Ziurys, \& Wyckoff}]{Milam_2005}
Milam, S.~N., Savage, C., Brewster, M.~A., Ziurys, L.~M., \& Wyckoff, S. 2005, \apj, 634, 1126

\bibitem[{{Miotello} {et~al.}(2014){Miotello}, {Bruderer, S.}, \& {van Dishoeck, E. F.}}]{Miotello_2014}
{Miotello}, {Bruderer, S.}, \& {van Dishoeck, E. F.} 2014, A\&A, 572, A96

\bibitem[{{Molano} \& {Kamp}(2012{\natexlab{a}})}]{Chaparro_1}
{Molano}, G. \& {Kamp}, I. 2012{\natexlab{a}}, \aap, 537, A138

\bibitem[{{Molano} \& {Kamp}(2012{\natexlab{b}})}]{Chaparro_2}
{Molano}, G. \& {Kamp}, I. 2012{\natexlab{b}}, \aap, 547, A7

\bibitem[{{Mordasini} {et~al.}(2012){Mordasini}, {Alibert}, {Benz}, {Klahr}, \& {Henning}}]{Mordasini_2012}
{Mordasini}, C., {Alibert}, Y., {Benz}, W., {Klahr}, H., \& {Henning}, T. 2012, \aap, 541, A97

\bibitem[{{Mulders} \& {Dominik}(2012)}]{Mulders_2012}
{Mulders} \& {Dominik}. 2012, A\&A, 539, A9

\bibitem[{{Ndugu} {et~al.}(2024){Ndugu}, {Bitsch}, \& {Lienert}}]{Ndugu_2024}
{Ndugu}, N., {Bitsch}, B., \& {Lienert}, J.~L. 2024, \aap, 691, A32

\bibitem[{{{\"O}berg} {et~al.}(2011){{\"O}berg}, {Murray-Clay}, \& {Bergin}}]{Oberg_2011}
{{\"O}berg}, K.~I., {Murray-Clay}, R., \& {Bergin}, E.~A. 2011, \apjl, 743, L16

\bibitem[{{Paine} {et~al.}(2025){Paine}, {Haworth}, \& {Nelson}}]{Paine_2025}
{Paine}, S., {Haworth}, T.~J., \& {Nelson}, R.~P. 2025, \mnras

\bibitem[{Pascucci {et~al.}(2013)Pascucci, Herczeg, Carr, \& Bruderer}]{Pascucci_2013}
Pascucci, I., Herczeg, G., Carr, J.~S., \& Bruderer, S. 2013, \apj, 779, 178

\bibitem[{{Pecaut} \& {Mamajek}(2013)}]{Pecaut_2013}
{Pecaut}, M.~J. \& {Mamajek}, E.~E. 2013, \apjs, 208, 9

\bibitem[{{Penzias}(1981)}]{Penzias_1981}
{Penzias}, A.~A. 1981, \apj, 249, 518

\bibitem[{{Perotti} {et~al.}(2023){Perotti}, {Christiaens}, {Henning}, {Tabone}, {Waters}, {Kamp}, {Olofsson}, {Grant}, {Gasman}, {Bouwman}, {Samland}, {Franceschi}, {van Dishoeck}, {Schwarz}, {G{\"u}del}, {Lagage}, {Ray}, {Vandenbussche}, {Abergel}, {Absil}, {Arabhavi}, {Argyriou}, {Barrado}, {Boccaletti}, {Caratti o Garatti}, {Geers}, {Glauser}, {Justannont}, {Lahuis}, {Mueller}, {Nehm{\'e}}, {Pantin}, {Scheithauer}, {Waelkens}, {Guadarrama}, {Jang}, {Kanwar}, {Morales-Calder{\'o}n}, {Pawellek}, {Rodgers-Lee}, {Schreiber}, {Colina}, {Greve}, {{\"O}stlin}, \& {Wright}}]{Perotti_2023}
{Perotti}, G., {Christiaens}, V., {Henning}, T., {et~al.} 2023, \nat, 620, 516

\bibitem[{{Pinilla} {et~al.}(2016){Pinilla}, {Klarmann, L.}, {Birnstiel, T.}, {Benisty, M.}, {Dominik, C.}, \& {Dullemond, C. P.}}]{Pinilla_2016}
{Pinilla}, {Klarmann, L.}, {Birnstiel, T.}, {et~al.} 2016, A\&A, 585, A35

\bibitem[{Piso {et~al.}(2015)Piso, Öberg, Birnstiel, \& Murray-Clay}]{Piso_2015}
Piso, A.-M.~A., Öberg, K.~I., Birnstiel, T., \& Murray-Clay, R.~A. 2015, \apj, 815, 109

\bibitem[{{Pontoppidan} \& {Blevins}(2014)}]{Pontoppidan_2014}
{Pontoppidan}, K.~M. \& {Blevins}, S.~M. 2014, Faraday Discussions, 168, 49

\bibitem[{Pontoppidan {et~al.}(2010)Pontoppidan, Salyk, Blake, Meijerink, Carr, \& Najita}]{Pontoppidan_2010}
Pontoppidan, K.~M., Salyk, C., Blake, G.~A., {et~al.} 2010, \apj, 720, 887

\bibitem[{Portilla-Revelo {et~al.}(2025)Portilla-Revelo, Getman, Ramírez-Tannus, Haworth, Waters, Bik, Feigelson, Kamp, van Terwisga, Frediani, Henning, Winter, Roccatagliata, Preibisch, Sabbi, Zeidler, \& Kuhn}]{Bayron_2025}
Portilla-Revelo, B., Getman, K.~V., Ramírez-Tannus, M.~C., {et~al.} 2025, \apj, 985, 72

\bibitem[{{Prochaska} {et~al.}(2020){Prochaska}, {Hennawi}, {Westfall}, {Cooke}, {Wang}, {Hsyu}, {Davies}, {Farina}, \& {Pelliccia}}]{pypeit:joss_arXiv}
{Prochaska}, J., {Hennawi}, J., {Westfall}, K., {et~al.} 2020, JOSS, 5, 2308

\bibitem[{{Qiao} {et~al.}(2023){Qiao}, {Coleman}, \& {Haworth}}]{Qiao_2023}
{Qiao}, L., {Coleman}, G. A.~L., \& {Haworth}, T.~J. 2023, \mnras, 522, 1939

\bibitem[{{Rab} {et~al.}(2018){Rab}, {G{\"u}del}, {Woitke}, {Kamp}, {Thi}, {Min}, {Aresu}, \& {Meijerink}}]{Rab2018}
{Rab}, C., {G{\"u}del}, M., {Woitke}, P., {et~al.} 2018, \aap, 609, A91

\bibitem[{{Ramirez-Tannus} {et~al.}(2021){Ramirez-Tannus}, {Backs}, {Bik}, {Bouwman}, {Brandner}, {Chevance}, {De Koter}, {Derkink}, {Feigelson}, {Geen}, {Getman}, {Henning}, {Kamp}, {Kaper}, {Kruijssen}, {Kuhn}, {Longmore}, {McLeod}, {Poorta}, {Povich}, {Preibisch}, {Roccatagliata}, {Sabbi}, {Sana}, {Waters}, {Winter}, {Zari}, \& {van Terwisga}}]{Ramirez-Tannus_2021}
{Ramirez-Tannus}, M.~C., {Backs}, F., {Bik}, A., {et~al.} 2021, JWST Proposal. Cycle 1, ID. \#1759

\bibitem[{{Ram{\'\i}rez-Tannus} {et~al.}(2020){Ram{\'\i}rez-Tannus}, {Poorta}, {Bik}, {Kaper}, {de Koter}, {De Ridder}, {Beuther}, {Brandner}, {Davies}, {Gennaro}, {Guo}, {Henning}, {Linz}, {Naylor}, {Pasquali}, {Ram{\'\i}rez-Agudelo}, \& {Sana}}]{Ramirez_2020}
{Ram{\'\i}rez-Tannus}, M.~C., {Poorta}, J., {Bik}, A., {et~al.} 2020, \aap, 633, A155

\bibitem[{Ramírez-Tannus {et~al.}(2023)Ramírez-Tannus, Bik, Cuijpers, Waters, Göppl, Henning, Kamp, Preibisch, Getman, Chaparro, Cuartas-Restrepo, de~Koter, Feigelson, Grant, Haworth, Hernández, Kuhn, Perotti, Povich, Reiter, Roccatagliata, Sabbi, Tabone, Winter, McLeod, van Boekel, \& van Terwisga}]{Ramirez-Tannus_2023}
Ramírez-Tannus, M.~C., Bik, A., Cuijpers, L., {et~al.} 2023, \apjl, 958, L30

\bibitem[{Ramírez-Tannus {et~al.}(2025)Ramírez-Tannus, Bik, Getman, Waters, Portilla-Revelo, Göppl, Winter, Frediani, Chaparro, Feigelson, Haworth, Henning, Hernández, Lemus-Nemocón, Preibisch, Roccatagliata, Sabbi, \& Zeidler}]{Ramirez-Tannus_2025}
Ramírez-Tannus, M.~C., Bik, A., Getman, K.~V., {et~al.} 2025

\bibitem[{{Reffert} {et~al.}(2015){Reffert}, {Bergmann}, {Quirrenbach}, {Trifonov}, \& {K{\"u}nstler}}]{Reffert_2015}
{Reffert}, S., {Bergmann}, C., {Quirrenbach}, A., {Trifonov}, T., \& {K{\"u}nstler}, A. 2015, \aap, 574, A116

\bibitem[{{Richert} {et~al.}(2018){Richert}, {Getman}, {Feigelson}, {Kuhn}, {Broos}, {Povich}, {Bate}, \& {Garmire}}]{Richert_2018}
{Richert}, A.~J.~W., {Getman}, K.~V., {Feigelson}, E.~D., {et~al.} 2018, \mnras, 477, 5191

\bibitem[{Rieke {et~al.}(2015)Rieke, Wright, Böker, Bouwman, Colina, Glasse, Gordon, Greene, Güdel, Henning, Justtanont, Lagage, Meixner, Nørgaard-Nielsen, Ray, Ressler, van Dishoeck, \& Waelkens}]{Rieke_2015}
Rieke, G.~H., Wright, G.~S., Böker, T., {et~al.} 2015, PASP, 127, 584

\bibitem[{Roccatagliata {et~al.}(2011)Roccatagliata, Bouwman, Henning, Gennaro, Feigelson, Kim, Sicilia-Aguilar, \& Lawson}]{Roccatagliata_2011}
Roccatagliata, V., Bouwman, J., Henning, T., {et~al.} 2011, \apj, 733, 113

\bibitem[{{Russeil} {et~al.}(2017){Russeil}, {Adami, C.}, {Bouret, J. C.}, {Hervé, A.}, {Parker, Q. A.}, {Zavagno, A.}, \& {Motte, F.}}]{Russeil_2017}
{Russeil}, {Adami, C.}, {Bouret, J. C.}, {et~al.} 2017, A\&A, 607, A86

\bibitem[{{Salyk} {et~al.}(2025){Salyk}, {Pontoppidan}, {Banzatti}, {Bergin}, {Arulanantham}, {Najita}, {Blake}, {Carr}, {Zhang}, \& {Xie}}]{Salyk_2025}
{Salyk}, C., {Pontoppidan}, K.~M., {Banzatti}, A., {et~al.} 2025, \apj, 169, 184

\bibitem[{Salyk {et~al.}(2008)Salyk, Pontoppidan, Blake, Lahuis, van Dishoeck, \& II}]{Salyk_2008}
Salyk, C., Pontoppidan, K.~M., Blake, G.~A., {et~al.} 2008, ApJ, 676, L49

\bibitem[{{Scally} \& {Clarke}(2001)}]{Scally_2001}
{Scally}, A. \& {Clarke}, C. 2001, \mnras, 325, 449

\bibitem[{{Schwarz} {et~al.}(2024){Schwarz}, {Henning}, {Christiaens}, {Gasman}, {Samland}, {Perotti}, {Jang}, {Grant}, {Tabone}, {Morales-Calder{\'o}n}, {Kamp}, {van Dishoeck}, {G{\"u}del}, {Lagage}, {Barrado}, {Caratti o Garatti}, {Glauser}, {Ray}, {Vandenbussche}, {Waters}, {Arabhavi}, {Kanwar}, {Olofsson}, {Rodgers-Lee}, {Schreiber}, \& {Temmink}}]{Schwarz_2024}
{Schwarz}, K.~R., {Henning}, T., {Christiaens}, V., {et~al.} 2024, \apj, 962, 8

\bibitem[{{Sellek} {et~al.}(2020){Sellek}, {Booth}, \& {Clarke}}]{Sellek_2020}
{Sellek}, A.~D., {Booth}, R.~A., \& {Clarke}, C.~J. 2020, \mnras, 492, 1279

\bibitem[{{Sellek} {et~al.}(2025){Sellek}, {Vlasblom}, \& {van Dishoeck}}]{Sellek_2024}
{Sellek}, A.~D., {Vlasblom}, M., \& {van Dishoeck}, E.~F. 2025, \aap, 694, A79

\bibitem[{{Shu} {et~al.}(2000){Shu}, {Najita}, {Shang}, \& {Li}}]{Shu00}
{Shu}, F.~H., {Najita}, J.~R., {Shang}, H., \& {Li}, Z.~Y. 2000, in Protostars and Planets IV, ed. V.~{Mannings}, A.~P. {Boss}, \& S.~S. {Russell}, 789--814

\bibitem[{{Smith} {et~al.}(2004){Smith}, {Herbst}, \& {Chang}}]{Smith_2004}
{Smith}, I. W.~M., {Herbst}, E., \& {Chang}, Q. 2004, \mnras, 350, 323

\bibitem[{Smith {et~al.}(2015)Smith, Pontoppidan, Young, \& Morris}]{Smith_2015}
Smith, R.~L., Pontoppidan, K.~M., Young, E.~D., \& Morris, M.~R. 2015, \apj, 813, 120

\bibitem[{Smith {et~al.}(2009)Smith, Pontoppidan, Young, Morris, \& van Dishoeck}]{Smith_2009}
Smith, R.~L., Pontoppidan, K.~M., Young, E.~D., Morris, M.~R., \& van Dishoeck, E.~F. 2009, \apj, 701, 163–175

\bibitem[{{Stapper} {et~al.}(2025){Stapper}, {Hogerheijde}, {van Dishoeck}, {Booth}, {Grant}, \& {van Terwisga}}]{Stapper_2025}
{Stapper}, L.~M., {Hogerheijde}, M.~R., {van Dishoeck}, E.~F., {et~al.} 2025, \aap, 693, A49

\bibitem[{{Stapper} {et~al.}(2022){Stapper}, {Hogerheijde}, {van Dishoeck}, \& {Mentel}}]{Stapper_2022}
{Stapper}, L.~M., {Hogerheijde}, M.~R., {van Dishoeck}, E.~F., \& {Mentel}, R. 2022, \aap, 658, A112

\bibitem[{Stevenson \& Lunine(1988)}]{Stevenson_1988}
Stevenson, D.~J. \& Lunine, J.~I. 1988, Icarus, 75, 146

\bibitem[{Stolte {et~al.}(2010)Stolte, Morris, Ghez, Do, Lu, Wright, Ballard, Mills, \& Matthews}]{Stolte_2010}
Stolte, A., Morris, M.~R., Ghez, A.~M., {et~al.} 2010, \apj, 718, 810

\bibitem[{{Tabone} {et~al.}(2023){Tabone}, {Bettoni}, {van Dishoeck}, {Arabhavi}, {Grant}, {Gasman}, {Henning}, {Kamp}, {G{\"u}del}, {Lagage}, {Ray}, {Vandenbussche}, {Abergel}, {Absil}, {Argyriou}, {Barrado}, {Boccaletti}, {Bouwman}, {Caratti o Garatti}, {Geers}, {Glauser}, {Justannont}, {Lahuis}, {Mueller}, {Nehm{\'e}}, {Olofsson}, {Pantin}, {Scheithauer}, {Waelkens}, {Waters}, {Black}, {Christiaens}, {Guadarrama}, {Morales-Calder{\'o}n}, {Jang}, {Kanwar}, {Pawellek}, {Perotti}, {Perrin}, {Rodgers-Lee}, {Samland}, {Schreiber}, {Schwarz}, {Colina}, {{\"O}stlin}, \& {Wright}}]{Tabone_2023}
{Tabone}, B., {Bettoni}, G., {van Dishoeck}, E.~F., {et~al.} 2023, Nat. Astron., 7, 805

\bibitem[{Taylor {et~al.}(1965)Taylor, Duke, Silver, \& Epstein}]{Taylor65}
Taylor, H.~P., Duke, M.~B., Silver, L.~T., \& Epstein, S. 1965, Geochim. Cosmochim. Acta, 29, 489

\bibitem[{{Temmink} {et~al.}(2024{\natexlab{a}}){Temmink}, {van Dishoeck}, {Gasman}, {Grant}, {Tabone}, {G{\"u}del}, {Henning}, {Barrado}, {Caratti o Garatti}, {Glauser}, {Kamp}, {Arabhavi}, {Jang}, {Kurtovic}, {Perotti}, {Schwarz}, \& {Vlasblom}}]{Temmink_water}
{Temmink}, M., {van Dishoeck}, E.~F., {Gasman}, D., {et~al.} 2024{\natexlab{a}}, \aap, 689, A330

\bibitem[{{Temmink} {et~al.}(2024{\natexlab{b}}){Temmink}, {van Dishoeck}, {Grant}, {Tabone}, {Gasman}, {Christiaens}, {Samland}, {Argyriou}, {Perotti}, {G{\"u}del}, {Henning}, {Lagage}, {Abergel}, {Absil}, {Barrado}, {Caratti o Garatti}, {Glauser}, {Kamp}, {Lahuis}, {Olofsson}, {Ray}, {Scheithauer}, {Vandenbussche}, {Waters}, {Arabhavi}, {Jang}, {Kanwar}, {Morales-Calder{\'o}n}, {Rodgers-Lee}, {Schreiber}, {Schwarz}, \& {Colina}}]{Temmink_CO}
{Temmink}, M., {van Dishoeck}, E.~F., {Grant}, S.~L., {et~al.} 2024{\natexlab{b}}, \aap, 686, A117

\bibitem[{{Thi} {et~al.}(2011){Thi}, {Woitke}, \& {Kamp}}]{Thi2011}
{Thi}, W.~F., {Woitke}, P., \& {Kamp}, I. 2011, \mnras, 412, 711

\bibitem[{Thiemens \& Heidenreich(1983)}]{Thiemans83}
Thiemens, M.~H. \& Heidenreich, J.~E. 1983, Science, 219, 1073

\bibitem[{Throop \& Bally(2005)}]{Throop_2005}
Throop, H.~B. \& Bally, J. 2005, \apj, 623, L149

\bibitem[{{Valegård} {et~al.}(2021){Valegård}, {Waters, L. B. F. M.}, \& {Dominik, C.}}]{Valegard_2021}
{Valegård}, {Waters, L. B. F. M.}, \& {Dominik, C.} 2021, A\&A, 652, A133

\bibitem[{{van Boekel} {et~al.}(2005){van Boekel}, {Min}, {Waters}, {de Koter}, {Dominik}, {van den Ancker}, \& {Bouwman}}]{Boekel_2005}
{van Boekel}, R., {Min}, M., {Waters}, L.~B.~F.~M., {et~al.} 2005, \aap, 437, 189

\bibitem[{{van Terwisga} {et~al.}(2020){van Terwisga}, {van Dishoeck, E. F.}, {Mann, R. K.}, {Di Francesco, J.}, {van der Marel, N.}, {Meyer, M.}, {Andrews, S. M.}, {Carpenter, J.}, {Eisner, J. A.}, {Manara, C. F.}, \& {Williams, J. P.}}]{vanTerwisga_2020}
{van Terwisga}, {van Dishoeck, E. F.}, {Mann, R. K.}, {et~al.} 2020, A\&A, 640, A27

\bibitem[{{van Terwisga} {et~al.}(2019){van Terwisga}, {Hacar}, \& {van Dishoeck}}]{vanTerwisga_2019}
{van Terwisga}, S.~E., {Hacar}, A., \& {van Dishoeck}, E.~F. 2019, \aap, 628, A85

\bibitem[{Van~Winckel(2003)}]{Van_Winckel_2003}
Van~Winckel, H. 2003, Annu. Rev. Astron., 41, 391

\bibitem[{{van Zadelhoff} {et~al.}(2001){van Zadelhoff}, {van Dishoeck, E. F.}, {Thi, W.-F.}, \& {Blake, G. A.}}]{van_Zadelhoff}
{van Zadelhoff}, {van Dishoeck, E. F.}, {Thi, W.-F.}, \& {Blake, G. A.} 2001, A\&A, 377, 566

\bibitem[{{Visser} {et~al.}(2009){Visser}, {van Dishoeck, E. F.}, \& {Black, J. H.}}]{Visser_2009}
{Visser}, {van Dishoeck, E. F.}, \& {Black, J. H.} 2009, A\&A, 503, 323

\bibitem[{{Visser} {et~al.}(2007){Visser}, {Geers}, {Dullemond}, {Augereau}, {Pontoppidan}, \& {van Dishoeck}}]{Visser_2007}
{Visser}, R., {Geers}, V.~C., {Dullemond}, C.~P., {et~al.} 2007, \aap, 466, 229

\bibitem[{{Vlasblom} {et~al.}(2025){Vlasblom}, {Temmink}, {Grant}, {Kurtovic}, {Sellek}, {van Dishoeck}, {G{\"u}del}, {Henning}, {Lagage}, {Barrado}, {Caratti o Garatti}, {Glauser}, {Kamp}, {Lahuis}, {Olofsson}, {Arabhavi}, {Christiaens}, {Gasman}, {Jang}, {Morales-Calder{\'o}n}, {Perotti}, {Schwarz}, \& {Tabone}}]{Vasblom_2024}
{Vlasblom}, M., {Temmink}, M., {Grant}, S.~L., {et~al.} 2025, \aap, 693, A278

\bibitem[{{Walborn} {et~al.}(2002){Walborn}, {Howarth}, {Lennon}, {Massey}, {Oey}, {Moffat}, {Skalkowski}, {Morrell}, {Drissen}, \& {Parker}}]{Walborn_2002}
{Walborn}, N.~R., {Howarth}, I.~D., {Lennon}, D.~J., {et~al.} 2002, \apj, 123, 2754

\bibitem[{{Walsh} {et~al.}(2015){Walsh}, {Nomura, Hideko}, \& {van Dishoeck, Ewine}}]{Walsh_2015}
{Walsh}, {Nomura, Hideko}, \& {van Dishoeck, Ewine}. 2015, A\&A, 582, A88

\bibitem[{{Walsh} {et~al.}(2013){Walsh}, {Millar}, \& {Nomura}}]{Walsh_2013}
{Walsh}, C., {Millar}, T.~J., \& {Nomura}, H. 2013, \apjl, 766, L23

\bibitem[{Wells {et~al.}(2015)Wells, Pel, Glasse, Wright, Aitink-Kroes, Azzollini, Beard, Brandl, Gallie, Geers, Glauser, Hastings, Henning, Jager, Justtanont, Kruizinga, Lahuis, Lee, Martinez-Delgado, Martínez-Galarza, Meijers, Morrison, Müller, Nakos, O’Sullivan, Oudenhuysen, Parr-Burman, Pauwels, Rohloff, Schmalzl, Sykes, Thelen, van Dishoeck, Vandenbussche, Venema, Visser, Waters, \& Wright}]{Wells_2015}
Wells, M., Pel, J.-W., Glasse, A., {et~al.} 2015, PASP, 127, 646

\bibitem[{{Wilson} \& {Rood}(1994)}]{Wilson_1994}
{Wilson}, T.~L. \& {Rood}, R. 1994, \araa, 32, 191

\bibitem[{Winn \& Fabrycky(2015)}]{winn_2015}
Winn, J.~N. \& Fabrycky, D.~C. 2015, ARA\&A, 53, 409

\bibitem[{{Winter} {et~al.}(2018){Winter}, {Clarke}, {Rosotti}, {Ih}, {Facchini}, \& {Haworth}}]{Winter_2018}
{Winter}, A.~J., {Clarke}, C.~J., {Rosotti}, G., {et~al.} 2018, \mnras, 478, 2700

\bibitem[{{Winter} \& {Haworth}(2022)}]{winter_2022}
{Winter}, A.~J. \& {Haworth}, T.~J. 2022, Eur. Phys. J. Plus, 137, 1132

\bibitem[{{Winter} {et~al.}(2020){Winter}, {Kruijssen}, {Chevance}, {Keller}, \& {Longmore}}]{Winter_2020}
{Winter}, A.~J., {Kruijssen}, J.~M.~D., {Chevance}, M., {Keller}, B.~W., \& {Longmore}, S.~N. 2020, \mnras, 491, 903

\bibitem[{{Woitke} {et~al.}(2016){Woitke}, {Min, M.}, {Pinte, C.}, {Thi, W.-F.}, {Kamp, I.}, {Rab, C.}, {Anthonioz, F.}, {Antonellini, S.}, {Baldovin-Saavedra, C.}, {Carmona, A.}, {Dominik, C.}, {Dionatos, O.}, {Greaves, J.}, {Güdel, M.}, {Ilee, J. D.}, {Liebhart, A.}, {Ménard, F.}, {Rigon, L.}, {Waters, L. B. F. M.}, {Aresu, G.}, {Meijerink, R.}, \& {Spaans, M.}}]{Woitke_2016}
{Woitke}, {Min, M.}, {Pinte, C.}, {et~al.} 2016, A\&A, 586, A103

\bibitem[{{Woitke} {et~al.}(2024{\natexlab{a}}){Woitke}, {Thi}, {Arabhavi}, {Kamp}, {K{\'o}sp{\'a}l}, \& {{\'A}brah{\'a}m}}]{Woitke2024}
{Woitke}, P., {Thi}, W.~F., {Arabhavi}, A.~M., {et~al.} 2024{\natexlab{a}}, \aap, 683, A219

\bibitem[{{Woitke} {et~al.}(2024{\natexlab{b}}){Woitke}, {Thi}, {Arabhavi}, {Kamp}, {K{\'o}sp{\'a}l}, \& {{\'A}brah{\'a}m}}]{Woitke_2024}
{Woitke}, P., {Thi}, W.~F., {Arabhavi}, A.~M., {et~al.} 2024{\natexlab{b}}, \aap, 683, A219

\bibitem[{Wright {et~al.}(2015)Wright, Wright, Goodson, Rieke, Aitink-Kroes, Amiaux, Aricha-Yanguas, Azzollini, Banks, Barrado-Navascues, Belenguer-Davila, Bloemmart, Bouchet, Brandl, Colina, Örs Detre, Diaz-Catala, Eccleston, Friedman, García-Marín, Güdel, Glasse, Glauser, Greene, Groezinger, Grundy, Hastings, Henning, Hofferbert, Hunter, Jessen, Justtanont, Karnik, Khorrami, Krause, Labiano, Lagage, Langer, Lemke, Lim, Lorenzo-Alvarez, Mazy, McGowan, Meixner, Morris, Morrison, Müller, rgaard Nielson, Olofsson, O’Sullivan, Pel, Penanen, Petach, Pye, Ray, Renotte, Renouf, Ressler, Samara-Ratna, Scheithauer, Schneider, Shaughnessy, Stevenson, Sukhatme, Swinyard, Sykes, Thatcher, Tikkanen, van Dishoeck, Waelkens, Walker, Wells, \& Zhender}]{Wright_2015}
Wright, G.~S., Wright, D., Goodson, G.~B., {et~al.} 2015, \pasp, 127, 595

\bibitem[{{Zannese} {et~al.}(2025){Zannese}, {Tabone}, {Habart}, {Dartois}, {Goicoechea}, {Coudert}, {Gans}, {Martin-Drumel}, {Jacovella}, {Faure}, {Godard}, {Tielens}, {Le Gal}, {Black}, {Vicente}, {Bern{\'e}}, {Peeters}, {Van De Putte}, {Chown}, {Sidhu}, {Schroetter}, {Canin}, \& {Kannavou}}]{Zannese_2025}
{Zannese}, M., {Tabone}, B., {Habart}, E., {et~al.} 2025, \aap, 696, A99

\bibitem[{{Zannese} {et~al.}(2024){Zannese}, {Tabone}, {Habart}, {Goicoechea}, {Zanchet}, {van Dishoeck}, {van Hemert}, {Black}, {Tielens}, {Veselinova}, {Jambrina}, {Menendez}, {Verdasco}, {Aoiz}, {Gonzalez-Sanchez}, {Trahin}, {Dartois}, {Bern{\'e}}, {Peeters}, {He}, {Sidhu}, {Chown}, {Schroetter}, {Van De Putte}, {Canin}, {Alarc{\'o}n}, {Abergel}, {Bergin}, {Bernard-Salas}, {Boersma}, {Bron}, {Cami}, {Dicken}, {Elyajouri}, {Fuente}, {Gordon}, {Issa}, {Joblin}, {Kannavou}, {Khan}, {Languignon}, {Le Gal}, {Maragkoudakis}, {Meshaka}, {Okada}, {Onaka}, {Pasquini}, {Pound}, {Robberto}, {R{\"o}llig}, {Schefter}, {Schirmer}, {Vicente}, \& {Wolfire}}]{Zannese_2024}
{Zannese}, M., {Tabone}, B., {Habart}, E., {et~al.} 2024, Nat. Astron., 8, 577

\end{thebibliography}
\begin{appendix}
\section{Local continuum normalization}\label{app:normalization}
\begin{figure*}[htp!]
    \centering
    \includegraphics[width=\linewidth]{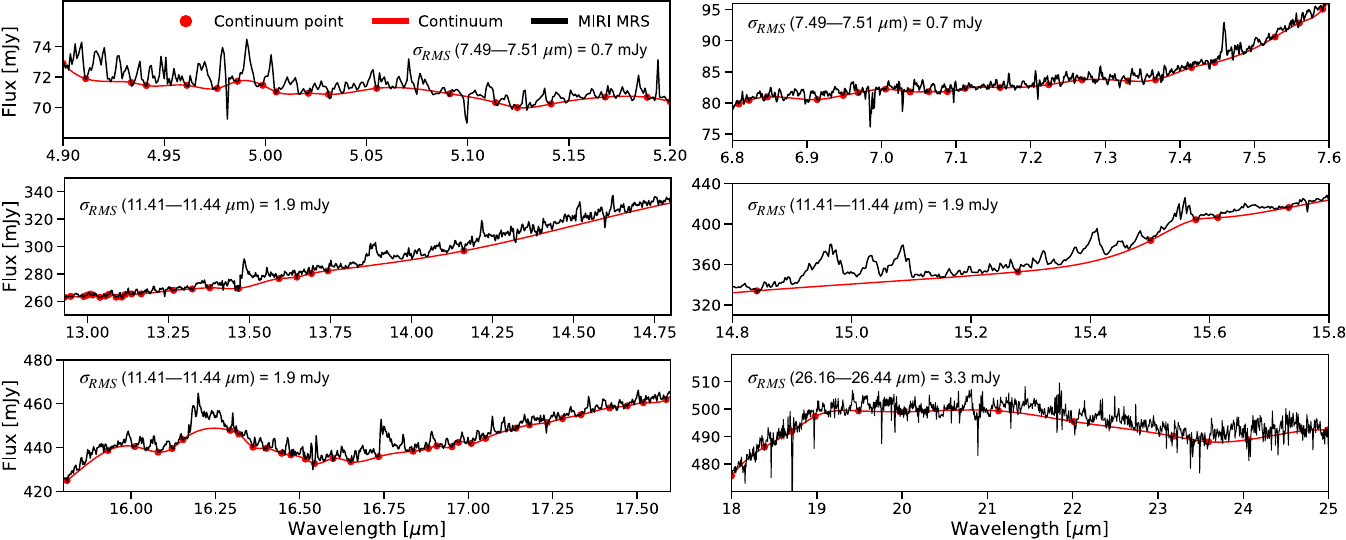}
    \caption{Overview of the local continuum subtraction of the MIRI MRS spectrum of XUE 10. The spectral uncertainty measured from line-free regions and used in the slab modeling is indicated in each panel.}
    \label{fig:continuum_normalization}
\end{figure*}

In order to fit the molecular emission in the observed spectrum, it is first needed to subtract the local dust continuum. We therefore fit and subtracted a spline function in each wavelength window of interest selecting line-free regions, as performed for example by \citet{Grant_2023}. We evaluated the spectral uncertainty as the standard deviation of the flux in a line-free region close to each fit range. The continuum points used to determine the spline in all windows and the estimated spectral uncertainty are reported in Figure~\ref{fig:continuum_normalization}. In the 12.9-17.6 $\mu$m wavelength region with detected emission from the CO$_2$ isotopologues, these species are so abundant to create a pseudo-continuum on top of the dust continuum, corresponding to their strongest emitting features, i.e., fundamental $\upsilon_2$ $\mathit{Q}$-branch and $\mathit{Q}$ hot bandheads, particularly those of $^{12}$CO$_2$ and $^{13}$CO$_2$. This makes it exceptionally challenging to subtract the local dust continuum. We therefore chose the continuum points based on templates of each isotopologue emission for different values of temperature, column density and equivalent emitting radius overlaid on the observed spectrum. We then continuum-subtracted the observed spectrum before fitting each species, selecting line-free regions or regions where the contamination from the other isotopologues was minimal. We note that in Figure~\ref{fig:continuum_normalization} it is only shown the interpolated continuum in the original spectrum between 12.9--17.6 $\mu$m before $^{12}$CO$_2$ fit and subtraction.
\section{On the broad non-continuum feature at 15.5--17 $\mu$m}\label{subsec:dust}
\begin{figure*}[htp!]
\centering
\includegraphics[width=0.8\linewidth]{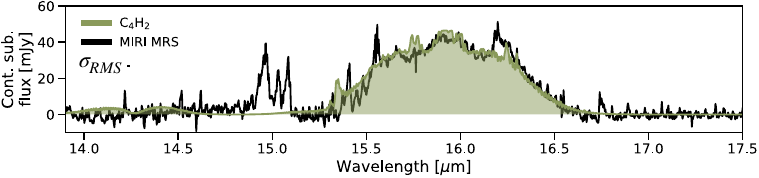}
\caption{Continuum-subtracted MIRI spectrum of XUE 10 (black) between 14 $\mu$m and 17.5 $\mu$m with overlaid best-fit slab model spectrum of diacetylene (C$_4$H$_2$). The spectral uncertainty is indicated in figure.
}\label{fig:C4H2bump}
\end{figure*}
We observe a peculiar \enquote{bump} in the MIRI spectrum of XUE 10 between 15.5 and 17 $\mu$m (see Figure~\ref{fig:MIRI_spec} and Figure~\ref{fig:C4H2bump}), which we had to subtract in order to fit the CO$_2$ isotopologues in this wavelength range. This feature has not been observed in any protoplanetary disk before \citep{Boersma_2010}. Here we explore possible candidate species that could explain this emission.

The only gas phase molecular species capable of creating such a strong feature is diacetylene (C$_4$H$_2$), as shown in Figure~\ref{fig:C4H2bump}. In order to reproduce this feature, the gas slab modeling approach requires C$_4$H$_2$ to emit with a very high column density of 3.3 $\times$ 10$^{19}$ cm$^{-2}$ at 221 K, and an equivalent emitting radius of 3.3 au ($\chi^2_{\rm red, min}$ = 8.1; Figure~\ref{fig:chisquaremaps}). However, the optically thin line emission expected to arise on top of the optically thick pseudo-continuum is not reproduced in the spectrum, nor the C$_4$H$_2$ line emission which should be visible at 14--14.5 $\mu$m. The overlap of an optically thick and an optically thin slab model component does not solve the issue. 

One may expect to see not only this species, but a \enquote{forest} of hydrocarbonic molecules, starting from the more common acetylene (C$_2$H$_2$) species. This has been observed in the carbon-rich inner regions of some disks around very low-mass stars \citep[e.g.,][]{Tabone_2023, Arabhavi_2024, long2024first, Kanwar_2024}. 
The emission of C$_4$H$_2$ may also be produced by ongoing thermal or UV photo-destruction of carbonaceous grains in the surface layer of XUE 10 \citep[e.g.,][]{Visser_2007, Kress_2010, Anderson_2017}. 

Another possibility may be the emission of PAHs or dust aggregates never seen before in IMTT or Herbig disks at \textit{Spitzer} resolution \citep[e.g.,][]{Meeus_2001, Keller_2008, Valegard_2021}. We tested the hypothesis of dust aggregates by fitting the JWST mid-IR dust continuum with the DuCK component of the Dust Continuum Kit with Line emission from Gas (DuCKLinG) modeling tool \citep{Kaeufer_2024}. We input the available dust grain opacities for Am Mg-olivine, Am Mg-pyroxene, silica, enstatite and forsterite of 0.1, 1.0, 2.0, 3.0, 4.0, and 5.0 $\mu$m in size. To model the stellar component, we gave in input the stellar parameters derived for XUE 10 (see Section \ref{subsec:results_star}). For the disk, we assumed standard parameter priors for the rim, the midplane and the surface, compatible with modeling ranges and observations of Herbig disks. Nonetheless, the \enquote{bump} is not reproduced by any combination of these common solid-state dust species. A more detailed analysis of the dust component and its mineralogy is beyond the scope of this paper, but it will be the subject of a forthcoming work.
\section{Spectral classification}\label{app:spectral_classification}
\begin{figure*}[htp!]
\centering
    \includegraphics[width=0.65\linewidth]{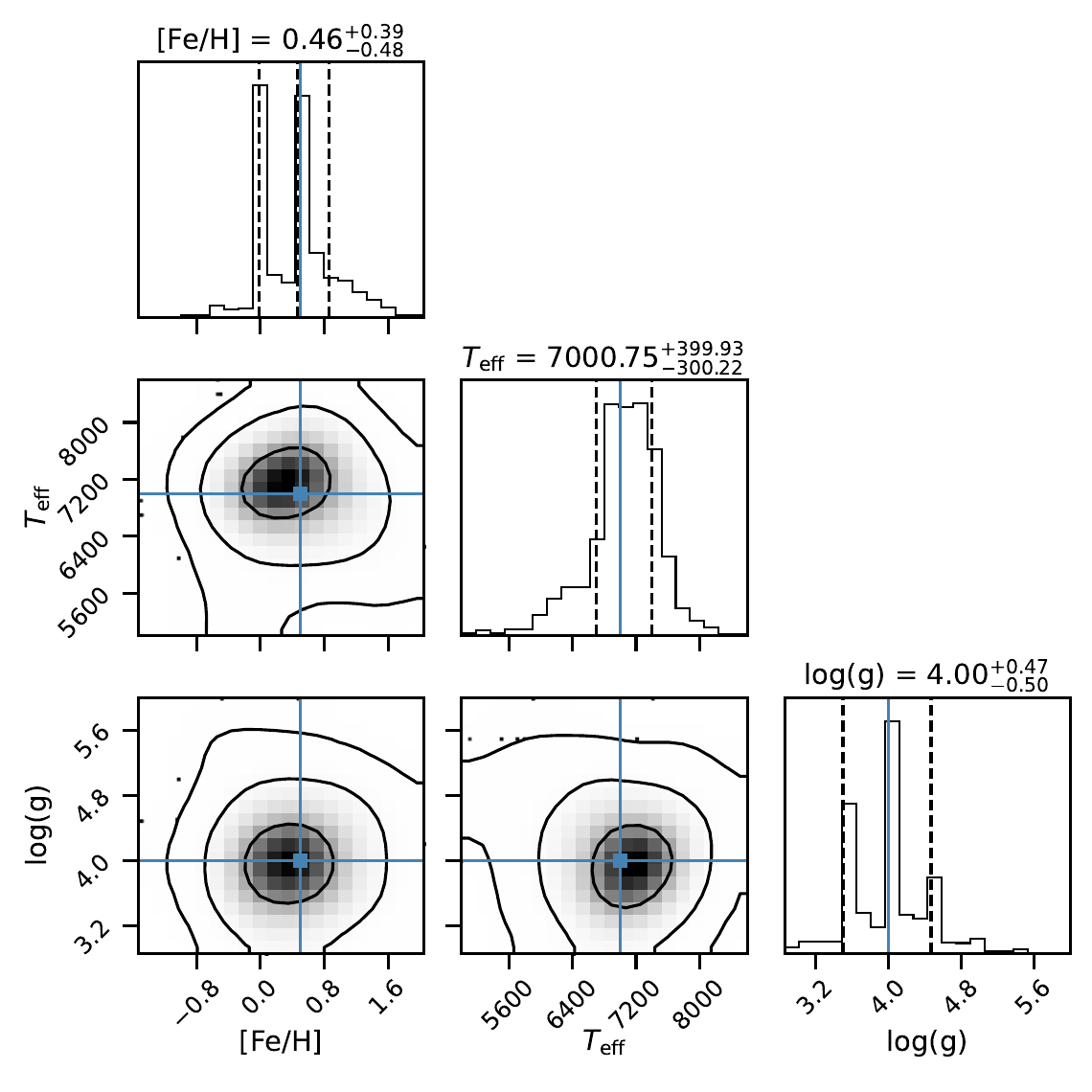}
    \caption{Corner plot showing the one and two smoothed dimensional projection of the posterior probability distribution of $T$$_{\rm eff}$, log(g) and [Fe/H]. The initial positions are drawn with a vertical solid blue line and a blue square respectively in the 1D and 2D projections. Vertical dashed lines in the diagonal histograms mark the 16\%, 50\% and 84\% percentiles. The solid line contours in the 2D projections are drawn at 68\% (1$\sigma$), 95\% (2$\sigma$), and 99\% (3$\sigma$) confidence levels.}\label{fig:cornerplots}
\end{figure*}
The spectral type of the XUE 10 source was determined using the PHOENIX stellar library \citep{Phoenix}, whose spectra cover the 3000--10\,000 $\AA$ range with a spectral resolution of $\Delta\lambda$ = 1 $\AA$ ($R = 100\,000$), which is approximately 4 times higher than in the FORS2 observed spectrum. We therefore degraded the template spectra according to the instrumental resolution of FORS2 in the observed grism setup. The library currently includes nearly 50\,000 spectra of dwarf, giant, and subgiant stars spanning a wide range in atmospheric parameters ($T$$_{\rm eff}$, log(g), [Fe/H], [$\alpha$/Fe]). We did not consider the line broadening due to stellar rotation, nor a possible enhancement of the $\alpha$-elements in the stellar photosphere, whose analysis is beyond the scopes of this work. Within the spectral resolution of the FORS2 observations ($\sim$ 120 km s$^{-1}$), we do not observe significant rotational broadening in the detected absorption lines. 

The adopted grid spans over 5000 $<$ $T$$_{\rm eff}$ $<$ 12\,000 K, in effective temperature, 2.0 $<$ log(g) $<$ 6.0 in surface gravity, and -2.0 $<$ [Fe/H] $<$ +1.0 in metallicity.
In order to control for dependencies from the luminosity class over the grid, we first normalized the spectra. To do so, we created a simple Python routine in which a spectrum is split in two parts, over 7730--8400 $\AA$ and 8400--9480 $\AA$ respectively, before and after the Paschen jump. Each part is individually normalized by subtracting a high-order polynomial, and then the two chunks are stitched together into the final normalized spectrum. We note that this routine has been written and optimized for the sole purpose of this analysis.

We proceeded by comparing the normalized (telluric-corrected) FORS2 spectrum with normalized PHOENIX templates, using the reduced $\chi^2$-test of the form
\begin{equation}\label{Eq. A.1}
\chi^{2}_{\rm red} = \frac{1}{N_{\rm obs}} ~ \sum_{i=1}^{N_{\rm obs}}{ \frac{(F_{\rm obs,i}-F_{\rm mod,i})^{2}}{\sigma^{2}}},
\end{equation}
where $N$$_{\rm obs}$ refers to the number of sampled wavelengths, $F$$_{\rm obs}$ is the observed FORS2 flux, $F$$_{\rm mod}$ the template flux, and $\sigma$ the error on the observed flux as derived from the data reduction pipeline. The best-fit template (see Figure~\ref{fig:FORS_spec}) corresponds to $T$$_{\rm eff}$ = 7000 K, log(g) = 4.0, and [Fe/H] = +0.5, with a minimum reduced chi-square of $\chi^{2}_{\rm red, min}$ = 12. A number of well-known absorption lines are present in the spectrum of XUE 10, whose line profile is well-matched by the best-fit template. In particular, the hydrogen Paschen lines Pa-9 (9229.7 $\AA$), Pa-10 (9015.3 $\AA$), and Pa-11 (8862.9 $\AA$), and the well-known calcium triplet CaII at 8498 $\AA$, 8542 $\AA$, and 8662 $\AA$. 

In order to evaluate the uncertainty on the fit quantities, we performed an MCMC simulation using the Python \texttt{emcee} package \citep{emcee}.  
We set up uniform priors in the adopted PHOENIX grid for the three fit parameters ($T$$_{\rm eff}$, log(g), [Fe/H]). 
We chose the initial positions of the MCMC walkers as the estimated chi-square best-fit values. We imposed the size of the random \enquote{kick} of the walkers from their initial position according to the dynamic range of each parameter space i.e., 500 K in temperature, and 0.5 for surface gravity and metallicity, to ensure a good parameter space sampling. 

We run the simulation with 300 walkers and 100\,000 steps, adjusted to minimize the autocorrelation times, and increase the number of effective samples in a reasonable amount of computational time. From the posterior distribution of samples, we discarded the first 20\,000 samples (20\% of the number of steps), and kept a sample from each walker every 20th step, i.e., thinning the posterior distribution by 20. The corner plot of the simulation is shown in Figure~\ref{fig:cornerplots}. The 1$\sigma$ and 2$\sigma$ 2D contours are well-defined around the initial positions (blue square). The statistical uncertainties ($\pm$) on the fit quantities are reported on top of the diagonal histograms. The median parameter values of the posterior distribution are 7000 K ($T$$_{\rm eff}$), 4.0 (log(g)), and +0.5 ([Fe/H]), in agreement with the chi-square fit within error bars.

According to our analysis, XUE 10 can be classified as a F0-F1 type PMS \citep{Gray_2009}. We combined the spectral information derived with PHOENIX with multi-band photometry to assess visual extinction ($A$$_{\rm V}$), age, stellar mass ($M$$_{\star}$), and bolometric luminosity ($L$$_{\rm bol}$) of XUE 10. The photometric data include PAN-STARRS, SDSS, Gaia, APASS, and HST in the optical range; DENIS, 2MASS, VISTA, and UKIRT in the mid-IR, and \textit{Spitzer} and WISE in the far-IR range. Assuming a total-to-selective extinction factor $R$$_{\rm V}$ = 3.3 \citep{Massi_2015, Russeil_2017, Fouesneau_2022}, we estimate with VOSA a visual extinction of $A$$_{\rm V}$ $\sim$ 4.8. This result is consistent with the $A$$_{\rm V}$ range of 4.6--4.8 derived using the XUE 10 Gaia color ($G$$_{\rm bp}$-$G$$_{\rm rp}$) where a range of spectral types A7--F1 is assumed.
We then compared the PHOENIX best-fit template with PMS isochrones of varying age (0.7, 1, 3, 5, 10 Myr) at solar metallicity and at the constrained $A$$_{\rm V}$, using the PARSEC database \citep{Parsec}\footnote{\url{http://stev.oapd.inaf.it/cgi-bin/cmd}}. We note that PARSEC isochrones are computed by default for an $R$$_{\rm V}$ = 3.1. However, the inferred uncertainties on $T$$_{\rm eff}$ and log(g) are too large to reliably locate XUE 10 on a specific isochrone in this range. XUE 10 is most likely consistent with a 1--3 Myr old PMS in NGC 6357. Starting from the VVV VISTA apparent $J$-band magnitude $m$$_{\rm J}$ = 12.004 $\pm$ 0.001, we derive a bolometric luminosity $L$$_{\rm bol}$ = 69 $\pm$ 5 $L$$_{\odot}$, adopting the bolometric correction in $J$-band for a F1 PMS of \citet{Pecaut_2013}, the constrained $A$$_{\rm V}$ ($\pm$ 0.2 from Gaia), and $R$$_{\rm V}$ = 3.3. Considering a PARSEC isochrone of 3 Myr extincted by a factor 4.8 in visual magnitude, the stellar counterpart of XUE 10 is constrained to a mass of 2.5--3.0 $M$$_{\odot}$.
\section{Fitting tests}\label{app:fitting_tests}
Assuming LTE conditions, the fitting procedure consists in comparing a grid of synthetic spectra computed for different values of ($T$$_{\rm gas}$, $N$$_{\rm tot}$, $R$$_{\rm em}$) with the observed continuum-subtracted spectrum through the reduced chi-square test of the form of Eq. \ref{Eq. A.1} \citep{Tabone_2023}.
Here, $N$$_{\rm obs}$ is the number of MIRI resolution elements covering species features in each fit wavelength window, $F$$_{\rm obs}$ and $F$$_{\rm mod}$ are respectively the observed and the model fluxes to compare, and $\sigma$ is the flux uncertainty estimated in a line-free region closed to the spectral region where the species is modeled (see Figure~\ref{fig:continuum_normalization}). This approach has been extensively used for the spectral fitting of many disk sources so far \citep[see, e.g.,][]{Grant_2023, Perotti_2023, Gasman_2023, Ramirez-Tannus_2023, Schwarz_2024, Temmink_water}, yielding best-fit parameters which are also consistent with the outcome of more sophisticated approaches, e.g., ProDiMo thermo-chemical models \citep{Woitke_2024} and MCMC simulations \citep{Temmink_water}. For all molecules, we run a grid of models varying $N$$_{\rm tot}$ (log$_{10}$-space) in steps of 0.2, $T$$_{\rm gas}$ in steps of 20 K, and $R$$_{\rm em}$ (log$_{10}$-space) in steps of 0.09. The dynamic ranges explored for each parameter are in Table~\ref{tab:fitting_ranges}. Given the limited spectral resolution of the MIRI MRS instrument \citep{Jones_2023}, we convolved the model spectra with a resolution $R$ = 3550 at 4.9--5.2 $\mu$m, $R$ = 3355 at 6.4--7.6 $\mu$m, $R$ = 2600 between 12.9 and 17.6 $\mu$m, and $R$ = 1730 at 18--25 $\mu$m, where we fit the different species. Then, we resampled the convolved models on the wavelength grid of the observations. 
\begin{table*}[htp!]
    \caption{Overview of the parameter ranges used for the LTE slab modeling procedure.}\label{tab:fitting_ranges}
    \centering
    {%
    \begin{tabular}{lccl}
 
\hline
\hline
\noalign{\vskip 1mm}
Species & log$_{10}$($N$$_{\rm tot}$) [cm$^{-2}$] & $T$$_{\rm gas}$ [K] & \multicolumn{1}{l}{Fitting ranges [$\mu$m]}\\
\noalign{\vskip 1mm}
\hline
\noalign{\vskip 0.5mm}
& & &\multicolumn{1}{l}{4.9--5.2 $\mu$m}\\
    \noalign{\vskip 0.5mm}
    \hline
    \noalign{\vskip 0.5mm}
    $^{12}$CO & 14--20 & 100--3400 & 4.903--4.911, 4.914--4.921, 4.924--4.933, 4.941--4.945, 4.976--4.984, 4.988--4.995\\
    & & & 5.116-- 5.120\\
\noalign{\vskip 0.5mm}
\hline
\noalign{\vskip 0.5mm}
& & &\multicolumn{1}{l}{12.9--17.6 $\mu$m}\\
    \noalign{\vskip 0.5mm}
    \hline
    \noalign{\vskip 0.5mm}
    $^{12}$CO$_2$ & 15--23 & 100--1200 & 13.045--13.055, 13.068--13.081, 13.096--13.103, 13.121--13.129, 13.147--13.154\\
    &&& 13.170--13.180, 13.196--13.238, 13.274--13.318, 13.466--13.554, 13.606--13.614\\
    &&& 13.635--13.642, 13.664--13.671, 13.743--13.802, 13.870--14.032, 14.114--14.122\\
    &&& 14.269--14.276, 14.368--14.525, 14.549--14.554, 14.607--14.625, 14.683--14.689\\
    &&& 14.718--14.724, 14.795--14.984, 15.121--15.133, 15.153--15.173, 15.191--15.207\\
    &&& 15.229--15.252, 15.281--15.346, 15.431--15.485, 16.178--16.290, 16.402--16.415\\
    &&& 16.444--16.454, 16.489--16.497, 16.736--16.803, 16.931--16.939, 16.974--16.986\\
    &&& 17.115--17.129, 17.162--17.169, 17.181--17.215, 17.395--17.409\\
    \noalign{\vskip 0.8mm}
    $^{13}$CO$_2$ & 15--23 & 100--1200 &12.961--12.966, 13.510--13.551, 13.852--13.940\\ 
    &&&15.268--15.425, 15.594--15.698, 15.733--15.761\\ 
    &&&15.789--15.858, 15.983--16.001, 16.192--16.224\\ 
    &&&16.255--16.266, 16.291--16.307, 16.361--16.396\\ 
    &&&16.417--16.427, 16.574--16.597, 16.670--16.709\\
    \noalign{\vskip 0.8mm}
    $^{16}$O$^{12}$C$^{18}$O & 15--23 & 100--1200 & 13.802--13.832, 14.209--14.272, 15.048--15.100, 16.749--16.835, 17.335--17.361\\
    \noalign{\vskip 0.8mm}
    $^{16}$O$^{12}$C$^{17}$O & 15--23 & 100--1200 &14.056--14.102, 14.979--15.044, 16.455--16.536\\ 
    \noalign{\vskip 1mm}
    \hline
    \end{tabular}
    }
    \begin{tablenotes}
    \footnotesize
    \item \textbf{Notes.} The equivalent emitting radius ($R$$_{\rm em}$) range is 0.01--10 au in log$_{10}$-space for all species.
    \end{tablenotes}
    \end{table*}
\section{Reduced chi-square maps}\label{app:chi-square}
\begin{figure*}
\centering
    \includegraphics[width=0.88\textwidth]{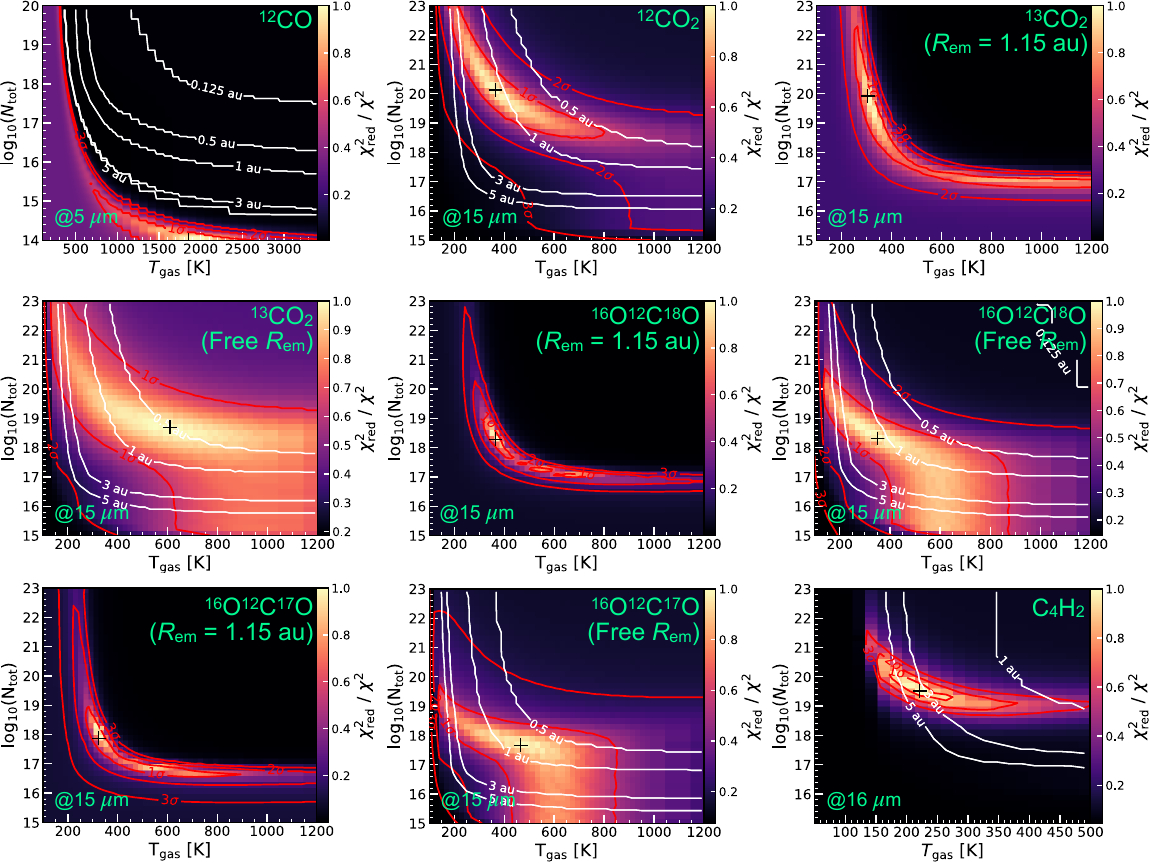}
    \caption{Reduced chi-square $N$$_{\rm tot}$--$T$$_{\rm gas}$ maps resulting from the fitting of 0D LTE gas slab models to the MIRI MRS spectrum for various molecular species at increasing wavelength.
    The fit of the carbon dioxide isotopologues, $^{13}$CO$_2$, $^{16}$O$^{12}$C$^{18}$O, and $^{16}$O$^{12}$C$^{17}$O with fixed/free emitting radius are labeled in figure. In colorbar are the $\chi^{2}_{red}/\chi^{2}$ values, with 1.0 corresponding to the best-fit model. The white contours show the range of fit emitting radii in au, while the red contours show the 1$\sigma$, 2$\sigma$, and 3$\sigma$ confidence intervals. The location of the best-fit models in the parameter space is indicated by a black cross marker.}
    \label{fig:chisquaremaps}
\end{figure*}

The $\chi^{2}_{red}$ maps for all fit species are shown in Figure~\ref{fig:chisquaremaps} in order of increasing wavelength in the spectrum. 
The results for $^{13}$CO$_2$, $^{16}$O$^{12}$C$^{18}$O, and $^{16}$O$^{12}$C$^{17}$O are reported assuming the best-fit emitting radius found for $^{12}$CO$_2$ ($R$$_{\rm em}$ = 1.15 au), and leaving $R$$_{\rm em}$ as a free parameter (see Section \ref{subsec:isotopic_ratios}).
\end{appendix}
\end{document}